\documentclass[prd,preprint,tightenlines,showpacs,preprintnumbers,nofootinbib,eqsecnum,superscriptaddress]{revtex4-1}

 \usepackage[dvips,final]{graphicx}
  \usepackage{amssymb}
   \usepackage{amsmath}
    \usepackage{amsfonts}
     \usepackage{epsfig}
      \usepackage{bm}% bold math
      \usepackage[dvipsnames,usenames]{color}
      \usepackage{comment}
       \usepackage[utf8]{inputenc}
        \usepackage{soul}

\usepackage[section]{placeins}

\usepackage{sidecap}
\usepackage{multirow}
\usepackage{booktabs}
\usepackage{array}
\usepackage{tabularx}
\usepackage{xcolor}
\usepackage{pstricks}
\def\frac#1#2{{\begingroup #1\endgroup\over #2}}
\def\Pom{\mathrm{I\!P}}

\newcommand{\half}{{1\over 2}}

\begin{document}

%%%%%%%%%%%%%%%%%%%%%%%%%%%%%%%%%%%%%%%%%%%%%%%%%%%%%%%%%%%%%%%%%%%%%%%%%%%
\title{Probing proton structure with $c \bar c$ correlations %\\ 
in ultraperipheral $pA$ collisions}
%%%%%%%%%%%%%%%%%%%%%%%%%%%%%%%%%%%%%%%%%%%%%%%%%%%%%%%%%%%%%%%%%%%%%%%%%%%

\author{Barbara Linek}
\email{barbarali@dokt.ur.edu.pl}
\affiliation{University of Rzeszow, ul. Pigonia 1, PL-35-959 Rzeszow, Poland}
\author{Agnieszka Łuszczak}
\email{Agnieszka.Luszczak@pk.edu.pl}
\affiliation{Cracow University of Technology, Department of Physics, PL-30-084 Krak\'ow, Poland}
\author{Marta Łuszczak}
\email{mluszczak@ur.edu.pl}
\affiliation{University of Rzeszow, ul. Pigonia 1, PL-35-959 Rzeszow, Poland}
\author{Roman Pasechnik}
\email{Roman.Pasechnik@hep.lu.se}
\affiliation{Department of Physics, Lund University, SE-223 62 Lund, Sweden}
\author{Wolfgang Sch\"afer}
\email{Wolfgang.Schafer@ifj.edu.pl} 
\affiliation{Institute of Nuclear
Physics, Polish Academy of Sciences, ul. Radzikowskiego 152, PL-31-342 
Krak{\'o}w, Poland}
\author{Antoni Szczurek}
\email{antoni.szczurek@ifj.edu.pl}
\affiliation{Institute of Nuclear Physics, Polish Academy of Sciences, 
ul. Radzikowskiego 152, PL-31-342 Krak{\'o}w, Poland}

\begin{abstract}
%\vspace{0.9cm} 
We study the exclusive diffractive $c \bar c$ photoproduction in ultraperipheral $pA$ collisions. The formalism makes use of off-diagonal generalizations of the unintegrated gluon distribution, the so-called generalized transverse momentum dependent distributions (GTMDs). We present two different formulations. The first one is based directly on gluon GTMD parametrizations in momentum space. Another option is the calculation of the GTMD as a Fourier transform of the dipole-nucleon scattering amplitude $N(Y,\vec{r}_{\perp},\vec{b}_{\perp})$. The latter approach requires some extra regularization discussed in the paper. Different dipole amplitudes from the literature are used. Compared to previous calculations in the literature, we integrate over the full phase space and therefore cross sections for realistic conditions are obtained. We present distributions in rapidity of $c$ or $\bar c$, transverse momentum of the $c \bar c$ pair, four-momentum transfer squared as well as the azimuthal correlation between a sum and a difference of the $c$ and $\bar c$ transverse momenta. The azimuthal correlations are partially due to the so-called elliptic gluon Wigner distribution. Different models lead to different modulations in the azimuthal angle. The modulations are generally smaller than 5\%. They depend on the range of transverse momentum selected for the calculation.
\end{abstract}
\maketitle

\section{Introduction}

%%%% 
 \begin{figure}
  \centering
  \includegraphics[width=.242\textwidth]{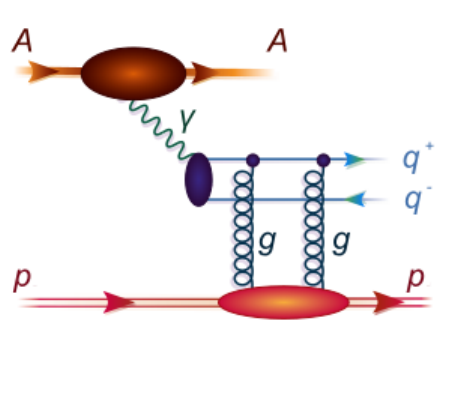}
  \includegraphics[width=.242\textwidth]{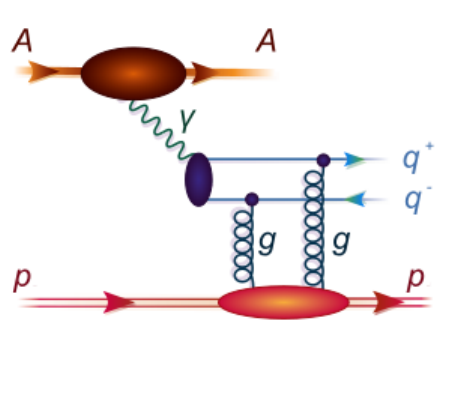}
  \includegraphics[width=.242\textwidth]{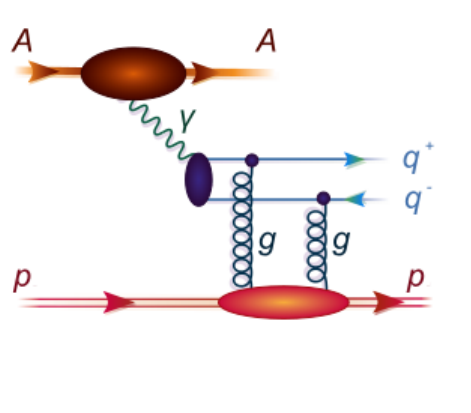}
  \includegraphics[width=.242\textwidth]{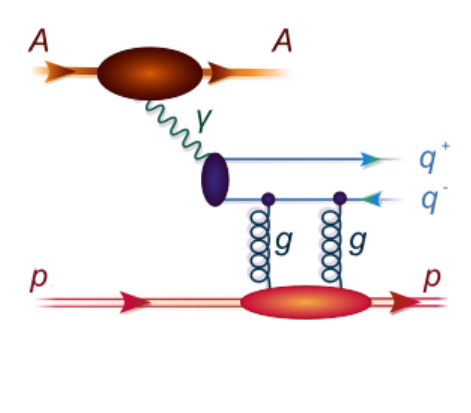}  
  \caption{Feynman diagrams for the diffractive photoproduction of $q \bar q$ pairs in nucleus-proton collisions, discussed in the present paper.}
\label{fig:diagrams}
\end{figure}
%%%%

The diffractive production of high momentum particles, such as quark-antiquark dijets, serves as a probe of various aspects of the hadron structure \cite{Ashery:2006zw}. For example, in the diffractive dissociation of photons or pions into dijets, the forward cross section at large jet transverse momenta maps out the unintegrated gluon distribution of the target \cite{Nikolaev:1994cd,Nikolaev:2000sh}. Recently, there has been much interest in a generalization of unintegrated (or transverse momentum dependent) parton distributions to a five-dimensional quasi-probability phase space distribution, known as the Wigner distribution \cite{Ji:2003ak,Belitsky:2003nz}, which depends on both the transverse momentum and impact parameter of a parton in the proton or nucleus. When transformed fully to momentum space these are equivalent to the generalized transverse momentum distributions (GTMDs), see e.g. \cite{Meissner:2009ww,Lorce:2013pza,Boussarie:2023izj}. In particular, these distributions encode the dependence on the transverse momentum transfer $\vec \Delta_\perp$ to the target and, therefore, are of relevance for the description of the forward cone in diffractive processes. 

The Wigner distribution depends on the transverse momentum $\vec k_\perp$ of partons as well as the impact parameter $\vec b_\perp$, including a dependence on the azimuthal angle between the two transverse vectors. In the GTMD approach this translates into a dependence on $\vec k_\perp$ and $\vec \Delta_\perp$. This dependence gives rise to a so-called elliptic Wigner function or GTMD \cite{Hagiwara:2016kam}. In the region of small-$x$ which is of interest in this paper, the equivalent color dipole approach for diffractive processes \cite{Kopeliovich:1981pz,Nikolaev:1991et} can be used. Here, the dipole amplitude depending on dipole size $\vec r_\perp$ and impact parameter $\vec b_\perp$ contains the same information as the Wigner function/GTMD. The azimuthal angle correlation discussed above translates then into a dependence on the dipole orientation with respect to the background color field of the target. In inclusive processes this dependence can give rise to flow-like azimuthal correlations, for example in prompt photons production \cite{Kopeliovich:2007fv}, or in inclusive two-particles' production \cite{Iancu:2017fzn}.

It was advocated that the elliptic gluon distributions could be studied in diffractive reactions, such as:
\begin{itemize}

\item Exclusive dijet production in $e p$ collisions \cite{Hatta:2016dxp,Boer:2021upt}. For calculations in the color dipole approach, see \cite{Altinoluk:2015dpi,Mantysaari:2019csc,Salazar:2019ncp};

\item Exclusive dijet photoproduction in $p A$ and $AA$ ultra-peripheral collisions (UPCs) \cite{Hagiwara:2017fye};

\item Exclusive $Q \bar Q$ (with $Q=c,b$) photoproduction in $p A$ and $AA$ UPCs \cite{ReinkePelicer:2018gyh}.

\end{itemize}
Here, we revisit the exclusive production of $c \bar c$ pairs in proton-lead collisions at LHC energies. The dominant reaction mechanism here is the diffractive photoproduction of the $c \bar c$ pair on a proton by a Weizs\"acker-Williams photon emitted by the lead nucleus, see the diagrams in Fig.~\ref{fig:diagrams}.
One is then interested in the correlations of
%%%
\begin{eqnarray}
    \vec P_\perp = \half( \vec p_{\perp Q} - \vec p_{\perp \bar Q} )\, ,
\end{eqnarray}
%%%
and the momentum transfer to the proton, which for the exclusive reaction fulfills
%%%
\begin{eqnarray}
    \vec \Delta_\perp = \vec p_{\perp Q} + \vec p_{\perp \bar Q} \, . 
\end{eqnarray}
%%%
In this work we account only for the contribution where the ion serves as a source of the photon. 
Here we exploit the fact, that the photon flux scales with the charge of the ion as $Z^2$. 
This factor will be missing when the photon is emitted by the proton. On the other hand, the integrated diffractive process on the nucleus will display a dependence $\sigma \propto A^\alpha$, with 
roughly $\alpha \sim 1.4 \div 1.5$ (see e.g. \cite{Ashery:2006zw,Nikolaev:2000sh}). While the diffractive photoproduction on the nucleus may therefore still be a sizeable effect, the extraction of the process discussed in this paper is possible if proton tagging \cite{Staszewski:2023chz,Trzebinski:2019tmk,Bossini:2023uwa} is available, due to the very sharp transverse momentum dependence of the electromagnetically scattered proton.

In distinction to dijet production, for the heavy quarks already the quark mass $m_Q$ can play the role of a hard scale, and also intermediate and small values of $\vec P_\perp$ are tractable. Here, we will calculate cross sections and distributions for realistic kinematic conditions at the LHC. We also discuss some subtleties regarding the correlations in the azimuthal angle
%%%
\begin{eqnarray}
\cos \phi  = \frac{\vec P_\perp \cdot \vec \Delta_\perp}{P_\perp \Delta_\perp} \, .
\end{eqnarray}
%%%

At large jet momenta $P_\perp$ the correlations in $\phi$ expected to be dominantly $\propto \cos(2 \phi)$ and are unambiguously expressed in terms of the elliptic Wigner function/GTMD.
At lower $P_\perp$, however, the azimuthal correlations are better understood as coming from the matrix element (or impact factor) at finite $\Delta_\perp$. While formally also these correlations can be absorbed into the Wigner function/GTMD, these would correspond to $\delta$-function terms in the latter. We discuss the relation between the different approaches and the alternative formulation in terms of a off-forward gluon density matrix introduced in Ref.~\cite{Nemchik:1997xb}.

This paper is organized as follows. In Section~\ref{sec:Formalism}, we explain how to calculate the diffractive cross section of interest and discuss the formalism starting from the dipole representation of the amplitude. Then, in Section~\ref{sec:distributions} we present our numerical results for differential cross sections as functions of various kinematical variables using a variety of the existing gluon GTMD models. Finally, in Section~\ref{sec:conclusions} the main conclusions are summarised.

%%%%%%%%%%%%%%%%%%%%%%%%%
\section{Formalism}
\label{sec:Formalism}
%%%%%%%%%%%%%%%%%%%%%%%%%

%------------------------------------------------------------------------------------
\subsection{Kinematics and cross section}
%------------------------------------------------------------------------------------

The cross section for the proton-nucleus reaction can be written in the following form
%%%%
\begin{eqnarray}
{d \sigma (p A \to Q \bar Q p A; s) \over dx_Q dx_{\bar Q}  d^2\vec P_\perp d^2 \vec \Delta_\perp} 
= {1 \over x_Q + x_{\bar Q}} f_{\gamma/A}(x_Q + x_{\bar Q}) 
\, {d \sigma (\gamma p \to Q \bar Q p; (x_Q + x_{\bar Q})s) \over dz d^2\vec P_\perp 
d^2 \vec  \Delta_\perp} \, ,
\end{eqnarray}
%%%%
with $z = x_Q/(x_Q + x_{\bar Q})$. Here, $x_Q, x_{\bar Q} $ are the fractions of the nucleus' Light-Front plus momentum carried by a heavy quark $Q$ and antiquark $\bar Q$ (with mass $m_Q$), respectively. The transverse momenta of quark and antiquark are
%%%%
\begin{eqnarray}
    \vec p_{\perp Q} = \vec P_\perp + \frac{\vec \Delta_\perp}{2} \, , \qquad 
    \vec p_{\perp \bar Q} = - \vec P_\perp + \frac{\vec \Delta_\perp}{2} \, , 
\end{eqnarray}
%%%
respectively, so that $\vec \Delta_\perp$ is the transverse momentum of the $Q \bar Q$ pair. As the photon is collinear to the incoming nucleus, $\vec \Delta_\perp$ is also equal, up to a sign, to the transverse momentum transfer to the proton target.

The Weizs\"acker-Williams photons carry the momentum fraction
%%%
\begin{eqnarray}
    x_A = x_Q + x_{\bar Q} \, . 
\end{eqnarray}
For the flux of quasireal photons,
%%%%
\begin{eqnarray}
f_{\gamma/A}(x_A) = {dN(x_A) \over dx_A} \, ,
\end{eqnarray}
%%%%
we use the well-known expression (see for example the review \cite{Baur:2001jj}),
%%%%
\begin{eqnarray}
\frac{dN(x_A)}{d x_A}=\frac{2Z^{2}\alpha_{\rm em}}{\pi x_A} \left[ \xi_{jA}K_{0}(\xi_{jA})K_{1}(\xi_{jA})-\frac{\xi_{jA}^2}{2}(K_{1}^{2}(\xi_{jA})-K_{0}^{2}(\xi_{jA})) \right],
\end{eqnarray}
where $Z$ correspond to the atomic number of the projectile particle, $\alpha_{\rm em}$ is the fine structure constant, $\xi_{jA}=x_A m_p (R_j+R_A)$ involves the target and nucleus radii ($R_j$ and $R_A$, respectively) and effectively excludes the overlap of the projectile and the target in impact parameter space, and $m_p$ is the proton mass.

Let us also quote a useful expression for the cross section which reads in terms of center-of-mass rapidities $y_Q, y_{\bar Q}$ of quarks,
%%%%
\begin{eqnarray}
{d \sigma (p A \to Q \bar Q p A; s) \over dy_Q dy_{\bar Q}  d^2\vec P_\perp d^2 \vec \Delta_\perp} = x_A {dN(x_A) \over dx_A}  \, \Big( z(1-z) {d \sigma (\gamma p \to Q \bar Q p; x_A s) \over dz d^2\vec P_\perp d^2 \vec  \Delta_\perp} \Big)\Big|_{z = {x_Q \over x_A}} \, ,
\end{eqnarray}
%%%%
where
%%%%
\begin{eqnarray}
x_Q = {\sqrt{p_{\perp Q}^2 + m_Q^2 \over s}} \, \exp(y_Q) , \qquad x_{\bar Q} = {\sqrt{p_{\perp \bar Q}^2 + m_Q^2 \over s}} \, \exp(y_{\bar Q}) \,.
\end{eqnarray}
%%%%
We also need the transverse mass of the $Q \bar Q$--pair, which square is given by
%%%%
\begin{eqnarray}
    M_\perp^2 = x_A
    \Big( \frac{p^2_{\perp Q} + m_Q^2}{x_Q} + 
    \frac{p^2_{\perp \bar Q} + m_Q^2}{x_{\bar Q}} \Big) \, .
\end{eqnarray}
%%%%
The rapidity of the $Q \bar Q$--pair in the cm-frame can then be calculated from
%%%%
\begin{eqnarray}
    Y_{\rm pair} = \half \log\Big( \frac{x_A}{x_B}\Big) =
    \log\Big( \frac{x_A \sqrt{s}}{M_\perp} \Big) \, , \qquad x_A x_B s = M^2_\perp.
\end{eqnarray}
%%%%
We will be also interested in the laboratory frame rapidities, which are obtained from the shift
%%%%
\begin{eqnarray}
    y_{Q,\bar Q}^{\rm LAB} = y_{Q, \bar Q} + \half \log\Big( \frac{Z}{A} \Big) \, , \quad Y_{\rm pair}^{\rm LAB} = Y_{\rm pair} + \half \log\Big( \frac{Z}{A} \Big),
\end{eqnarray}
%%%
where for the $^{208}{\rm Pb}$ nucleus, we have $A=208, Z=82$.
%------------------------------------------------------------------------------------
\subsection{Color dipole representation of the diffractive amplitude}
%------------------------------------------------------------------------------------

We start our discussion from a basic description of the color dipole approach to diffractive processes \cite{Kopeliovich:1981pz,Nikolaev:1991et}. Here, the cross section for the $\gamma p \to Q \bar Q$ diffractive dissociation process is written as
%%%
\begin{eqnarray}
    {d \sigma (\gamma p \to Q \bar Q p; s_{\gamma p}) \over dz d^2\vec P_\perp d^2 \vec  \Delta_\perp} =\overline{\sum_{\lambda_\gamma, \lambda, \bar \lambda}}
    \Big|\int  \frac{d^2 \vec b_\perp d^2\vec r_\perp}{(2\pi)^2}
     e^{-i \vec \Delta_\perp \cdot \vec b_\perp} 
    e^{-i \vec P_\perp \cdot \vec r_\perp} 
    N(Y,\vec r_\perp, \vec b_\perp) \, \Psi^{\lambda_\gamma}_{\lambda \bar \lambda}(z,\vec r_\perp) \, \Big|^2 \, . \nonumber 
    \\ \label{eq:diffractive_xsec}
\end{eqnarray}
%%%
Above, $z, 1-z$ are the Light-Front momentum fractions carried by quark/antiquark in the $\gamma \to Q \bar Q$ transition. The corresponding Light-Front wave function for the $\gamma \to Q \bar Q$,
%%%
\begin{eqnarray}
\Psi^{\lambda_\gamma}_{\lambda \bar \lambda}(z,\vec r_\perp) = \frac{1}{\sqrt{4 \pi z(1-z)}} \int \frac {d^2 \vec l_\perp}{(2 \pi)^2} \, e^{i \vec r_\perp \cdot \vec l_\perp} \, \Psi^{\lambda_\gamma}_{\lambda \bar \lambda}(z,\vec l_\perp) \, ,
\end{eqnarray}
%%%
depends on the Light-Front helicities of quarks, $\lambda/2, \bar \lambda/2$ and photon, $\lambda_\gamma$. Its explicit form in transverse momentum space can be found for example in Ref.~\cite{Kovchegov.2012}.

Our main interest is in the dipole scattering amplitude $N(Y,\vec r_\perp, \vec b_\perp)$. Its energy dependence is encoded through the ``rapidity'' $Y$. We define the latter as $Y = \ln(x_0/x_\Pom)$, with $x_0 = 0.01$. Here,
%%%
\begin{eqnarray}
x_\Pom \equiv x_B = {M_\perp^2 \over s_{\gamma p}} = {M_\perp^2 \over x_A s} \, ,
\end{eqnarray}
%%%%
where $M_\perp$ is the transverse mass of the $Q \bar Q$ system. The dipole amplitude is related to the familiar color dipole cross section  (see e.g. the textbook \cite{Barone:2002cv}) via
%%%%
\begin{eqnarray}
    \sigma(x_\Pom,\vec r_\perp) = 2 \int d^2\vec b _\perp \, N(Y,\vec r_\perp, \vec b_\perp) \, .
    \label{eq:dipole_xsec}
\end{eqnarray}
%%%%
It is related to an off-diagonal generalization of the unintegrated gluon distribution -- a gluon density matrix through the relation \cite{Nemchik:1997xb}:
%%%
\begin{eqnarray}
     N(Y,\vec r_\perp, \vec b_\perp) &=& \int d^2\vec q_\perp d^2 \vec \kappa_\perp \, f\Big(Y,\frac{\vec q_\perp}{2} + \vec \kappa_\perp, \frac{\vec q_\perp}{2} - \vec \kappa_\perp\Big) \exp[i \vec q_\perp \cdot \vec b_\perp]  \nonumber \\
    &\times&\Big\{ \exp\Big[i \half \vec q_\perp \cdot \vec r_\perp\Big] + \exp\Big[-i \half \vec q_\perp \cdot \vec r_\perp\Big] - \exp[i \vec \kappa_\perp \cdot \vec r_\perp] - \exp[-i \vec \kappa_\perp \cdot \vec r_\perp] \Big\} \, . \nonumber \\
    \label{eq:dipole_vs_GTMD}
\end{eqnarray}
%%%
Quark and antiquark move at impact parameters, 
%%%
\begin{eqnarray}
\vec b_{\perp Q} = \vec b_\perp + \frac{\vec r_\perp}{2} \, , \qquad 
\vec b_{\perp \bar Q} = \vec b_\perp - \frac{\vec r_\perp}{2} \, , 
\end{eqnarray}
%%%
which allow us to relate the four phase factors in the curly bracket in Eq.~(\ref{eq:dipole_vs_GTMD})
to the four diagrams of Fig.~\ref{fig:diagrams} in an obvious fashion. We write the unintegrated gluon density matrix in the following form,
%%%
\begin{eqnarray}
 f\Big(Y,\frac{\vec q_\perp}{2} + \vec \kappa_\perp, \frac{\vec q_\perp}{2} - \vec \kappa_\perp\Big) = \frac{\alpha_s}{4 \pi N_c} \, \frac{{\mathcal F}\Big(x_\Pom, \frac{\vec q_\perp}{2} + \vec \kappa_\perp, \frac{\vec q_\perp}{2} - \vec \kappa_\perp\Big)}
 {(\frac{\vec q_\perp}{2} + \vec \kappa_\perp)^2
 (\frac{\vec q_\perp}{2} - \vec \kappa_\perp)^2} \, ,
 \label{eq:f_Y}
\end{eqnarray}
%%%
where $N_c=3$ for the number of colors in QCD, and we put in evidence the strong coupling constant $\alpha_s$ and gluon propagators here.

Integrating Eq.~(\ref{eq:dipole_vs_GTMD}) over $\vec b_\perp$, we obtain for the dipole cross section the well-known representation \cite{Nikolaev:1994ce}
%%%
\begin{eqnarray}
\sigma(x_\Pom,\vec{r}_\perp) = \frac{2 \pi}{N_c} \, \int \frac{d^2 \vec \kappa_\perp}{\kappa_\perp^4} \alpha_s {\mathcal F}(x_\Pom, \vec \kappa_\perp, - \vec\kappa_\perp) \, \Big\{ 2 - e^{i \vec \kappa_\perp \cdot \vec r_\perp} -
e^{-i\vec \kappa_\perp \cdot \vec r_\perp} \Big\} \, ,
\label{eq:dipole_xsec}
\end{eqnarray}
%%%
so that indeed $\mathcal{F}(x,\vec \kappa_{\perp 1}, \vec \kappa_{\perp 2})$ is the proper off-forward generalization of the standard unintegrated gluon distribution, which relates to its collinear counterpart via
%%%
\begin{eqnarray}
    x g(x,\mu^2) = \int \frac{d^2 \vec \kappa_\perp}{\pi \kappa_\perp^2} \, \theta(\mu^2 - \kappa_\perp^2) \, {\mathcal F}(x, \vec \kappa_\perp, - \vec\kappa_\perp) \, .
\end{eqnarray}
%%%
Below, we will also use the non-perturbative parameter,
%%%%
\begin{eqnarray}
    \sigma_0(x_\Pom) = \frac{4\pi}{N_c} \int \frac{d^2 \vec \kappa_\perp}{\kappa_\perp^4} \alpha_s {\mathcal F}(x_\Pom, \vec \kappa_\perp, - \vec\kappa_\perp) \, ,
    \label{eq:sigma_0}
\end{eqnarray}
%%%%
which has the interpretation of the dipole cross section for large dipoles.

Now, inserting the representation of the dipole amplitude given in Eq.~(\ref{eq:dipole_vs_GTMD}) into Eq.~(\ref{eq:diffractive_xsec}), we obtain for the diffractive amplitude the following convolution structure:
%%%
\begin{eqnarray}
{\mathcal A}(Y,\vec P_\perp, \vec \Delta_\perp)  &\propto&   \int d^2\vec \kappa_\perp \, f\Big(Y, \frac{\vec \Delta_\perp}{2} + \vec \kappa_\perp, \frac{\vec \Delta_\perp}{2} - \vec \kappa_\perp\Big) 
  \, \Big\{ \Psi_{\lambda \bar \lambda}^{\lambda_\gamma}\Big(z,\vec P_\perp + \frac{\vec \Delta_\perp}{2}\Big) + \Psi_{\lambda \bar \lambda}^{\lambda_\gamma}\Big(z,\vec P_\perp - \frac{\vec \Delta_\perp}{2}\Big)
  \nonumber \\
  &-& \Psi_{\lambda \bar \lambda}^{\lambda_\gamma}(z,\vec P_\perp + \vec \kappa_\perp) - 
  \Psi_{\lambda \bar \lambda}^{\lambda_\gamma}(z,\vec P_\perp - \vec \kappa_\perp)
  \Big\} \,. \nonumber \\
  \label{eq:amplitude_structure}
\end{eqnarray}
%%%
Here, in the terminology of small-$x$ (or BFKL) factorization, one would refer to the structure in brackets as the impact factor for the coupling of two off-shell gluons to the $\gamma \to Q \bar Q$ amplitude. The two $t$-channel gluons carry the transverse momenta,
%%%
\begin{eqnarray}
    \vec \kappa_{\perp 1} = \frac{\vec \Delta_\perp}{2} + \vec \kappa_\perp , \quad \vec \kappa_{\perp 2} = \frac{\vec \Delta_\perp}{2} - \vec \kappa_\perp  \,,
\end{eqnarray}
%%%
and the impact factor has the property that it vanishes when either of the gluon transverse momenta goes to zero, i.e.~for $\vec \kappa_\perp = \pm \vec \Delta_\perp/2$.

Finally, we obtain for our diffractive photoproduction cross section:
%%%
\begin{eqnarray}
       {d \sigma (\gamma p \to Q \bar Q p; s_{\gamma p}) \over dz d^2\vec P_\perp d^2 \vec  \Delta_\perp} = e_f^2 \alpha_{\rm em} \, 2 N_c (2 \pi)^2 \Big\{ 
       \Big(z^2 + (1-z)^2 \Big) \Big| {\vec {\mathcal M}_0} \Big|^2 + m_Q^2 \Big| { {\mathcal M}_1} \Big|^2
       \Big\} \,.
\end{eqnarray}
Here, ${\mathcal M}_1$ and $\vec {\mathcal M}_0$ are amplitudes for the sum of quark helicities equal to one or zero, respectively. Explicitly, we find (see e.g.~Ref.~\cite{Nikolaev:1998wf}):
%%%%
\begin{eqnarray}
\mathcal{\vec M}_{0}(\vec P_\perp ,\vec \Delta_{\perp}) &=&  \int \frac{d^2\vec k_\perp}{2\pi} \, f\Big(Y,\frac{\vec \Delta_\perp}{2} + \vec k_\perp, \frac{\vec \Delta_\perp}{2} - \vec k_\perp\Big)
\Big\{ 
\frac{\vec P_\perp - \vec \Delta_\perp/2 } 
{(\vec P_\perp - \vec \Delta_\perp/2)^{2}+m_Q^2}
\nonumber \\
&+& \frac{\vec P_\perp + \vec \Delta_\perp/2 }{(\vec P_\perp + \vec \Delta_\perp/2)^{2}+m_Q^2} 
- \frac{\vec P_\perp -\vec k_\perp}{(\vec P_\perp -\vec k_\perp)^{2}+m_Q^2} - \frac{\vec P_\perp + \vec k_\perp}{(\vec P_\perp +\vec k_\perp)^{2}+m_Q^2} \Big\}
\, , \nonumber \\
\mathcal{M}_{1}(\vec P_\perp ,\vec \Delta_{\perp}) &=&  \int \frac{d^2\vec k_\perp}{2\pi} \, f\Big(Y,\frac{\vec \Delta_\perp}{2} + \vec k_\perp, \frac{\vec \Delta_\perp}{2} - \vec k_\perp\Big)
\Big\{ 
\frac{1} 
{(\vec P_\perp - \vec \Delta_\perp/2)^{2}+m_Q^2}
\nonumber \\
&+& \frac{1}{(\vec P_\perp + \vec \Delta_\perp/2)^{2}+m_Q^2} 
- \frac{1}{(\vec P_\perp -\vec k_\perp)^{2}+m_Q^2} - \frac{1}{(\vec P_\perp +\vec k_\perp)^{2}+m_Q^2} \Big\} \,.
\label{eq:diffractive_amplitudes}
\end{eqnarray}
%%%%%

In order to understand the origin of azimuthal correlations, it is useful to decompose our amplitude. To this end, let us introduce
%%%
\begin{eqnarray}
    \vec {\mathcal J}_0(\vec P_\perp, \vec q_\perp) &=& \frac{\vec P_\perp - \vec q_\perp}{ (\vec P_\perp - \vec q_\perp)^2 + m_Q^2} +\frac{\vec P_\perp + \vec q_\perp}{ (\vec P_\perp + \vec q_\perp)^2 + m_Q^2} -  \frac{2 \vec P_\perp }{\vec P_\perp^2 + m_Q^2},
    \nonumber \\
    {\mathcal J}_1(\vec P_\perp, \vec q_\perp) &=& \frac{1}{ (\vec P_\perp - \vec q_\perp)^2 + m_Q^2} +\frac{1}{ (\vec P_\perp + \vec q_\perp)^2 + m_Q^2} -  \frac{2 }{\vec P_\perp^2 + m_Q^2} \,.
\end{eqnarray}
%%%
Then, our matrix elements take the form:
%%%
\begin{eqnarray}
\vec {\mathcal M}_0(\vec P_\perp, \vec \Delta_\perp) &=&  \vec {\mathcal J_0}(\vec P_\perp, \half \vec \Delta_\perp) \, 
    C(Y,\vec \Delta_\perp) 
 - \int \frac{d^2 \vec k_\perp}{2 \pi} \vec {\mathcal J_0}(\vec P_\perp, \vec k_\perp) \, f\Big(Y,\frac{\vec \Delta_\perp}{2} + \vec k_\perp, \frac{\vec \Delta_\perp}{2} - \vec k_\perp\Big)
  \, , \nonumber \\
    {\mathcal M}_1(\vec P_\perp, \vec \Delta_\perp) &=&  {\mathcal J_1}(\vec P_\perp, \half \vec \Delta_\perp) \, 
    C(Y,\vec \Delta_\perp) 
 - \int \frac{d^2 \vec k_\perp}{2 \pi} {\mathcal J_1}(\vec P_\perp, \vec k_\perp) \, f\Big(Y,\frac{\vec \Delta_\perp}{2} + \vec k_\perp, \frac{\vec \Delta_\perp}{2} - \vec k_\perp\Big)
  \, . \nonumber \\
\end{eqnarray}
%%%%
There emerges a rapidity-dependent form factor,
%%%%
\begin{eqnarray}
    C(Y,\vec \Delta_\perp) = \int \frac{d^2 \vec k_\perp}{2 \pi} \,f\Big(Y,\frac{\vec \Delta_\perp}{2} + \vec k_\perp, \frac{\vec \Delta_\perp}{2} - \vec k_\perp\Big) \, ,
    \label{eq:C_Y_delta}
\end{eqnarray}
%%%
which is a non-perturbative parameter, as is obvious from its form in the forward limit as an integral,
%%%
\begin{eqnarray}
    C(Y,0) = \frac{1}{4\pi N_c}\int \frac{d^2 \vec k_\perp}{2 \pi k_\perp^4} \, \alpha_s {\cal F}
    (x_\Pom,\vec k_\perp, -\vec k_\perp) \, ,
\end{eqnarray}
%%%%
that converges at soft, non-perturbative values of $k_\perp$. Indeed, as can be seen from Eq.~(\ref{eq:sigma_0}), it is directly proportional to the dipole cross section for large dipoles, $\sigma_0(x_\Pom)$.

%------------------------------------------------------------------------------------
\subsection{GTMD representation}
%------------------------------------------------------------------------------------

In the literature, often a different momentum-space representation of the diffractive amplitude is used (see, for example, Refs.~\cite{Hagiwara:2016kam,Hagiwara:2017fye,ReinkePelicer:2018gyh}). Namely, one introduces the Fourier transform of the dipole amplitude (using the normalization and notation of Ref.~\cite{ReinkePelicer:2018gyh}),
%%%
\begin{eqnarray}
    T(Y,\vec k_\perp, \vec \Delta_\perp)
    = \int \frac{d^2 \vec b_\perp}{(2 \pi)^2} \frac{d^2 \vec r_\perp}{(2 \pi)^2} \, e^{-i \vec \Delta_\perp \cdot \vec b_\perp} \, e^{- i \vec k_\perp \cdot \vec r_\perp} \, 
     N(Y,\vec r_\perp, \vec b_\perp) \, .
     \label{eq:T_definition}
\end{eqnarray}
%%%
Here, $T(Y,\vec k_\perp, \vec \Delta_\perp)$ is often referred to as the generalized transverse momentum distribution (GTMD) of gluons in the proton target. Certainly, just like the gluon density matrix $f\Big(Y,\frac{\vec q_\perp}{2} + \vec \kappa_\perp, \frac{\vec q_\perp}{2} - \vec \kappa_\perp\Big)$, it encodes the same information as the dipole amplitude. What is the relation between these two momentum space distributions?

To answer this question, let us perform the Fourier transform by inserting Eq.~(\ref{eq:dipole_vs_GTMD}) into Eq.~(\ref{eq:T_definition}), which yields
%%%%
\begin{eqnarray}
    T(Y,\vec k_\perp, \vec \Delta_\perp) &=& C(Y,\vec \Delta_\perp) \Big( \delta^{(2)}(\vec k_\perp -\frac{\vec \Delta_\perp}{2}) + 
    \delta^{(2)}(\vec k_\perp + \frac{\vec \Delta_\perp}{2}) \Big) \nonumber \\
    && - f\Big(Y,\frac{\vec \Delta_\perp}{2} + \vec k_\perp, \frac{\vec \Delta_\perp}{2} - \vec k_\perp\Big) 
    - f\Big(Y,\frac{\vec \Delta_\perp}{2} - \vec k_\perp, \frac{\vec \Delta_\perp}{2} + \vec k_\perp\Big) \, ,
    \label{eq:T_vs_f}
\end{eqnarray}
%%%%
where $C(Y,\vec \Delta_\perp)$ of Eq.~(\ref{eq:C_Y_delta}) multiplies a combination of $\delta$-functions. Now, evidently $T$ is essentially equal to the $f$, up to the term containing $\delta$-functions. The latter, however, will not contribute when convoluted with the impact factor in Eq.~(\ref{eq:amplitude_structure}). Formally, we might therefore replace
%%%
\begin{eqnarray}
    f\Big(Y,\frac{\vec \Delta_\perp}{2} + \vec k_\perp, \frac{\vec \Delta_\perp}{2} - \vec k_\perp\Big) \to - \half \, T(Y,\vec k_\perp, \vec \Delta_\perp) \, .
\end{eqnarray}
%%%%
In practical numerical applications, however, this equivalence is not that obvious. In diffractive interactions the values of $\Delta_\perp$ are bounded by the diffractive slope $B_D$, so that at large values of $k_\perp$ the delta-functions are irrelevant, and the two gluon distributions, $f$ and $T$ are equal to each other, up to a factor of $1/2$.

An immediate corollary of the representation in Eq.~({\ref{eq:T_vs_f}}) is the sum rule
%%%
\begin{eqnarray}
    \int d^2 \vec k_\perp \, T(Y,\vec k_\perp, \vec \Delta_\perp) = 0 \, ,
    \label{eq:sum_rule}
\end{eqnarray}
%%%%
which encodes the fact, that $N(Y,\vec r_\perp, \vec b_\perp) \to 0$ for $r_\perp \to 0$.
Generally speaking, as evidenced by the presence of $\delta$-function terms, the Fourier transform is a non-convergent integral and does not exist as a function. One therefore needs to regularize the Fourier transform which is often done by inserting a Gaussian cutoff function \cite{Hagiwara:2016kam,Hagiwara:2017fye,ReinkePelicer:2018gyh}:
%%%%
\begin{eqnarray}
    T(Y,\vec k_\perp, \vec \Delta_\perp) = \int \frac{d^2 \vec b_\perp}{(2 \pi)^2} \frac{d^2 \vec r_\perp}{(2 \pi)^2}  
    e^{-i \vec \Delta_\perp \cdot \vec b_\perp} e^{-i \vec k_\perp \cdot \vec r_\perp} \, N(Y,\vec r_\perp, \vec b_\perp) \, e^{-  \varepsilon r_\perp^2}\, .
     \label{eq:reg}
\end{eqnarray}
%%%%
Now, inserting the representation in Eq.~(\ref{eq:dipole_vs_GTMD}), we obtain the cutoff-dependent $T$ as
%%%
\begin{eqnarray}
     T(Y,\vec k_\perp, \vec \Delta_\perp) &=&C(Y,\vec \Delta_\perp) \Big( \delta_\varepsilon^{(2)}(\vec k_\perp -\frac{\vec \Delta_\perp}{2}) + 
    \delta_\varepsilon^{(2)}(\vec k_\perp + \frac{\vec \Delta_\perp}{2}) \Big) \nonumber \\
    && - f_\varepsilon \Big(Y,\frac{\vec \Delta_\perp}{2} + \vec k_\perp, \frac{\vec \Delta_\perp}{2} - \vec k_\perp\Big) 
    - f_\varepsilon \Big(Y, \frac{\vec \Delta_\perp}{2} - \vec k_\perp, \frac{\vec \Delta_\perp}{2} + \vec k_\perp\Big) \, ,
\end{eqnarray}
%%%%
where
%%%%
\begin{eqnarray}
    \delta_\varepsilon^{(2)}(\vec k_\perp) = \frac{1}{4 \pi \varepsilon} \, \exp\Big(-\frac{k_\perp^2}{4 \varepsilon} \Big) \, ,
\end{eqnarray}
%%%
is a ``smeared out'' delta-distribution, and
%%%
\begin{eqnarray}
   f_\varepsilon \Big(Y, \frac{\vec \Delta_\perp}{2} - \vec k_\perp, \frac{\vec \Delta_\perp}{2} + \vec k_\perp\Big) &=& \int d^2 \vec \kappa_\perp \,  f\Big(Y, \frac{\vec \Delta_\perp}{2} - \vec \kappa_\perp, \frac{\vec \Delta_\perp}{2} + \vec \kappa_\perp\Big) \, \delta_\varepsilon^{(2)}(\vec k_\perp - \vec \kappa_\perp)
\end{eqnarray}
%%%
is a smeared out version of the gluon density matrix. The regularized $T$-matrix also fulfills the sum rule of Eq.~(\ref{eq:sum_rule}).

Let us finally quote the expressions of matrix elements $\vec {\mathcal M}_0$, 
${\mathcal M}_1$ in terms of $T$, which read:
%%%
\begin{eqnarray}
  \vec {\mathcal M}_0 &=& \int \frac{d^2 \vec k_\perp}{2 \pi} \, T(Y, \vec k_\perp, \vec \Delta_\perp) 
  \, \, \Big \{ \frac{\vec P_\perp - \vec k_\perp}{(\vec P_\perp - \vec k_\perp)^2 + m_Q^2} - \frac{\vec P_\perp}{\vec P_\perp^2 + m_Q^2}
  \Big \} \, ,\nonumber \\
 {\mathcal M}_1 &=& \int \frac{d^2 \vec k_\perp}{2 \pi} \, T(Y, \vec k_\perp, \vec \Delta_\perp) 
  \, \, \Big\{ \frac{1}{(\vec P_\perp - \vec k_\perp)^2 + m_Q^2} 
  -  \frac{1}{\vec P_\perp^2 + m_Q^2} 
  \Big \} 
  \, .
\end{eqnarray}
%%%%
In its derivation we made use of the sum rule of Eq.~(\ref{eq:sum_rule}). Notice, that in effect here we use the impact factor for forward scattering, and all dependence on $\vec \Delta_\perp$ has been absorbed into the GTMD $T(Y,\vec k_\perp, \vec \Delta_\perp)$.

%------------------------------------------------------------------------------------
\subsection{Benchmark gluon GTMDs}
%------------------------------------------------------------------------------------
%%%%
 \begin{figure}
  \centering
  \includegraphics[width=.4\textwidth]{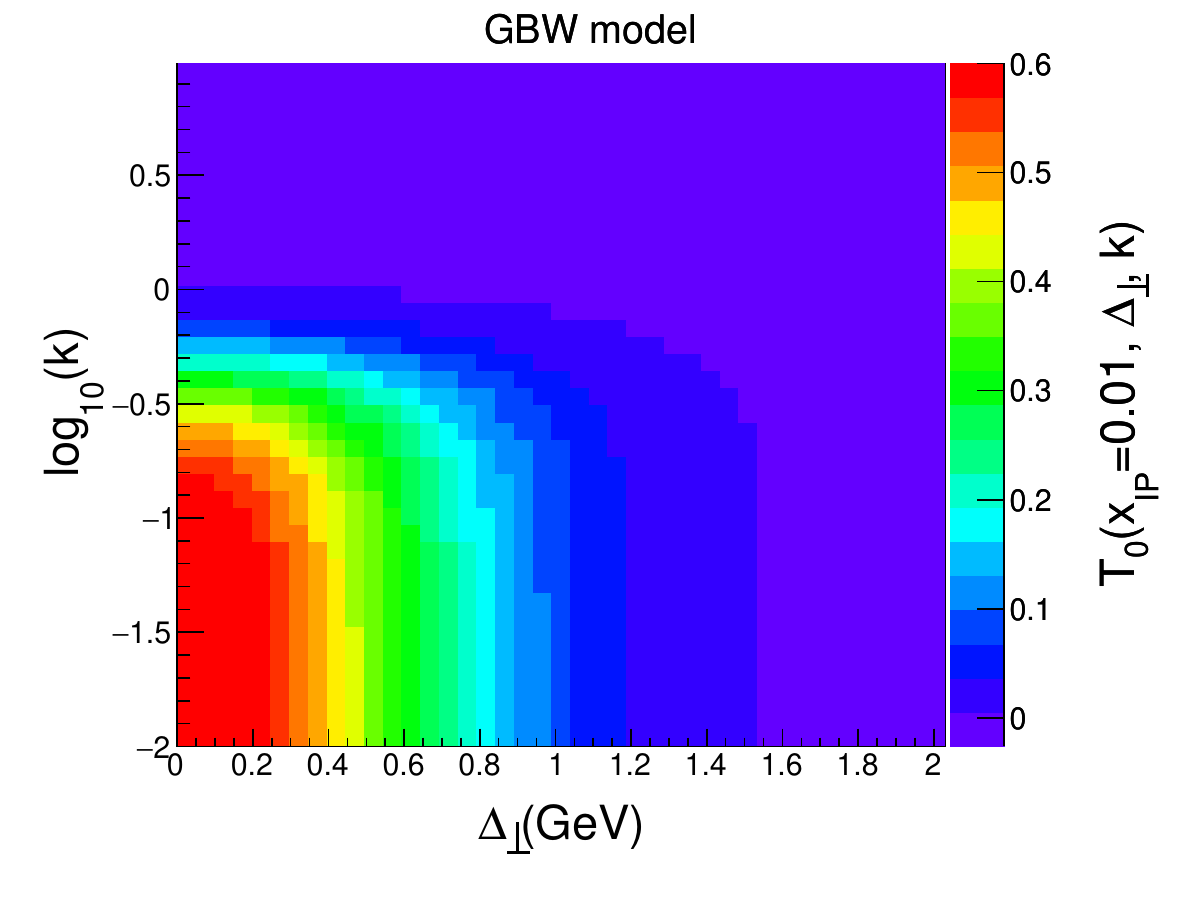}
  \includegraphics[width=.4\textwidth]{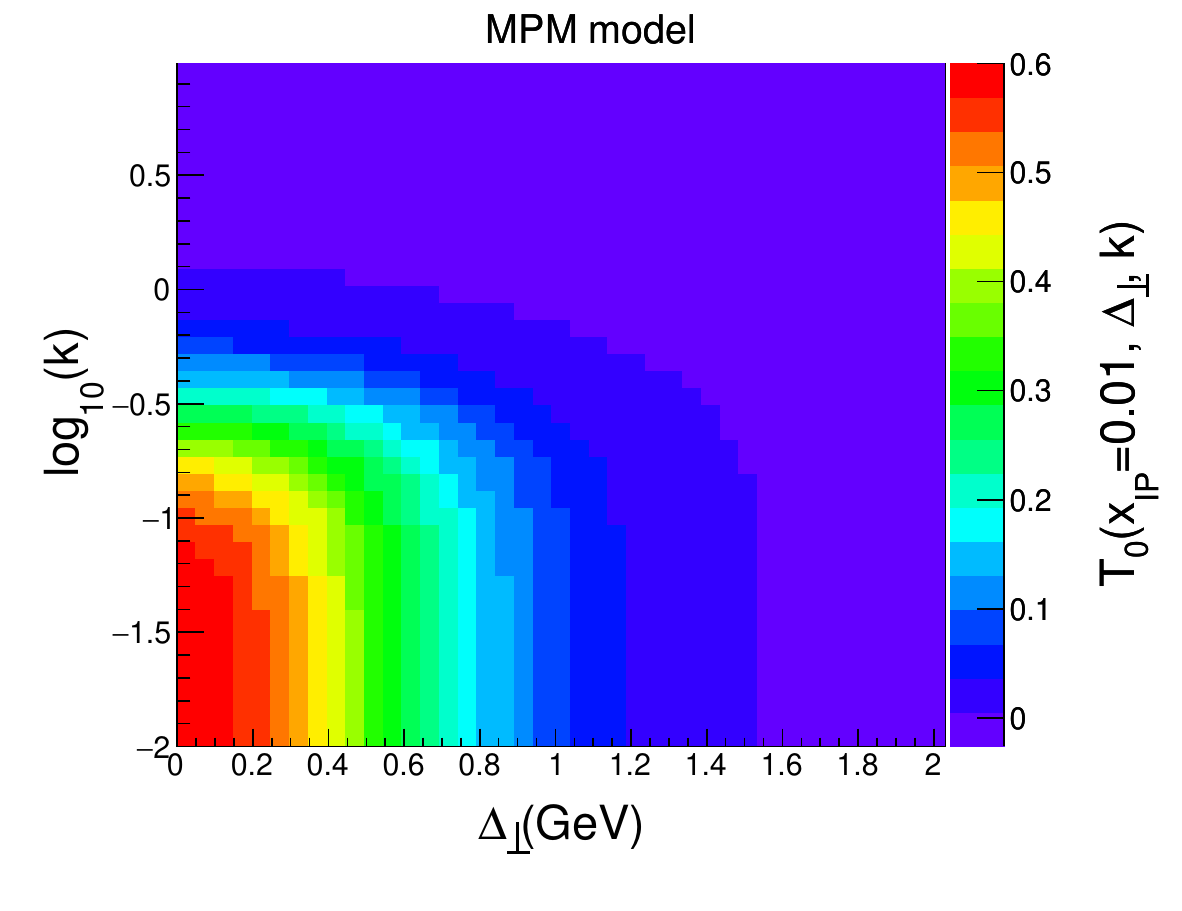}
  \includegraphics[width=.4\textwidth]{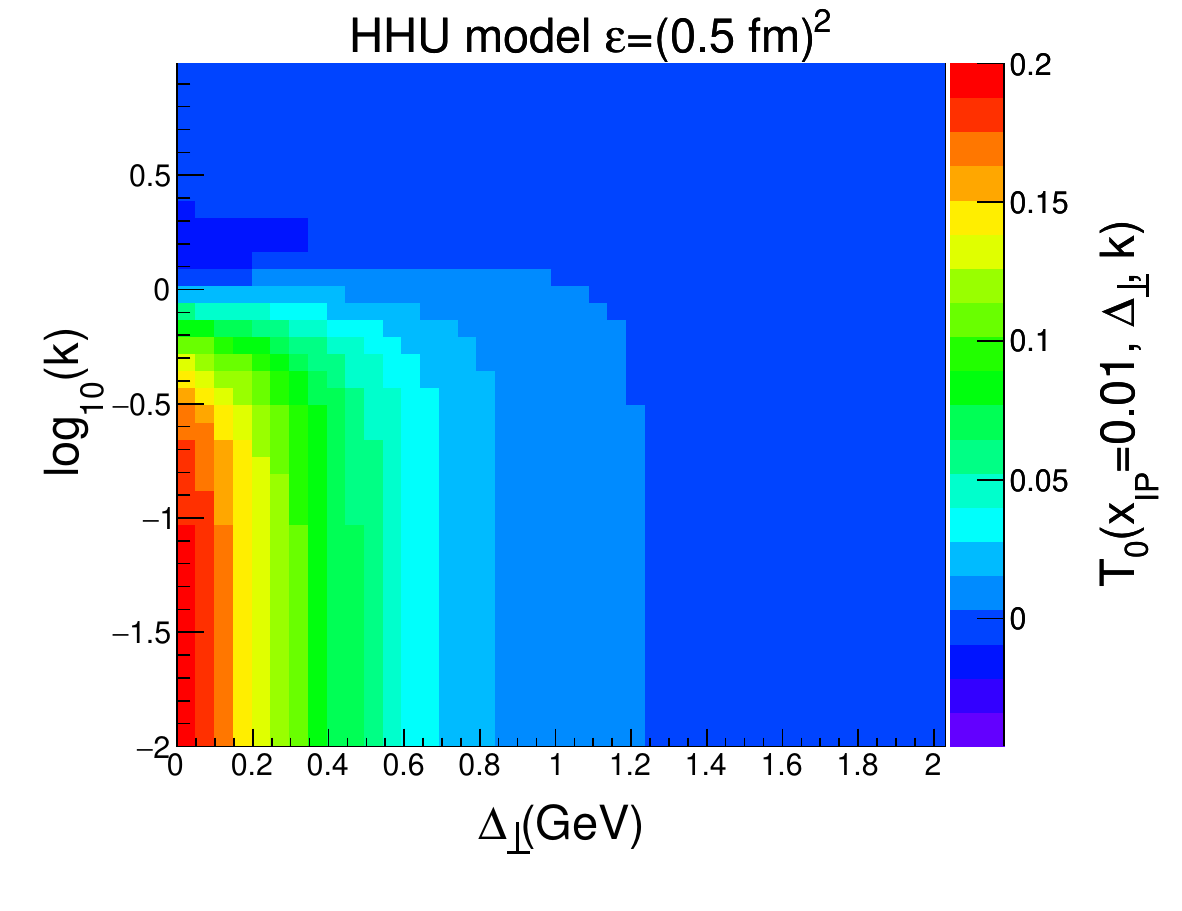}
  \includegraphics[width=.4\textwidth]{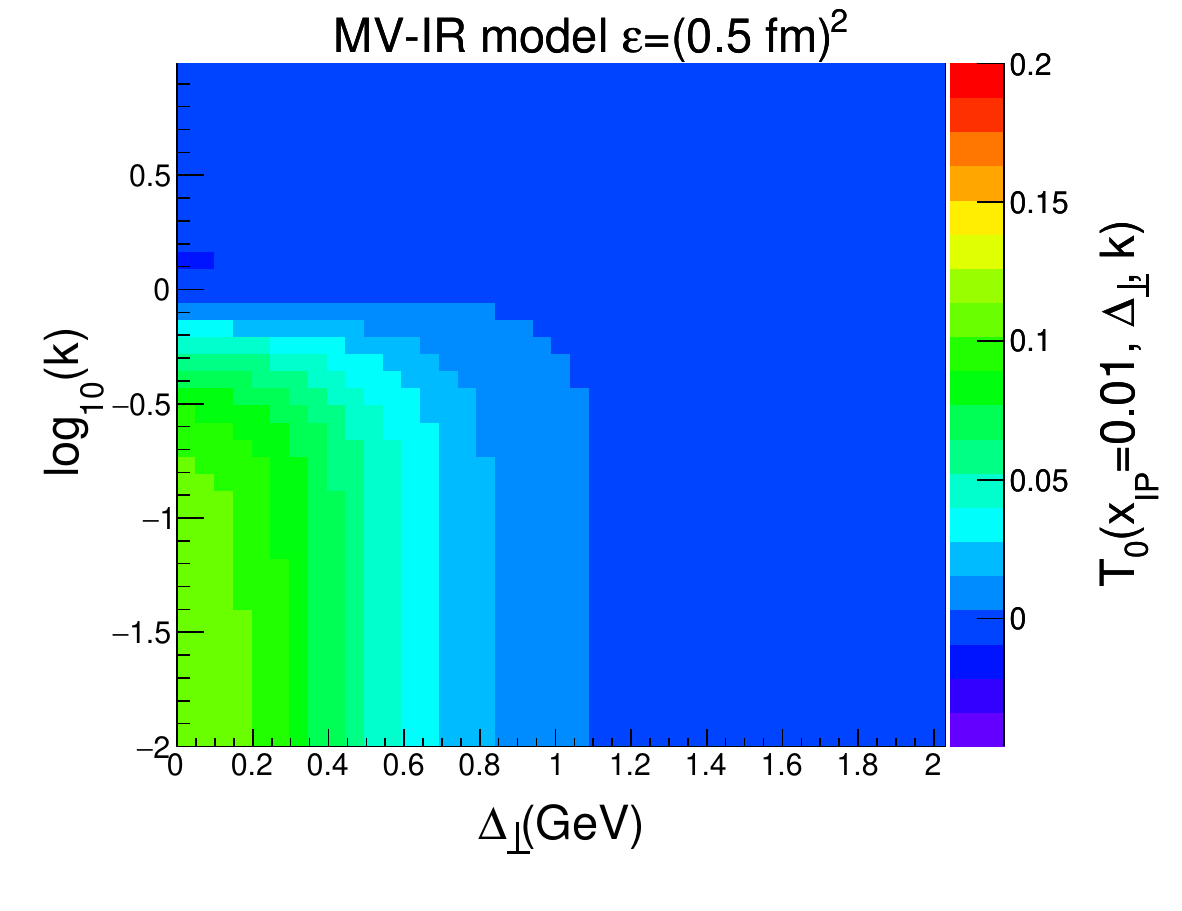}
  \includegraphics[width=.4\textwidth]{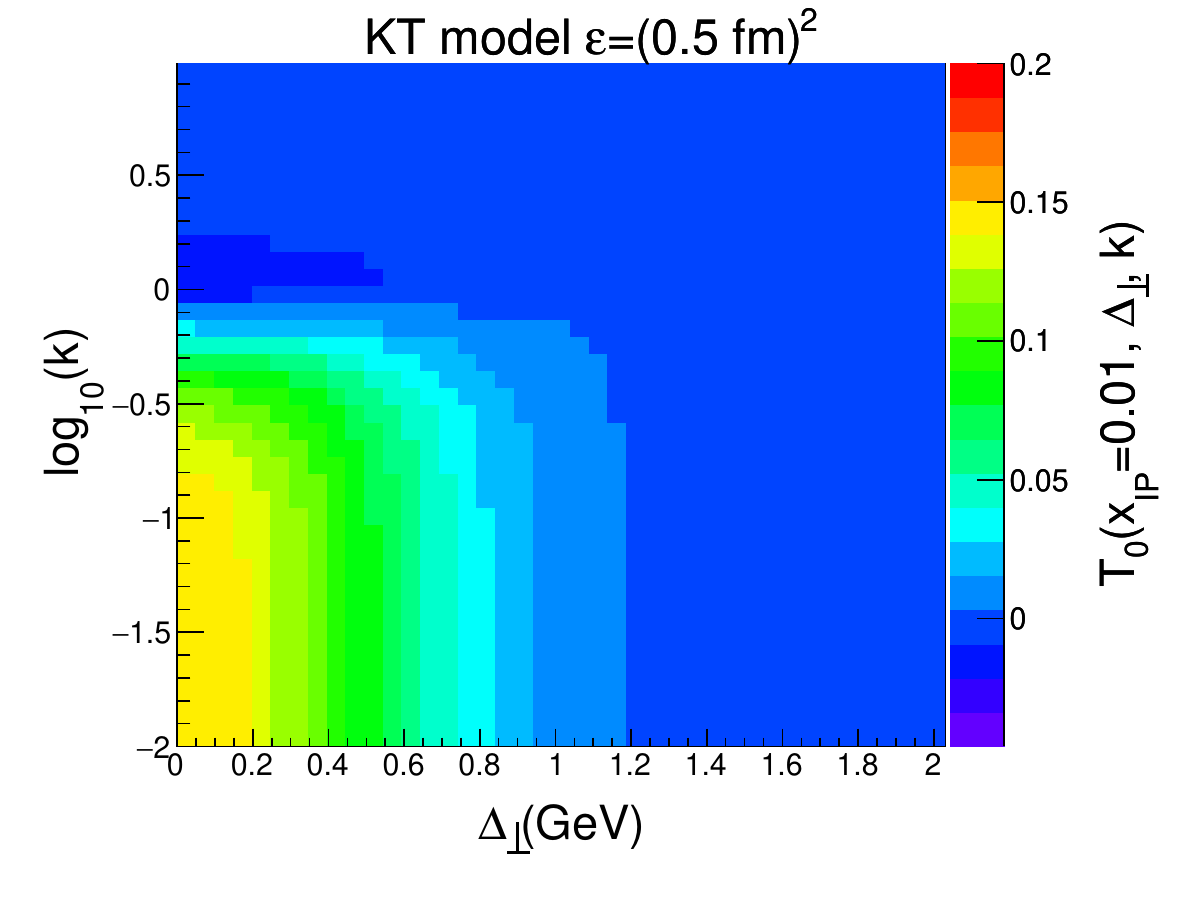}
  \caption{Different parametrizations for the gluon distribution in the proton.}
\label{fig:TkD}
\end{figure}
%%%%
Let us briefly discuss the different GTMD models used in this work. We follow both approaches presented above, and consider a total of five different models as benchmarks in our analysis of the differential cross section of charm photoproduction in $pA$ UPCs.

First, we use two different parametrizations of the off-forward gluon density matrix $f$ of Eq.~(\ref{eq:f_Y}). Here, we choose to write
%%%
\begin{eqnarray}
     f\Big(Y,\frac{\vec \Delta_\perp}{2} + \vec k_\perp, \frac{\vec \Delta_\perp}{2} - \vec k_\perp\Big) = \frac{\alpha_s}{4 \pi N_c} \, \frac{{\cal F}(x_\Pom, \vec k_\perp, -\vec k_\perp)}{k_\perp^4} \, \exp\Big[ - \half B \vec \Delta^2 \Big] \, . 
     \label{eq:model_f}
\end{eqnarray}
%%%
Such a form has been suggested in Ref.~\cite{Ivanov:2004ax} for the case of vector meson production (for a recent use, see Ref.~\cite{Cisek:2022yjj}). The same approach has also been taken in Ref.~\cite{Kopeliovich:2007fv}. For the diffractive slope, we use $B = 4 \, \rm{GeV}^{-2}$, and the diagonal unintegrated gluon distribution $\cal F$ is taken from two different models: the Golec-Biernat--W\"usthoff (GBW) model \cite{Golec-Biernat:1998zce}, and the Moriggi-Peccini-Machado (MPM) parametrization \cite{Moriggi:2020zbv}.

We also consider models based on the regularized Fourier transform of dipole amplitudes as in Eq.~(\ref{eq:reg}). The first model is based on a  %Ref.~\cite{Hagiwara:2016kam} by 
numerical solution of the Balitsky-Kovchegov equation for the dipole $S$-matrix with impact parameter dependence \cite{Balitsky:1995ub,Kovchegov:1999yj}.
Such solutions have been obtained, for example in Refs. \cite{Golec-Biernat:2003naj,Berger:2010sh,Berger:2011ew}. In our work we use the results of Ref. \cite{Hagiwara:2016kam}.
This solution is based on exploiting a symmetry first observed in \cite{Gubser:2011qva}. We label it as HHU below. 

The second model, based on the original effective McLerran-Venugopalan model of Ref.~\cite{McLerran:1993ka}, that has been generalised for the proton target and extended to incorporate inhomogeneities in the transverse-plane distribution of gluons in Ref.~\cite{Iancu:2017fzn}. The latter model has been applied for exclusive diffractive light and heavy quarks' photoproduction in Refs.~\cite{Hagiwara:2017fye} and \cite{ReinkePelicer:2018gyh}, respectively. We denote this model as MV-IR in what follows. A third model is the so-called bSat model of Kowalski and Teaney \cite{Kowalski:2003hm}. The Kowalski-Teaney (KT) model has been adjusted to the proton structure function data and hence it gives a rather realistic dipole amplitude. We use parameters from Table I fit 3 of \cite{Kowalski:2003hm} for the gluon distribution. In this regard, the HHU and MV-IR models can be considered as rather toy models, they do, however, incorporate non-trivial dipole orientation effects, which is not the case for the KT model.

The leading dependence on dipole orientation is quantified by the elliptic part $N_\epsilon$ of the dipole amplitude in the Fourier expansion,
%%%
\begin{eqnarray}
    N(Y,\vec r_\perp, \vec b_\perp) = N_0(Y, r_\perp, b_\perp) + 2 \, \cos (2 \phi_{br}) \, N_\epsilon(Y,r_\perp, b_\perp) + \dots \,. 
\end{eqnarray}
%%%
We translate the isotropic and elliptic parts of the dipole amplitude to GTMDs by the appropriate Fourier-Bessel transforms \cite{Hatta:2016dxp,Hagiwara:2017fye,ReinkePelicer:2018gyh}:
%%%
\begin{eqnarray}
T_0(Y,k_\perp,\Delta_\perp) &=& \frac{1}{4\pi^{2}} \int_0^\infty b_\perp db_\perp J_0(\Delta_\perp b_\perp) \int_0^\infty r_\perp dr _\perp   J_0(k_\perp r_\perp) \, N_0(Y,r_\perp, b_\perp) e^{- \varepsilon r_\perp^2}, \nonumber \\
T_\epsilon(Y,k_\perp,\Delta_\perp) &=& \frac{1}{4\pi^{2}}  \int_0^\infty b_\perp db_\perp J_2(\Delta_\perp b_\perp) \int_0^\infty r_\perp dr _\perp   J_2(k_\perp r_\perp) \, N_\epsilon(Y,r_\perp, b_\perp) e^{-\varepsilon r_\perp^2} \,.
\label{eq:Fourier_transform}
\end{eqnarray}
%%%
The explicit form of the matrix element for the elliptic glue for arbitrary quark mass $m_Q$ has been found in Ref.~\cite{ReinkePelicer:2018gyh}. We briefly review its derivation in Appendix~\ref{sec:appendix_me}.

It is interesting to note that the gluon density matrices, constructed according to the prescription of Eq.~(\ref{eq:model_f}), do also lead to a dipole amplitude that depends on dipole orientation, as can be seen by plugging the expression of Eq.~(\ref{eq:model_f}) into Eq.~(\ref{eq:dipole_vs_GTMD}). It gives us the expression
%%%
\begin{eqnarray}
    N(Y,\vec r_\perp, \vec b_\perp) = \frac{1}{4} \Big\{ t_N\Big(\vec b_\perp + \frac{\vec r_\perp}{2} \Big)  + t_N\Big(\vec b_\perp - \frac{\vec r_\perp}{2} \Big) - 2 t_N(\vec b_\perp) \Big\} \sigma_0(x_\Pom)  + \half t_N(\vec b_\perp) \sigma(x_\Pom,\vec r_\perp) \, , \nonumber \\
    \label{eq:N_from_f}
\end{eqnarray}
%%%
where 
%%%
\begin{eqnarray}
    t_N(\vec b_\perp) = \int \frac{d^2 \vec q_\perp}{(2 \pi)^2} \, \exp(-i \vec q_\perp \cdot \vec b_\perp) \exp\Big[ - \half B  q_\perp^2 \Big] \, . 
\end{eqnarray}
Again, we see that the dipole orientation dependence appears together with the non-perturbative parameter $\sigma_0$ -- the dipole cross section for large dipoles.

We want to stress, that the correlation between $\vec r_\perp$ and $\vec b_\perp$ emerges, although Eq.~(\ref{eq:model_f}) does not contain any correlation between $\vec k_\perp$ and $\vec \Delta_\perp$!
The $\vec r_\perp \cdot \vec b_\perp$ correlation is in fact of a simple geometric origin, as it singles out the contributions of diagrams where only the quark or antiquark interact and probe the matter density at their respective impact parameters. There is no obvious way how to construct an off-forward glue that leads to totally isotropic dipole amplitude. For this to happen, the relevant dipole sizes simply have to be small enough for the matter density to be constant over distances $\sim r_\perp$. Then the shifts in the curly brackets do not matter and only the last term in Eq.~(\ref{eq:N_from_f}) effectively contributes. Of course, in general one would expect Eq.~(\ref{eq:f_Y}) to have nontrivial azimuthal correlations between $\vec k_\perp$ and $\vec \Delta_\perp$, in which case the dipole amplitude will have a genuinely dynamical elliptic piece that can contribute also at hard momenta $\vec P_\perp$.

\begin{figure}
  \centering
  \includegraphics[width=.49\textwidth]{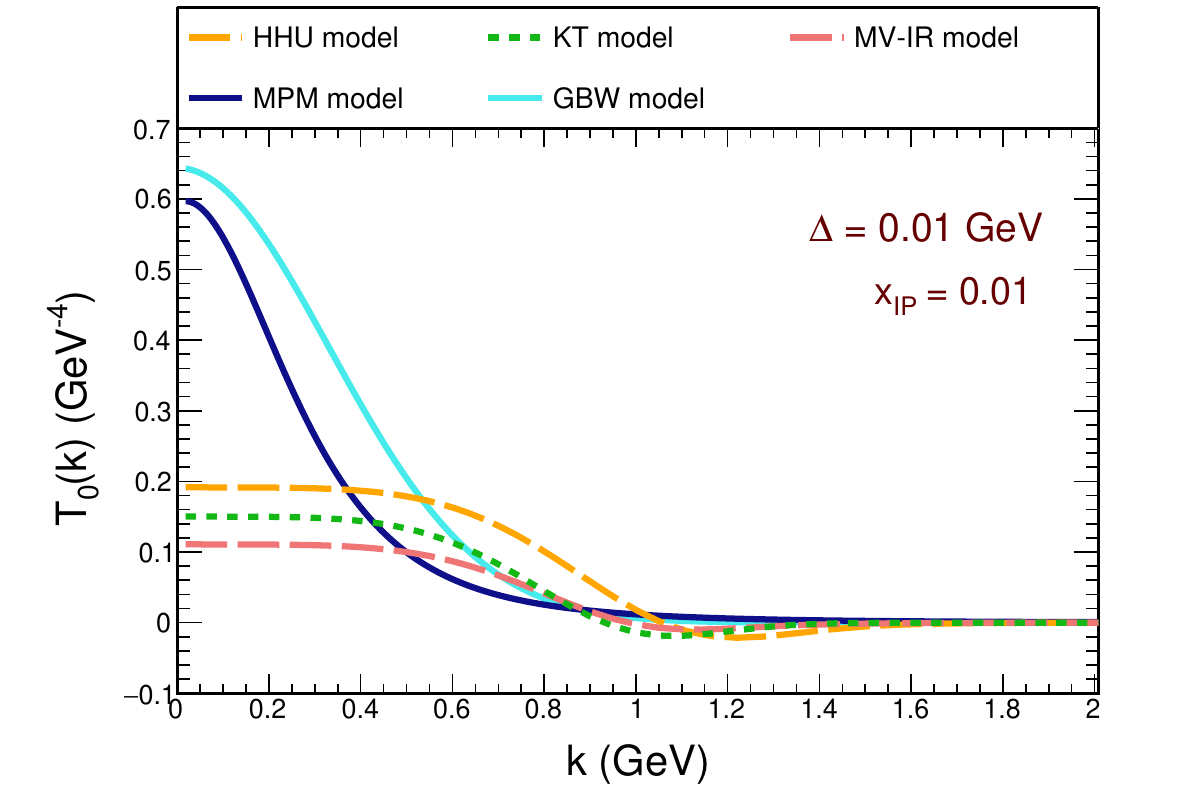}
  \includegraphics[width=.49\textwidth]{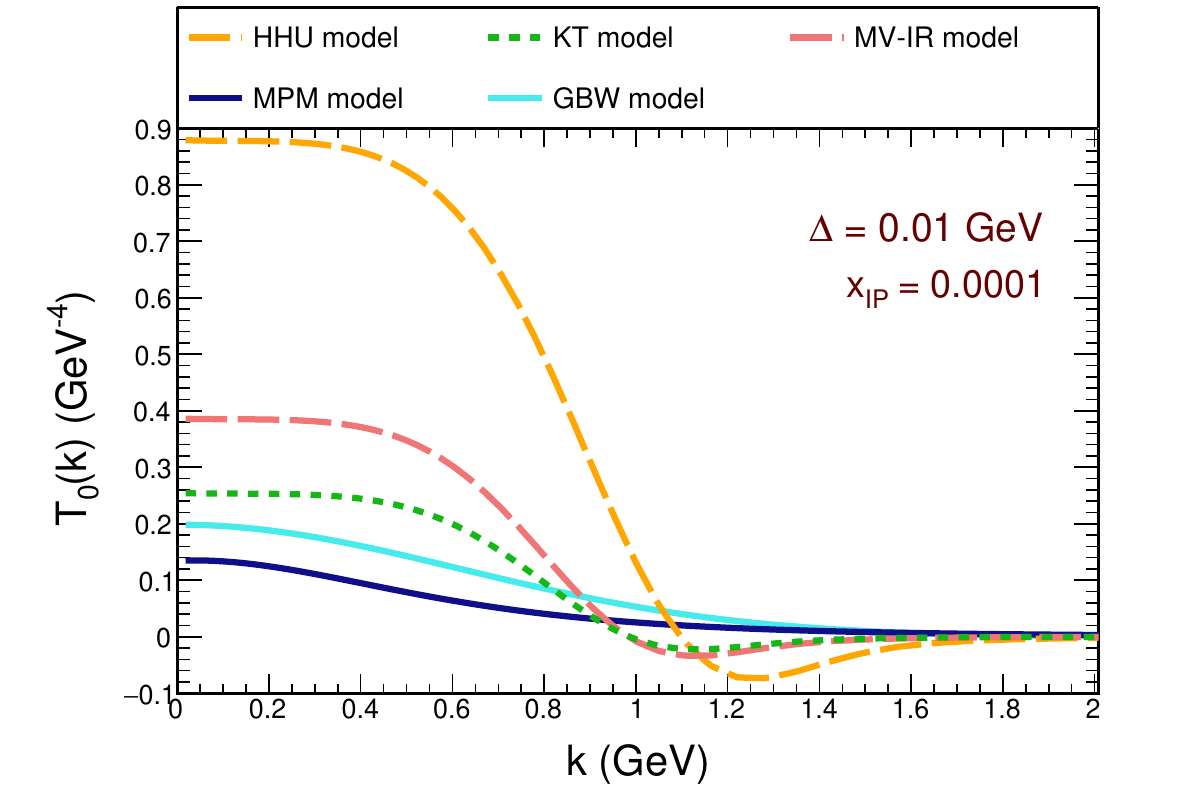}
  \caption{Different parametrizations for the gluon distribution in the proton for $\Delta = 0.01$ GeV and $x_{\Pom}=0.01$ (on the left side) as well as $x_{\Pom}=0.0001$ (on the right side). For the HHU, KT and MV-IR models, the regularization parameter $\varepsilon = (0.5 \, \rm{fm})^{-2}$ was chosen.}
\label{fig:Tk0}
\end{figure}

Finally, to wrap up this section, in Fig.~\ref{fig:TkD} we show the $(\vec{k}_\perp, \vec{\Delta}_\perp)$ maps for various dipole-nucleon amplitudes $T(Y, \vec{k}_\perp,\vec{\Delta}_\perp)$ for comparison.  In Fig. \ref{fig:Tk0} we show for $x_\Pom = 0.01$ and $\Delta_\perp=0.01$ GeV the function of Eq.~ (\ref{eq:model_f}) for the GBW and MPM models, and the result of the Fourier transform Eq~(\ref{eq:Fourier_transform}). Notice that the latter do change sign, in accordance with the sum rule Eq.~(\ref{eq:sum_rule}).

%%%%
%%%%
%%%%%%%%%%%%%%%%%%%%%%%%%%%%%%%%%%%%%%%%%%%%%%%%%%%%%%%%%%%%%%%%%%%%%%%%%%%%
\section{Numerical results}
\label{sec:distributions}
%%%%%%%%%%%%%%%%%%%%%%%%%%%%%%%%%%%%%%%%%%%%%%%%%%%%%%%%%%%%%%%%%%%%%%%%%%%%
%-------------------------------------------------------
\begin{table}
\centering
\begin{tabular}{| l | r | r | r | r |}
\hline
  & \multicolumn{2}{|c|}{$x_\Pom < 1.0$} & \multicolumn{2}{|c|}{$x_\Pom < 0.05$} \\\hline
GTMD approaches & $\sigma$ [$\rm \mu b$] & $\sigma_{P_{\perp}>5.0 GeV}$ [$\rm \mu b$] & $\sigma$ [$\rm \mu b$] & $\sigma_{P_{\perp}>5.0 GeV}$ [$\rm \mu b$] \\\hline
GBW & 335.199 & 0.051 & 330.046 & 0.046 \\ 
MPM & 321.141 & 0.201 & 293.300 & 0.173 \\ \hline
\multicolumn{5}{|c|}{$\varepsilon = (0.5 \, {\rm fm})^{-2}$} \\  \hline
HHU & 520.691 & 4.573 & 520.691 & 4.573\\
KT & 66.699 & 0.111 & 65.023 & 0.110 \\
MV-IR & 136.675 & 0.606 & 129.883 & 0.526\\  \hline
\multicolumn{5}{|c|}{$\varepsilon = \frac{1}{2} (0.5 \, {\rm fm})^{-2}$} \\  \hline
HHU & 743.411 & 4.348 & 743.411 & 4.348\\
KT & 85.487 & 0.106 & 83.039 & 0.105\\
MV-IR & 169.561  & 0.587  & 161.360 & 0.510\\
\hline
\end{tabular}
\caption{\label{tab:widgets}Total cross section for $0.01 < P_{\perp} < 10.0$ GeV and for $5.0 < P_{\perp} < 10.0$ GeV and different approaches. In this table, the cross sections are integrated over $\Delta_\perp < 3\, \rm{GeV}$ and $-8 < y_{c}<8$.}
\label{tab:cross_sections}
\end{table}
%-------------------------------------------------------
We now turn to our numerical results for the production of $c \bar c$ pairs. In this work, we use the charm quark mass value of $m_c = 1.5 \, \rm{GeV}$.
In this section, we show the cross-section distributions for $c\bar c$ photoproduction in $pA$ UPCs differential in $y^{\rm LAB}_c, Y^{\rm LAB}_{\rm pair}, x_\Pom$, as well as $P_\perp, \Delta_\perp, t$ and $\phi$, for several benchmark models of the gluon GTMD in the proton discussed above. We will show numerical results for differential distributions integrated over $0.01 < P_\perp < 10 \, \rm{GeV}$, as well as over $5 < P_\perp < 10 \, \rm{GeV}$ domains of the phase space. For all distributions calculated as Fourier transforms of $N(Y, \vec{r}_{\perp}, \vec{b}_{\perp})$ (see Eq.~(\ref{eq:reg})) we use the same regularization parameter $\varepsilon = (0.5 \, {\rm fm})^{-2}$ as in Ref.~\cite{ReinkePelicer:2018gyh}. In Table~\ref{tab:cross_sections}, we show our results for the integrated cross section for $P_\perp < 10 \, \rm{GeV}$ as well as for $5 < P_\perp < 10 \, \rm{GeV}$. In addition we show results for $x_\Pom < 0.05$. The results with the extra cut are only slightly smaller, except of the MPM model. In Table 1 we also varied the parameter $\varepsilon$ by a factor two in order to give an estimate on the dependence on $\varepsilon$, which turns out to be rather large. 

We notice considerable differences for the different GTMDs at large transverse momenta.  For instance, the GBW distribution drops off with transverse momentum uch faster than other GTMDs. 
 This is due to the fact that the GBW UGD does not possess the perturbative power-law tail at large momenta, but rather features a Gaussian distribution in transverse momenta with an $x$-dependent width.
It should also be noted that the original MV-IR distribution is formulated in Ref.~\cite{Iancu:2017fzn} without $x_\Pom$ (or $Y$) dependence. To get semi-realistic results for the differential distributions, the original MV-IR has been modified as 
%%%
\begin{eqnarray}
    T_{\rm MV-IR}^{\rm mod}(Y,\vec k_\perp, \vec \Delta_\perp) = T_{\rm MV-IR}(\vec k_\perp, \vec \Delta_\perp) \, e^{\lambda Y}, \quad \, Y = \ln\left(\frac{0.01}{x_\Pom}\right) \,,
\end{eqnarray}
%%%
with $\lambda = 0.277$. It should be noted that more realistic extensions of the MV-IR model have been proposed in the literature, see e.g. \cite{Mantysaari:2020lhf}.

\begin{figure}
  \centering
  \includegraphics[width=.49\textwidth]{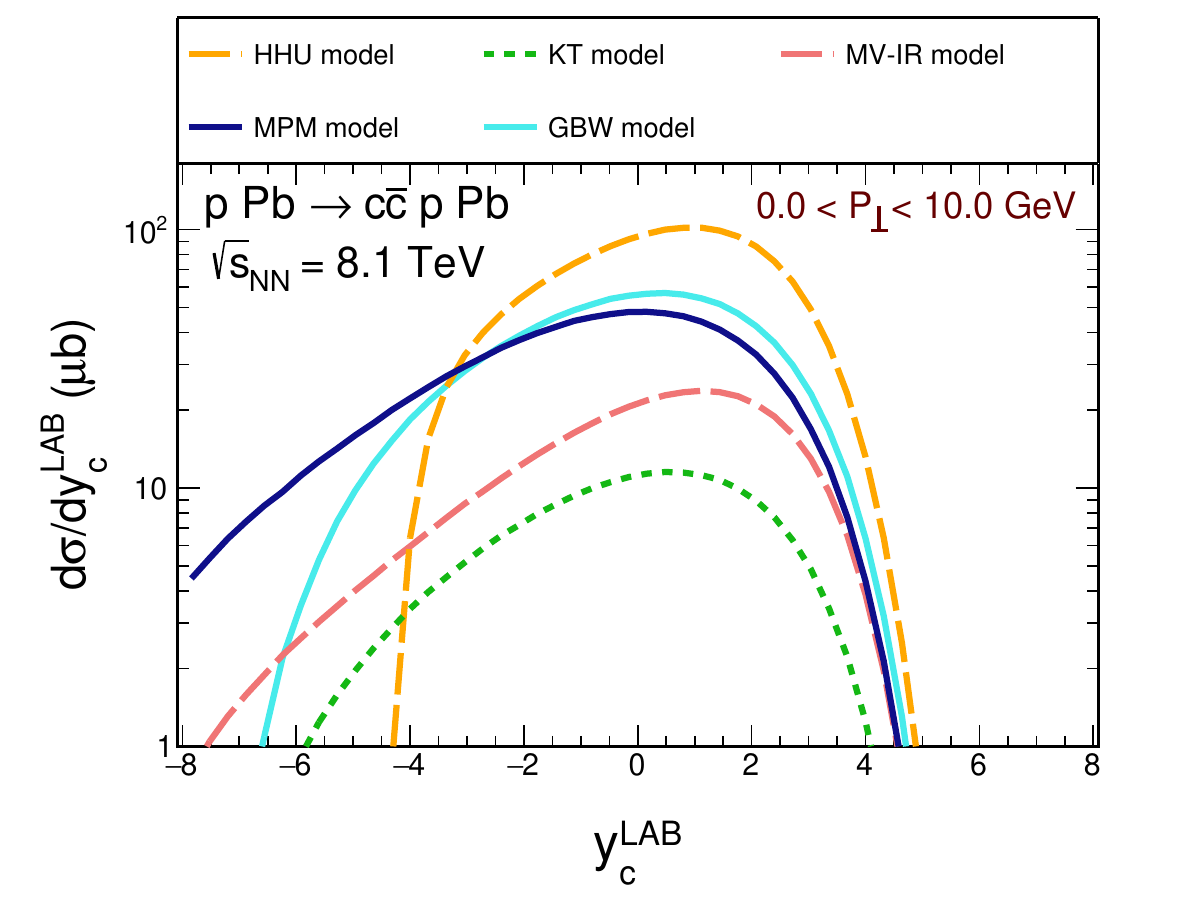}
  \includegraphics[width=.49\textwidth]{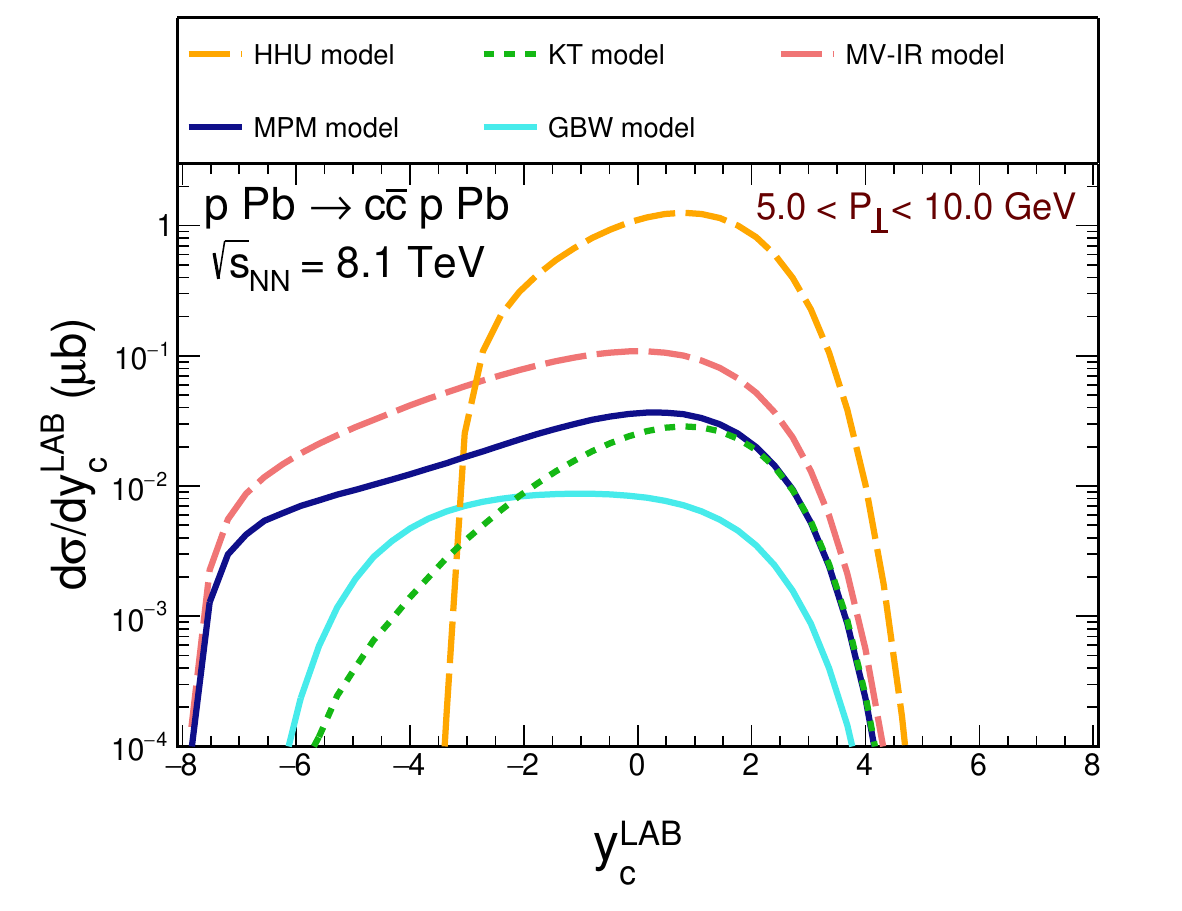}
  \caption{Distributions in lab-frame rapidity of the charm quark $y^{\rm LAB}_c$ for $0.01 < P_{\perp} < 10.0$ GeV on the left and for $5.0 < P_{\perp} < 10.0$ GeV on the right.}
\label{fig:y}
\end{figure}

\begin{figure}
  \centering
  \includegraphics[width=.49\textwidth]{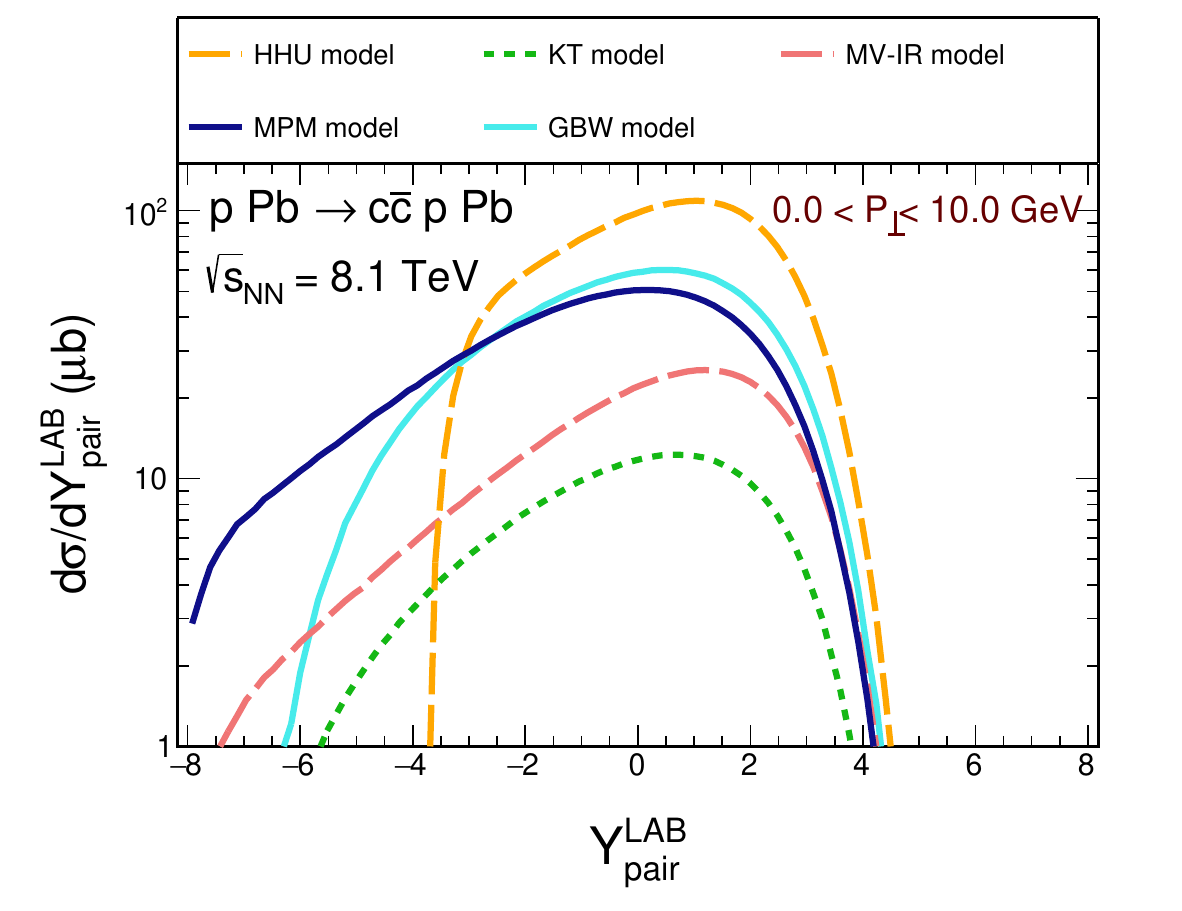}
  \includegraphics[width=.49\textwidth]{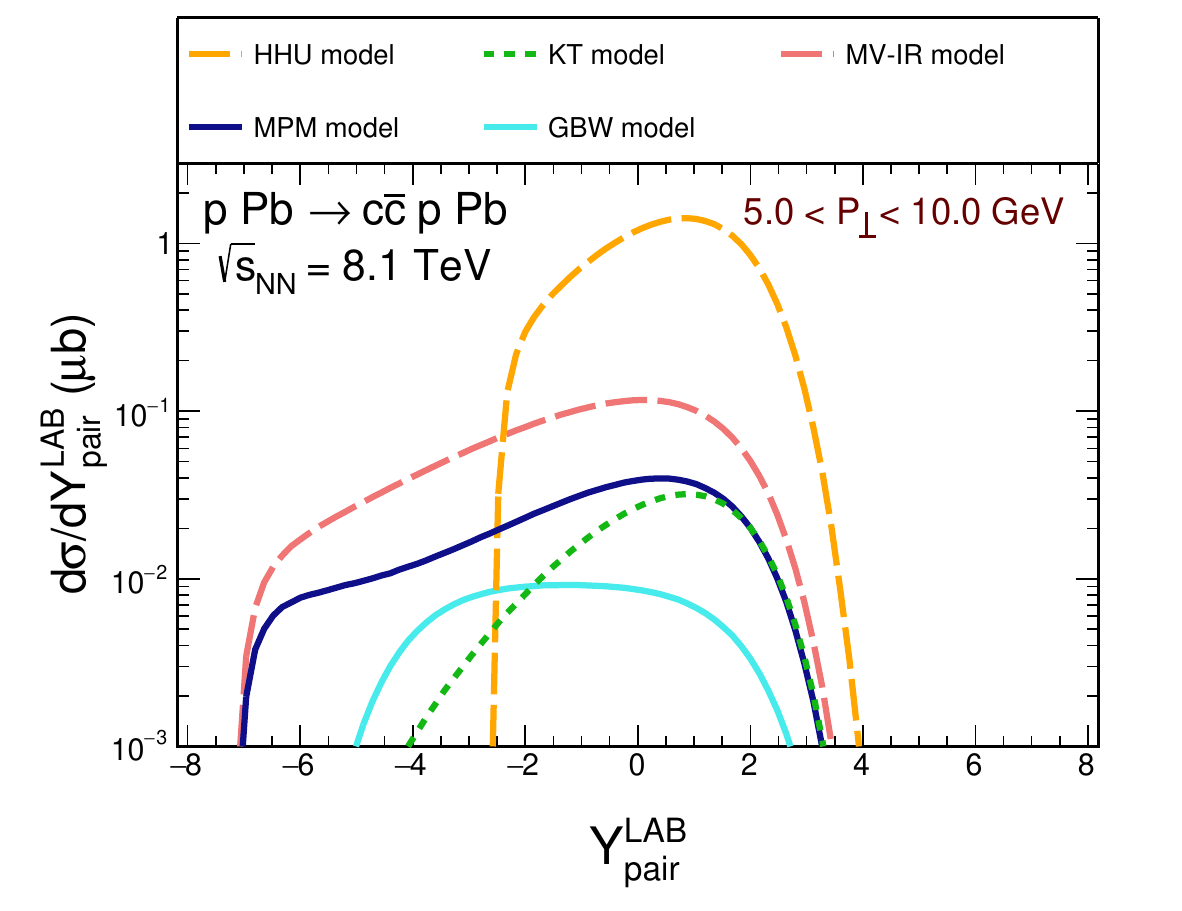}
  \caption{Distributions in rapidity of the $c\bar c$ pair $Y_{\rm pair}$ for $0.01 < P_{\perp} < 10.0$ GeV on the left and for $5.0 < P_{\perp} < 10.0$ GeV on the right.}
\label{fig:ypair}
\end{figure}

\begin{figure}
  \centering
  \includegraphics[width=.49\textwidth]{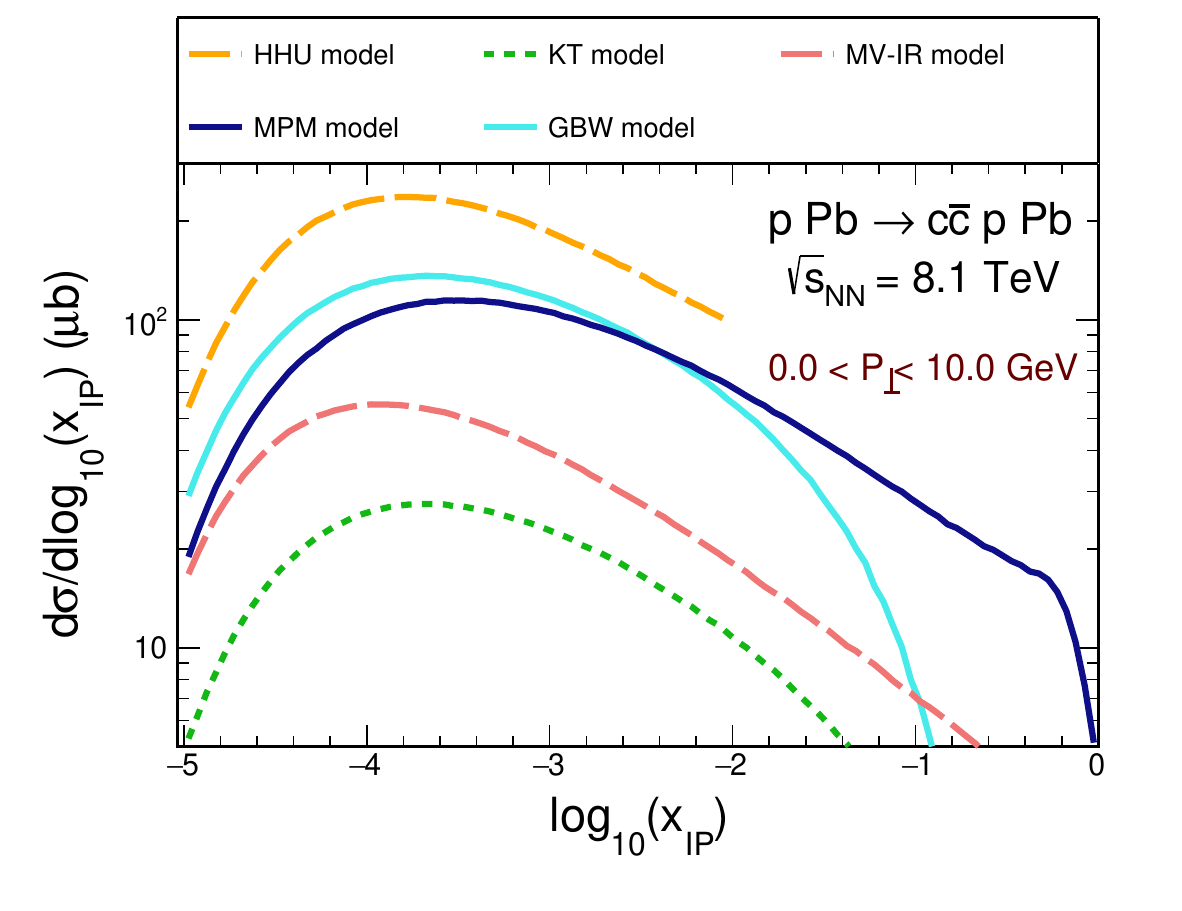}
  \includegraphics[width=.49\textwidth]{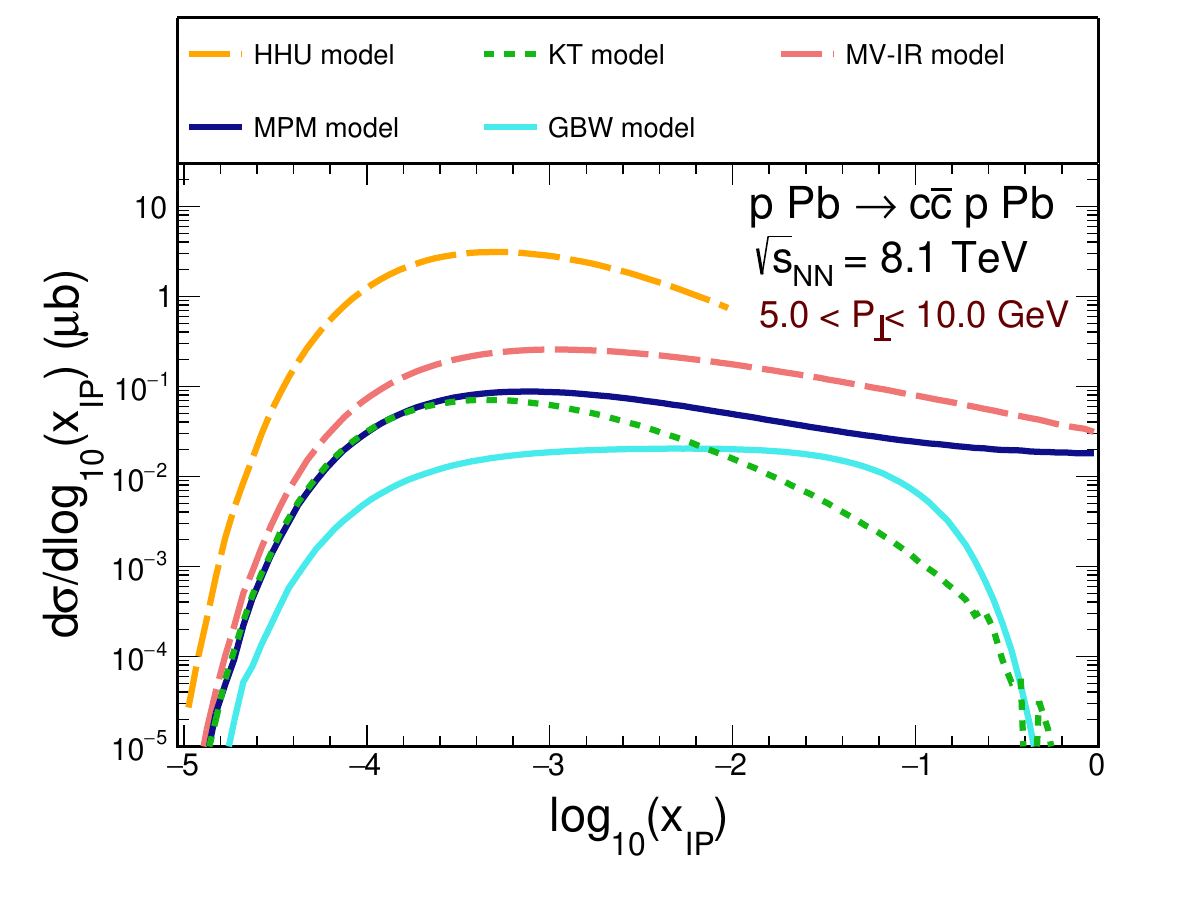}
  \caption{Distributions in $\log_{10}(x_{\Pom})$ for $0.01 < P_{\perp} < 10.0$ GeV on the left and for $5.0 < P_{\perp} < 10.0$ GeV on the right.}
\label{fig:logxb}
\end{figure}

We start our presentation from the lab-frame rapidity distributions of charm quarks in Fig.~\ref{fig:y}. Here, the incoming nucleus has a large positive rapidity, and the proton -- a large negative rapidity. The range of the rapidity distribution at large $y^{\rm LAB}_c$ is essentially controlled by the photon flux. We see that the HHU GTMD leads to a significantly higher peak in the charm rapidity distribution than other models, and extends at the negative side only to $y^{\rm LAB}_c \sim - 4$. This is due to the fact that it has a support only for small $x_\Pom < 0.01$. The GBW distribution gives results consistent with an earlier calculation of Gon\c{c}alves at al \cite{Goncalves:2019jdl}, where only rapidity distributions were studied. In addition, the distributions in lab-frame rapidity of the $c \bar c$-pair are shown in Fig.~\ref{fig:ypair}. They rather closely resemble the single-quark rapidity distributions. The related $x_{\Pom}$ distributions are presented in Fig.~\ref{fig:logxb}. Values of $x_{\Pom}$ as small as 10$^{-5}$ enter the calculation. Note that in the case of HHU GTMD, we assumed that this distribution only applies for $x_\Pom < 0.01$. 

A comment on the range of the $x_\Pom$ distributions is in order. Notice, that the diffractively scattered proton will carry a fraction $\xi \approx 1- x_\Pom$ of the beam 
momentum. Also, the rapidity gap between the products of photon dissociation rises at small $x_\Pom$ as $\propto \log(1/(1-\xi)) \sim \log(1/x_\Pom)$.
It is therefore clear that an experiment, which tags protons carrying a sizeable fraction of the beam momentum, would provide an upper limit on $x_\Pom$. For a discussion of relevant forward detectors at the LHC, see \cite{Staszewski:2023chz,Trzebinski:2019tmk,Bossini:2023uwa}.
This goes together with considerations on the applicability of the dipole amplitudes used in this paper. This will in general require a large rapidity gap as mentioned above. There will be no sharp value of $x_\Pom$ where the dipole approach breaks down, but one would expect it to require $x_\Pom \ll 0.1$, with a conservative upper limit of $x_\Pom < 0.05$. In absence of concrete experimental conditions, below we will integrate over the whole phase space, i.e. up to $x^{\rm max}_\Pom = 1$, expecting that the impact from large $x_\Pom$ to the integrals is in general small. An exception are the distributions in rapidity, where large negative rapidities are close to the proton beam, and therefore associated with small gaps. We return to this issue at the end of this section, where we demonstrate the role of a cutoff $x_\Pom^{\rm max}$ on the various distributions.

We observe that the cut due to $\varepsilon$ in Eq.(\ref{eq:Fourier_transform}) has a different impact  for different models. In particular, it has a strong effect on the KT model in the region of small $P_T$, where the bulk of the cross section resides. This makes the distributions integrated over $P_T$ for the KT model very small. For $P_T > 2 \, \rm{GeV}$, however the results are comparable to other models. Eventually the value of the $\varepsilon$ could be adjusted to experimental data.
%%%%
%%%%
%%%%
%%%%
%%%% 
%%%%
%%%%
 \begin{figure}
  \centering
  \includegraphics[width=.49\textwidth]{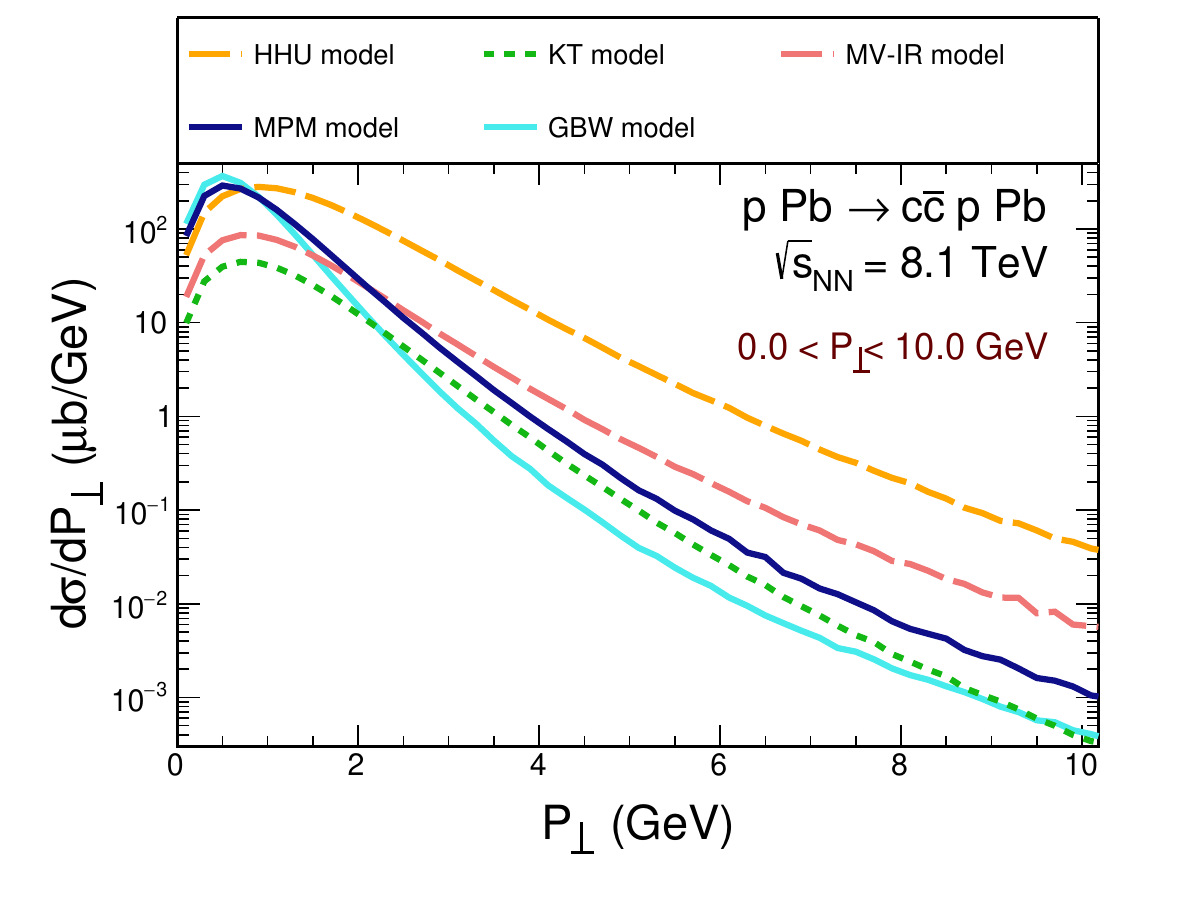}
  \caption{Distributions in $P_{\perp}$.}
\label{fig:pt}
\end{figure}
%%%%

\begin{figure}
  \centering
  \includegraphics[width=.49\textwidth]{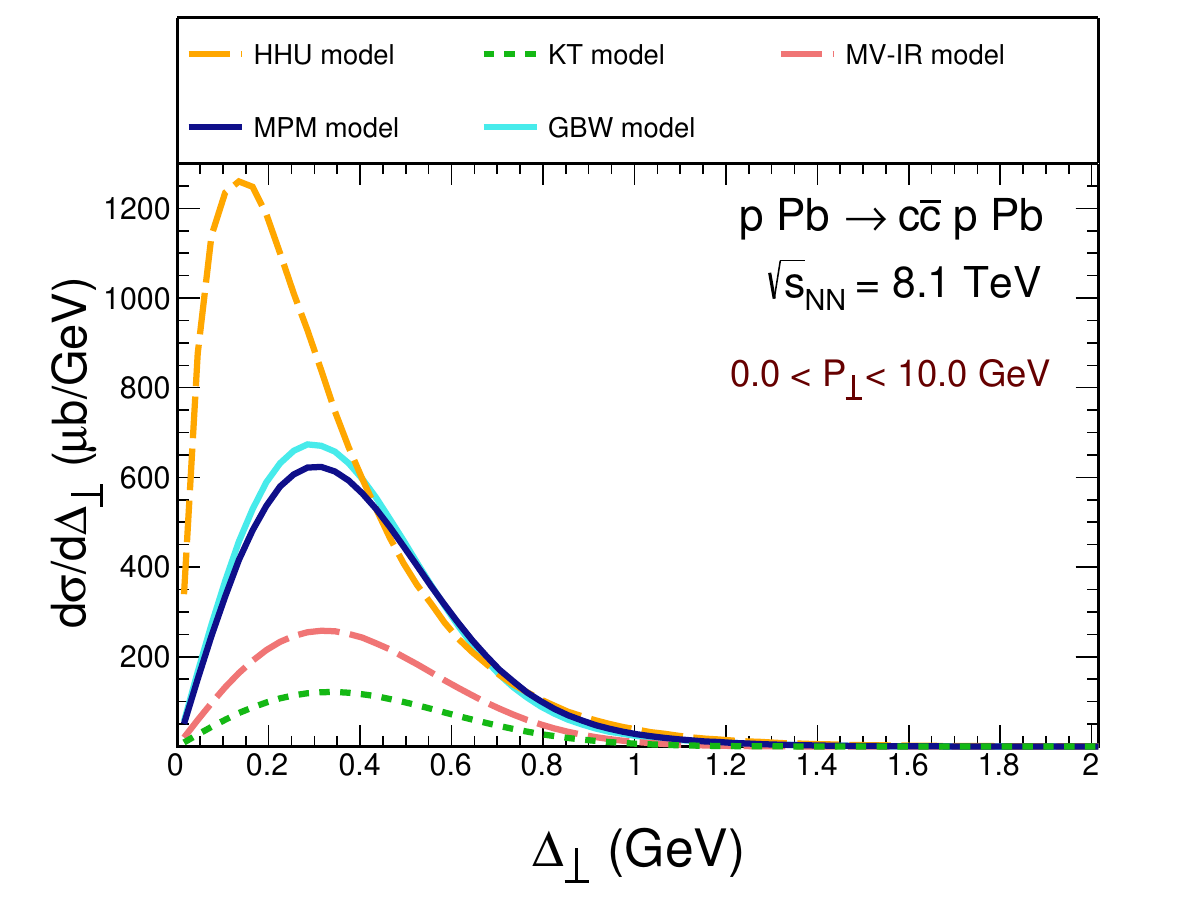}
  \includegraphics[width=.49\textwidth]{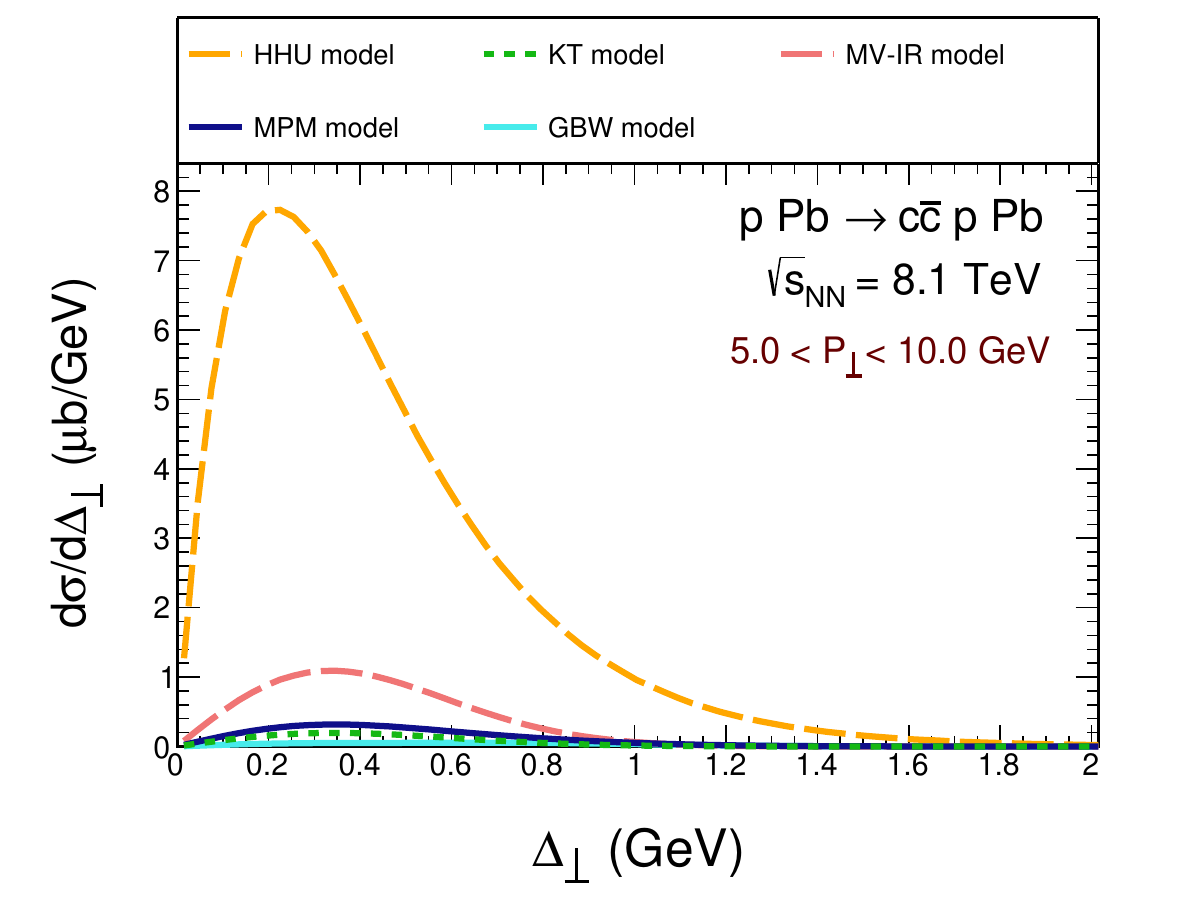}
   \caption{Distributions in $\Delta_{\perp}$ for $0.01 < P_{\perp} < 10.0 \, \rm{GeV}$ on the left and for $5.0 < P_{\perp} < 10.0 \, \rm{GeV}$ on the right.}
\label{fig:Delta}
\end{figure}

Let us now term to transverse momentum distributions of the produced charm quarks. In Fig.~\ref{fig:pt} we show the differential distributions in $P_\perp$. We notice a considerable difference in the tail of the distributions, with the GBW glue giving the softest behaviour as expected. In Fig.~\ref{fig:Delta} we present the charm distribution in $\Delta_\perp$. As generically anticipated for diffractive scattering, such a distribution is peaked at low values of $\Delta_\perp \sim 1/\sqrt{B}$, with $B$ being the diffractive slope. The HHU model somewhat stands out as it gives rise to a peak in $\Delta_\perp$ distribution that is shifted to a softer value than that for the other GTMD models considered in this work. These distributions are the result of the dynamics of gluons encoded in the BK equation.
Presumably the behaviour of the $\Delta_\perp$ distribution could be made more realistic by introducing a similar regulator function in the impact parameter variable $b_\perp$ as in Eq.~(\ref{eq:Fourier_transform}) for the dipole size $r_\perp$. Here however we restrain from introducing additional parameters.

\begin{figure}
  \centering
  \includegraphics[width=.49\textwidth]{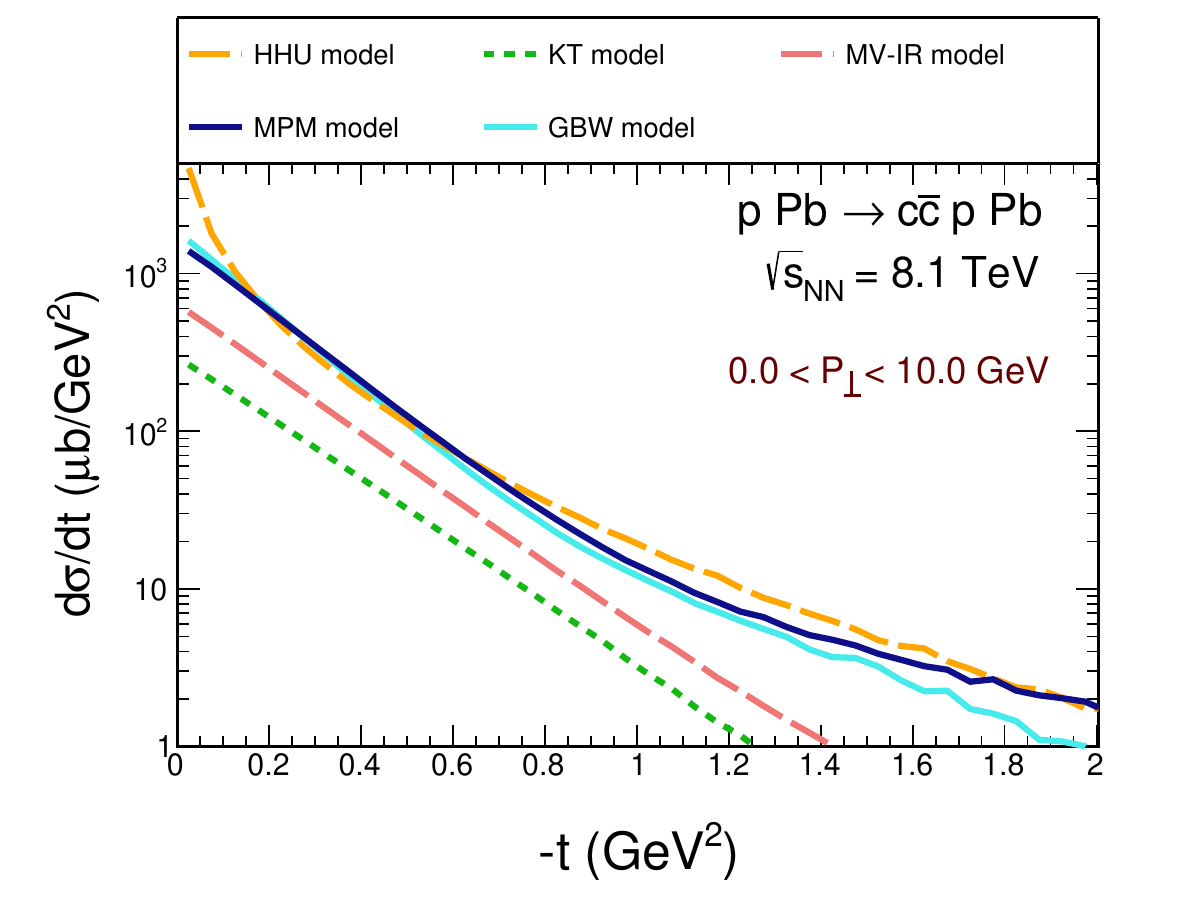}
  \includegraphics[width=.49\textwidth]{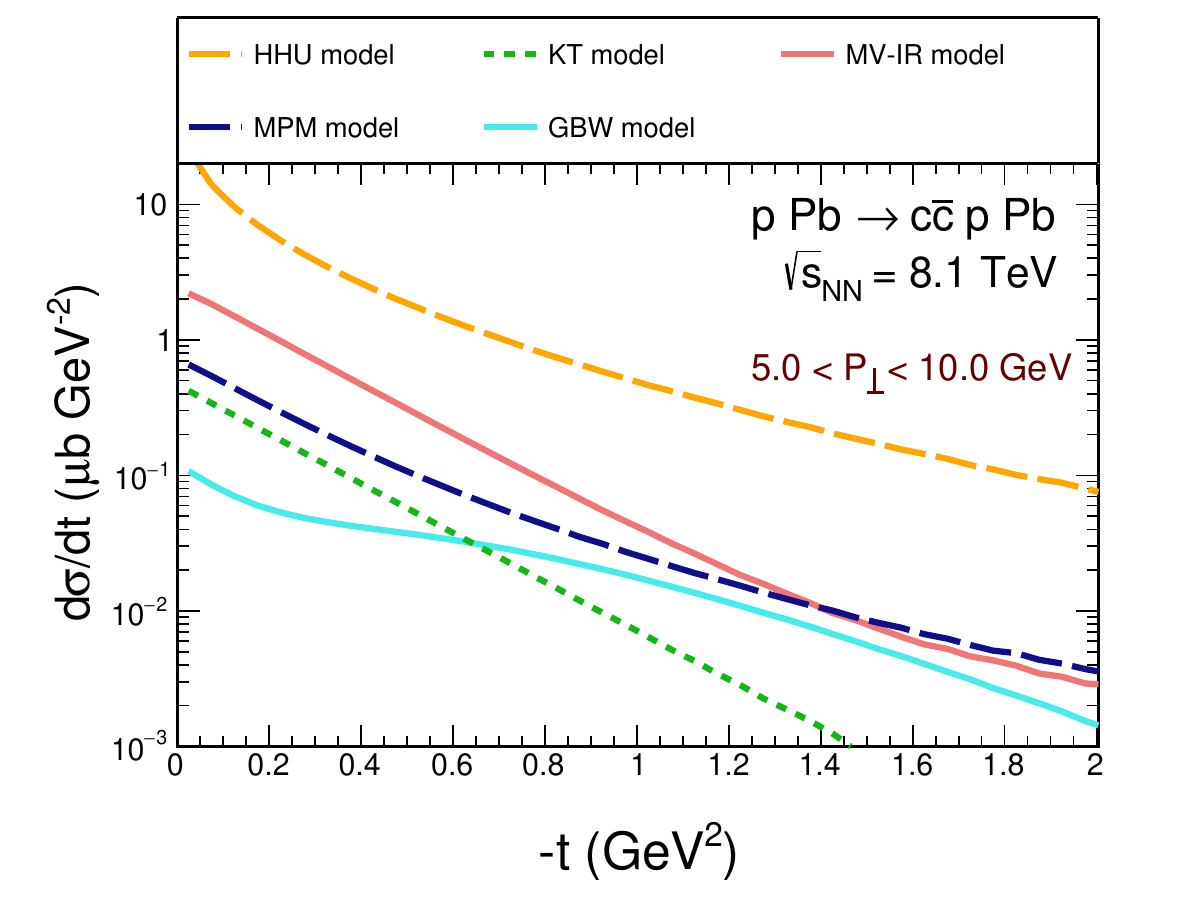}
   \caption{Distributions in $-t$ for $0.01 < P_{\perp} < 10.0$ GeV on the left and for $5.0 < P_{\perp} < 10.0$ GeV  on the right.}
\label{fig:t}
\end{figure}

Next, in Fig.~\ref{fig:t}, we show the distribution in the Mandelstam variable
\begin{eqnarray}
    t = -\frac{\Delta_\perp^2 + x_\Pom^2 m_p^2}{1-x_\Pom}
\end{eqnarray}
at the proton side. We see the typical diffractive peak for the $P_\perp$-integrated case. Again, the HHU  model stands out with a noticeable curvature and very sharp forward peaking. In the large-$P_\perp$ tail, the $t$-distribution flattens out considerably for all considered GTMD benchmark models. Notice that this case corresponds to a rather large diffractive mass and, therefore, to a sizeable longitudinal momentum transfer.
%%%%
%%%%
%%%%
%%%%

\begin{figure}
  \centering
  \includegraphics[width=.49\textwidth]{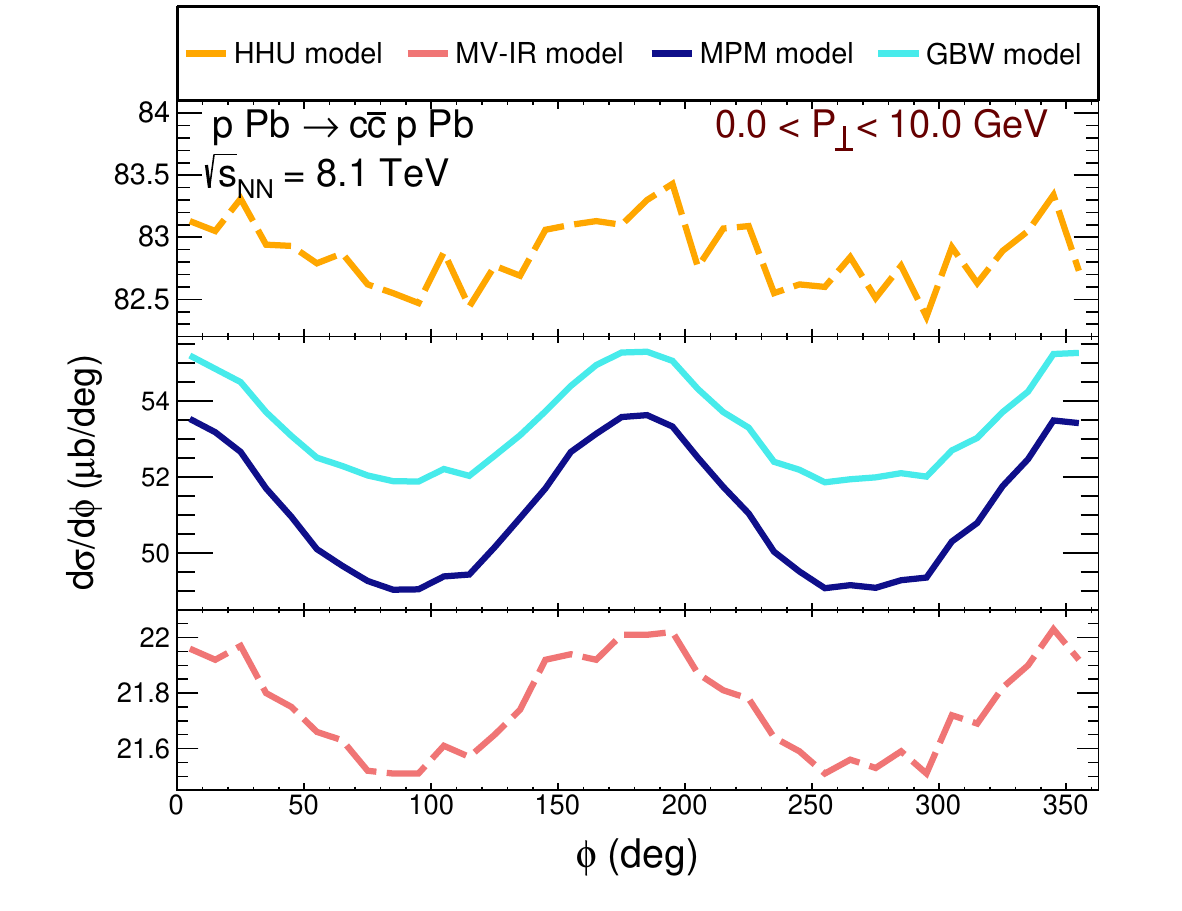}
  \includegraphics[width=.49\textwidth]{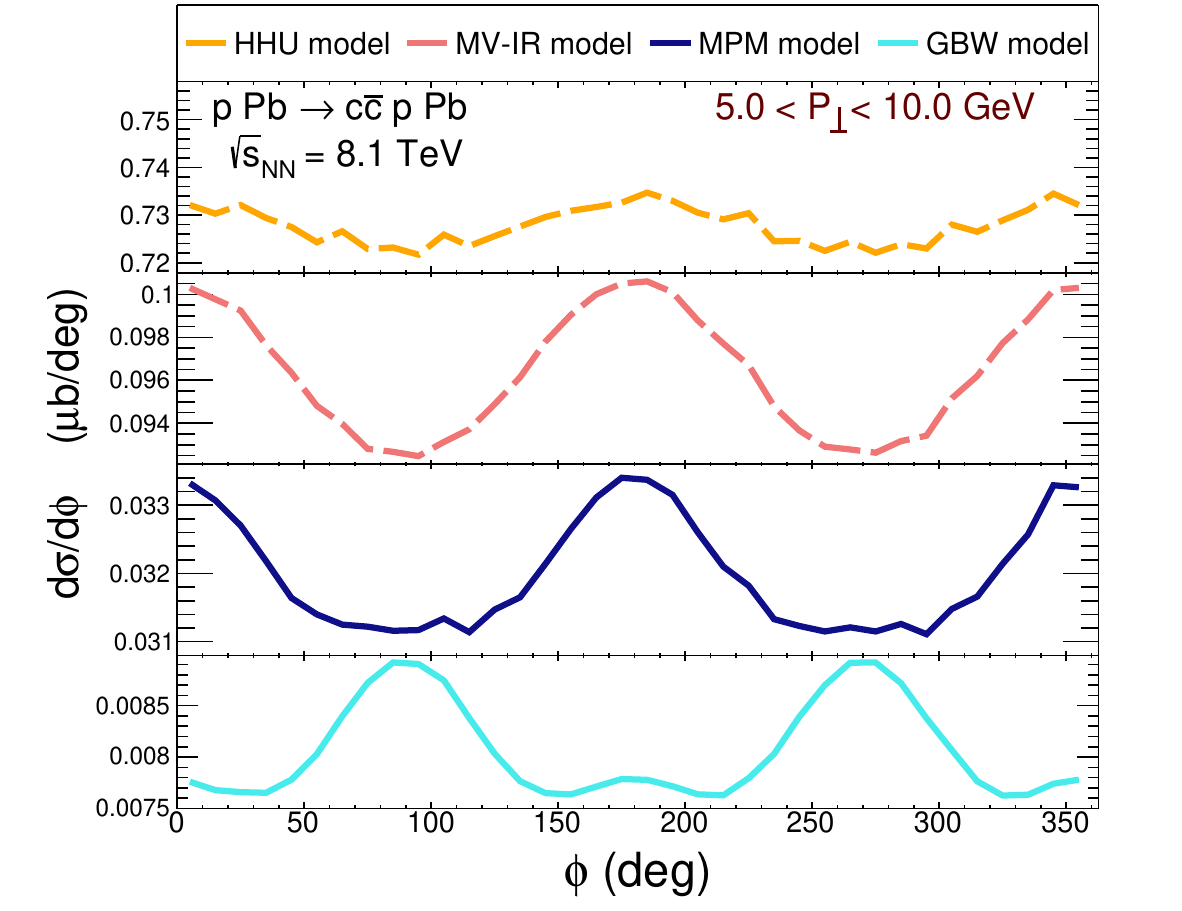}
   \caption{Distributions in the azimuthal angle $\phi$ between $\vec P_\perp$ and $\vec \Delta_\perp$ for $0.01 < P_{\perp} < 10.0 \, \rm{GeV}$ on the left and for $5.0 < P_{\perp} < 10.0 \, \rm{GeV}$ on the right.}
\label{fig:phi}
\end{figure}

We now turn to the angular correlations in azimuthal angle $\phi$ between $\vec P_\perp$ and $\vec \Delta_\perp$, shown in Fig.~\ref{fig:phi}, which are one of the main results of this work. There are in general clearly visible correlations. Notice that for the GBW and MPM GTMDs these correlations are of the ``geometric'' origin, such that in momentum space they are fully generated by the matrix element. The angular modulations obtained from the GBW or MPM GTMDs are of order 10 \%. We observe that at large $P_\perp$ they quickly drop and change shape. Notice, that the calculations with these two GTMDs include all possible harmonics, not only $\cos 2 \phi$. In the case when these correlations are computed accounting for the elliptic gluon distribution only (see Eq.~(\ref{eq:reg})), i.e.~when only $\cos 2 \phi$ modulations are included, they appear to be at the level of about 1 - 5\%.

\begin{figure}
  \centering
  \includegraphics[width=.49\textwidth]{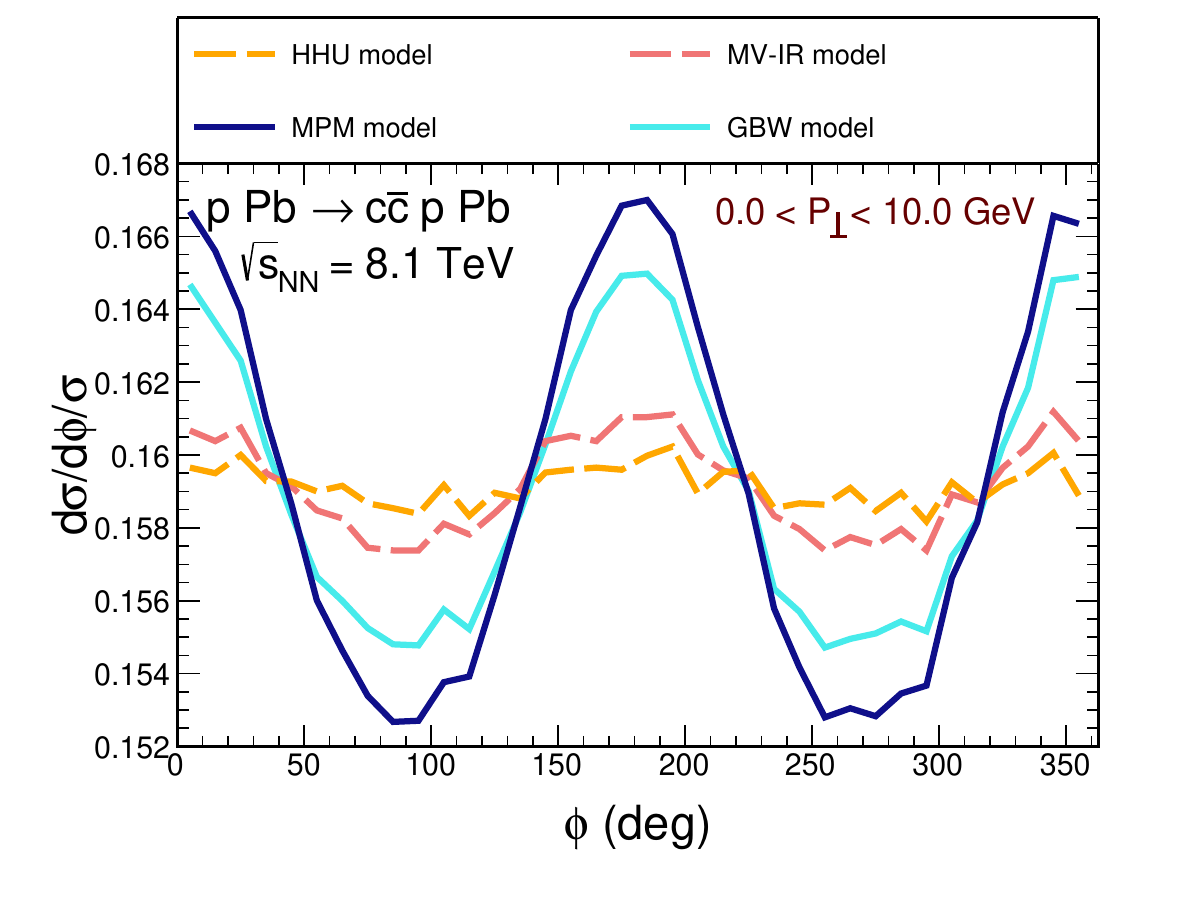}
  \includegraphics[width=.49\textwidth]{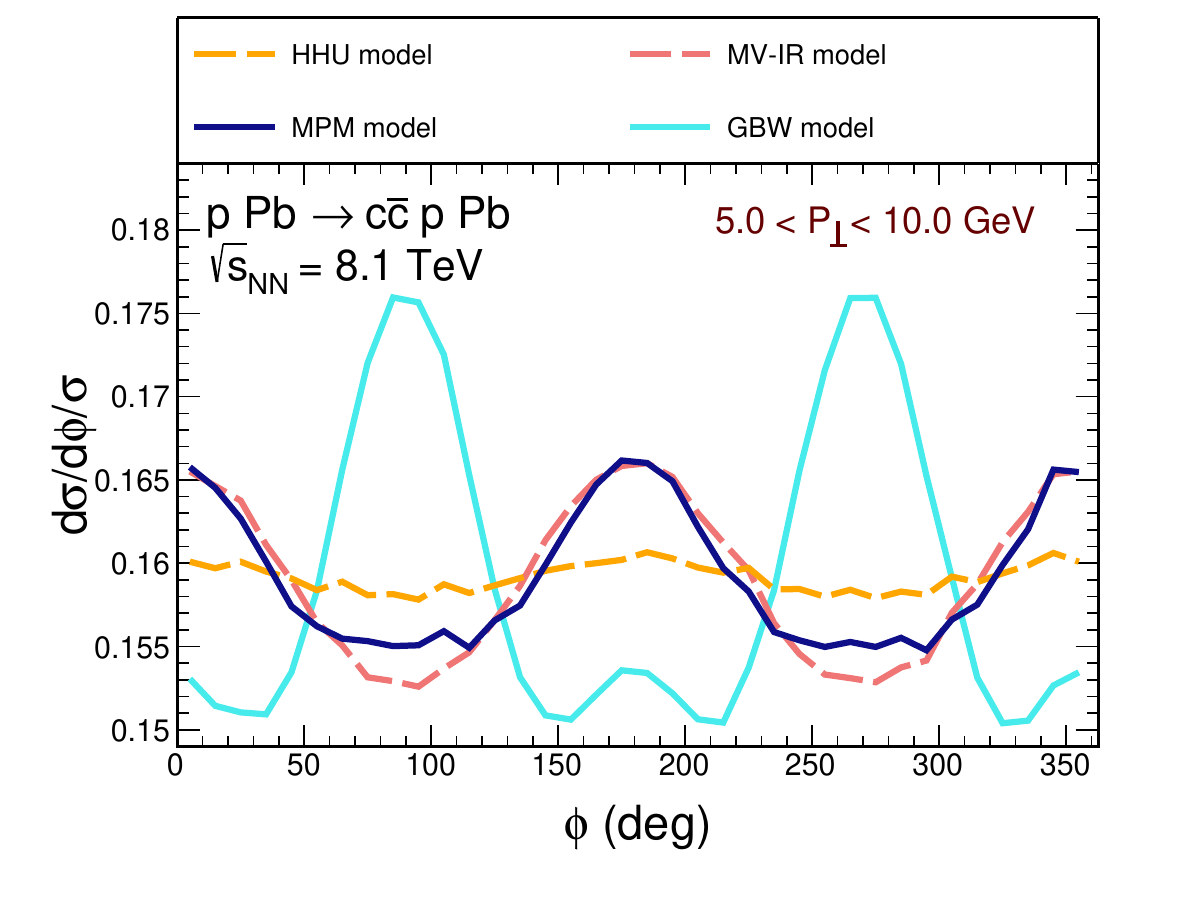}
   \caption{Distributions in the azimuthal angle $\phi$ between $\vec P_\perp$ and $\vec \Delta_\perp$ normalised to the total cross section for $0.01 < P_{\perp} < 10.0 \, \rm{GeV}$ on the left and for $5.0 < P_{\perp} < 10.0 \, \rm{GeV}$ on the right.}
\label{fig:wphi}
\end{figure}

In order to better visualize the strength of the azimuthal correlations in Fig.~\ref{fig:wphi}, we present also the angular distributions divided by the integrated cross section for a given GTMD model. Such normalised distributions are easier to compare for different GTMDs. Again, we show the results for two ranges of $P_{\perp}$ as explained in the figure. The azimuthal modulations are of the order of a few percent.

\begin{figure}
  \centering
  \includegraphics[width=0.45\textwidth]{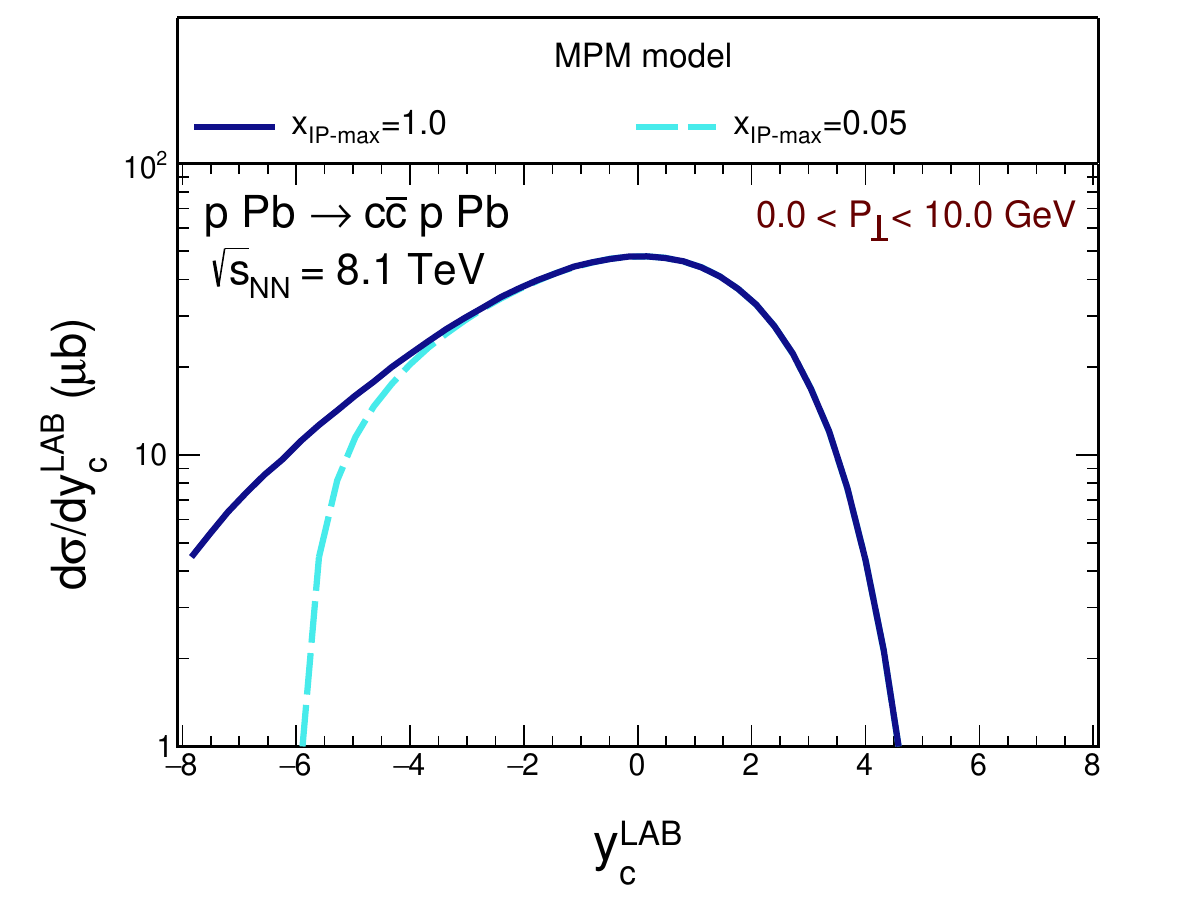}
  \includegraphics[width=0.45\textwidth]{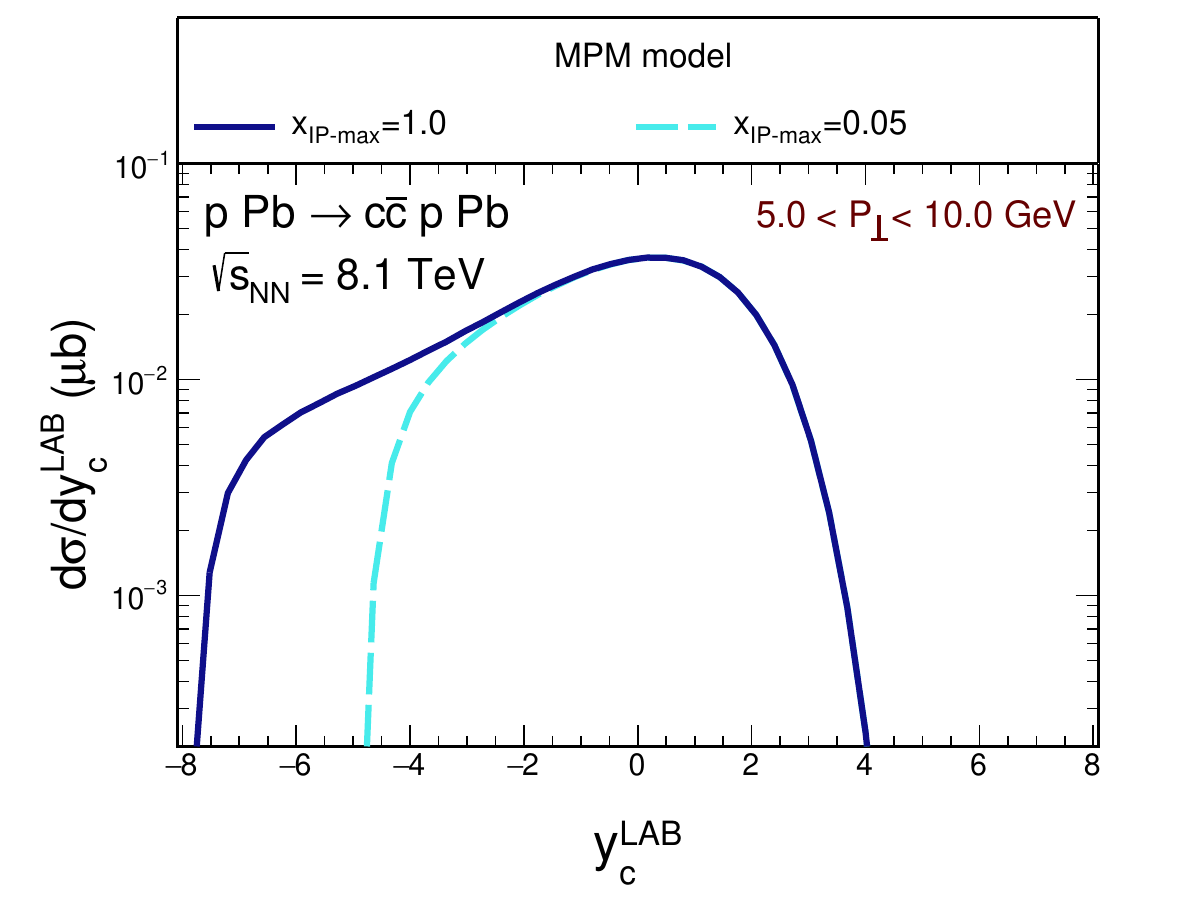}
  \caption{Distributions of MPM model in the $y^{\rm LAB}_{c}$ for $x_{\Pom}^{\rm  max}= 1.0$ and $x_{\Pom}^{\rm  max}= 0.05$.}
  \label{fig:y_xpomcut}
\end{figure}

\begin{figure}
  \centering
  \includegraphics[width=0.45\textwidth]{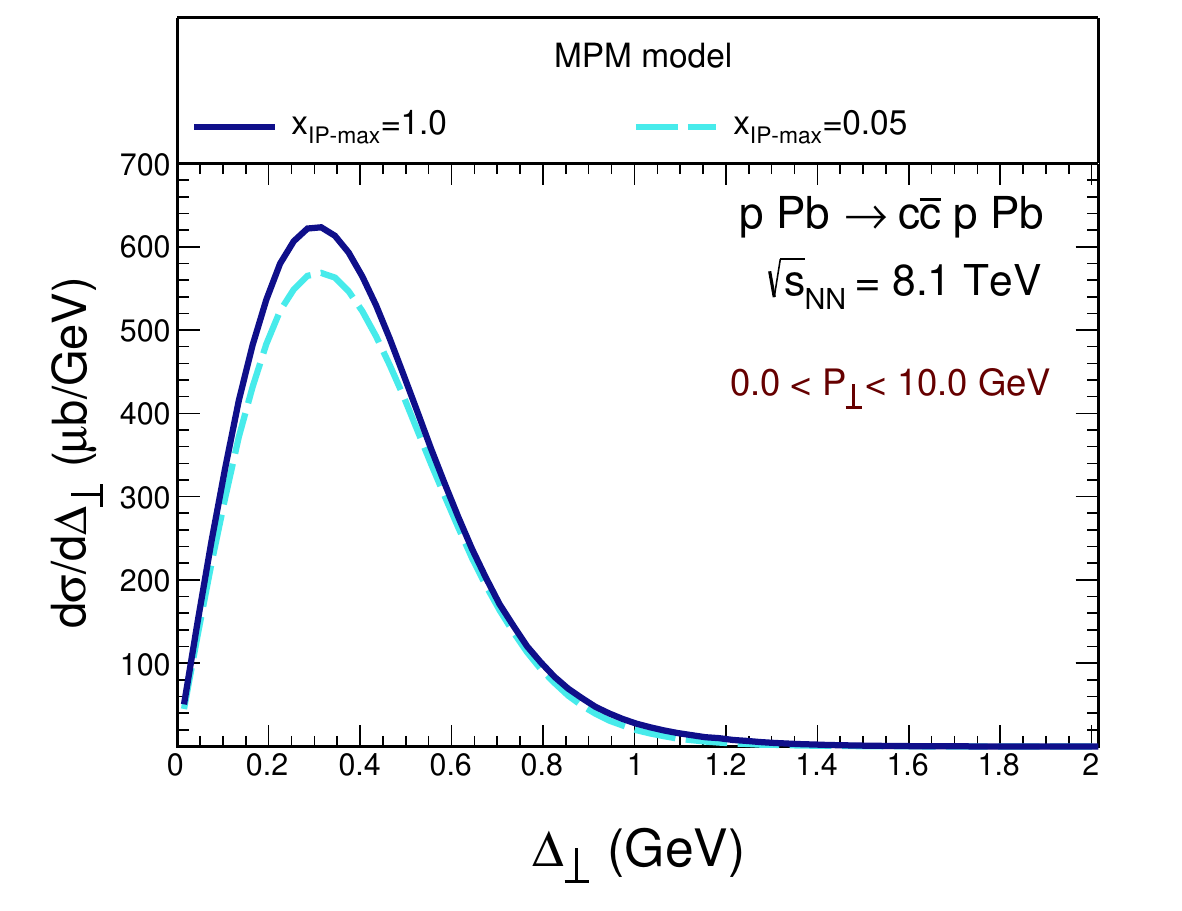}
  \includegraphics[width=0.45\textwidth]{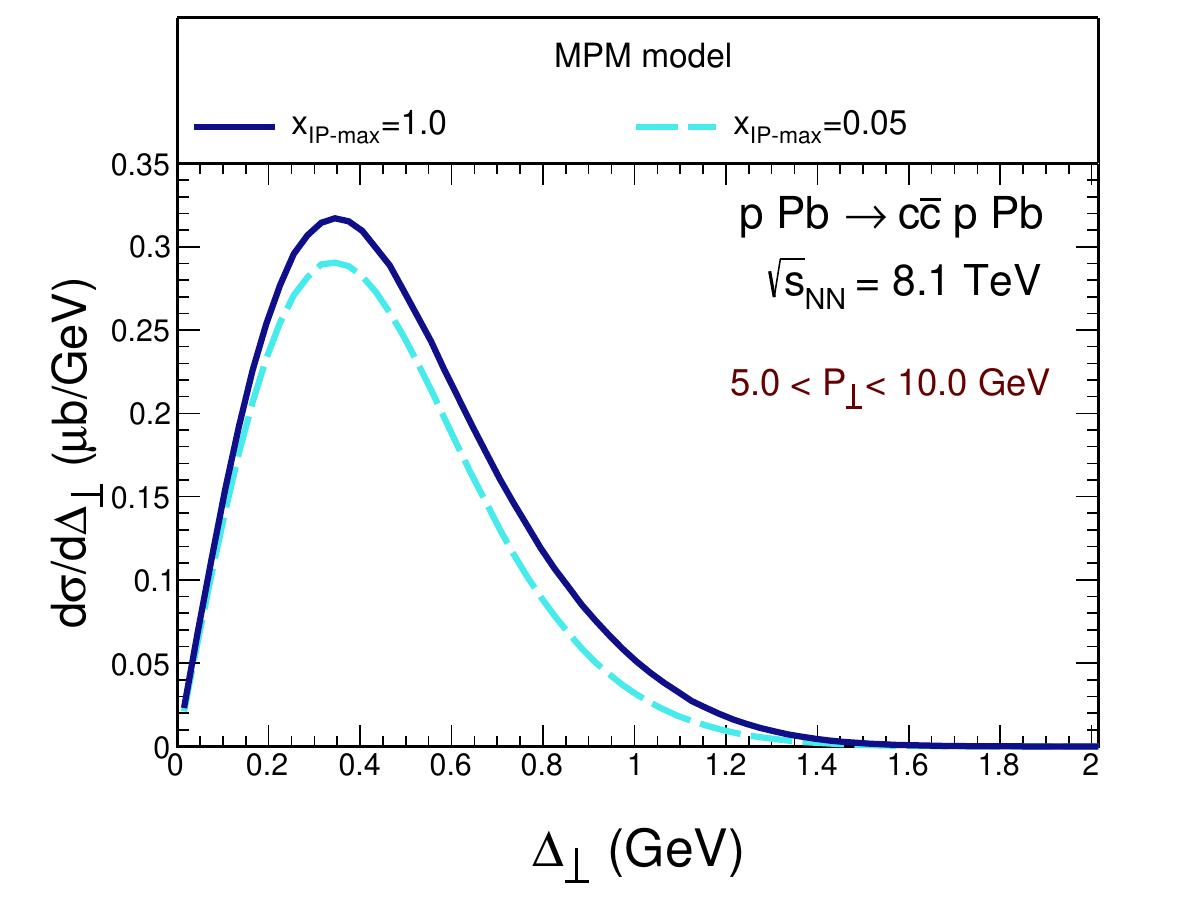}
  \caption{Distributions of MPM model in the $\Delta_{\perp}$ for $x_{\Pom}^{ \rm  max}= 1.0$ and $x_{\Pom}^{\rm max}= 0.05$.}
  \label{fig:delta_xpomcut}
\end{figure}

\begin{figure}
  \centering
  \includegraphics[width=0.45\textwidth]{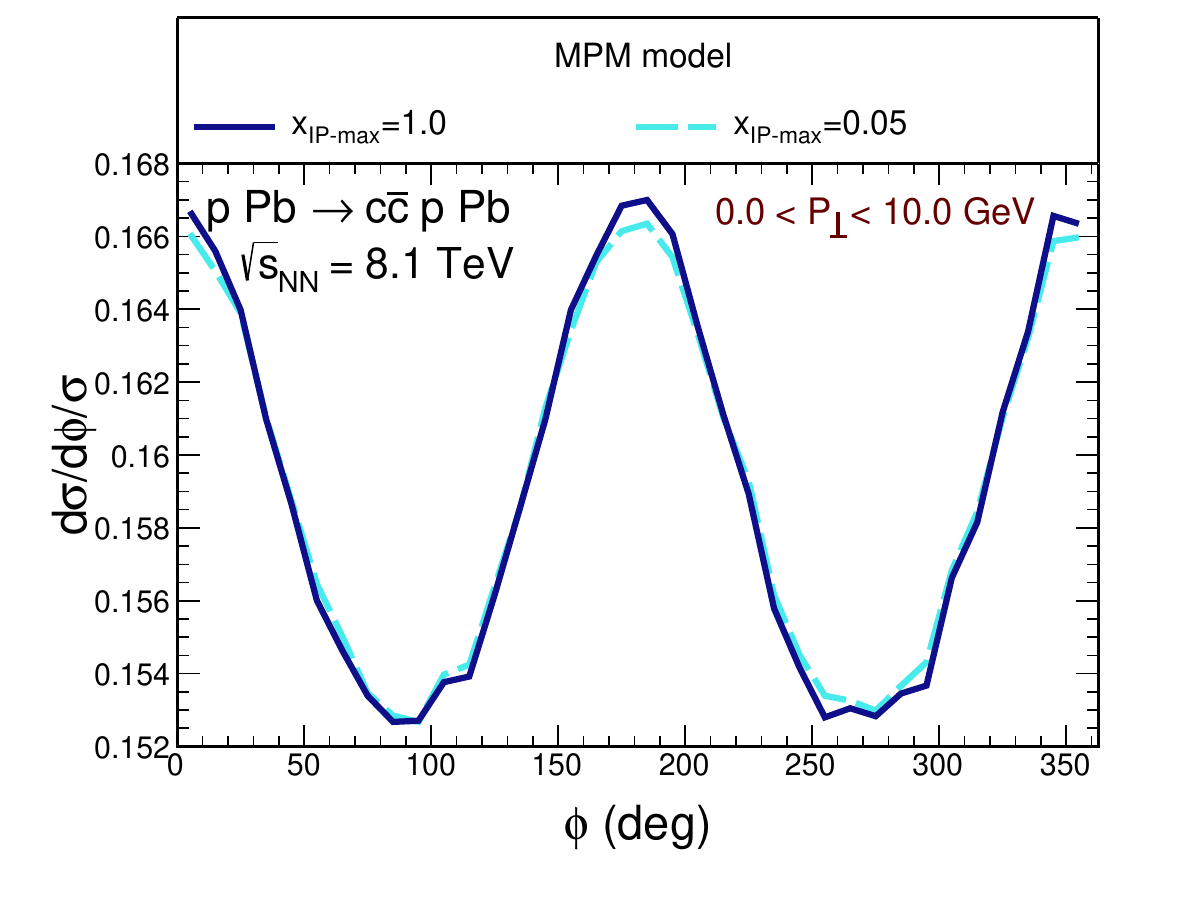}
  \includegraphics[width=0.45\textwidth]{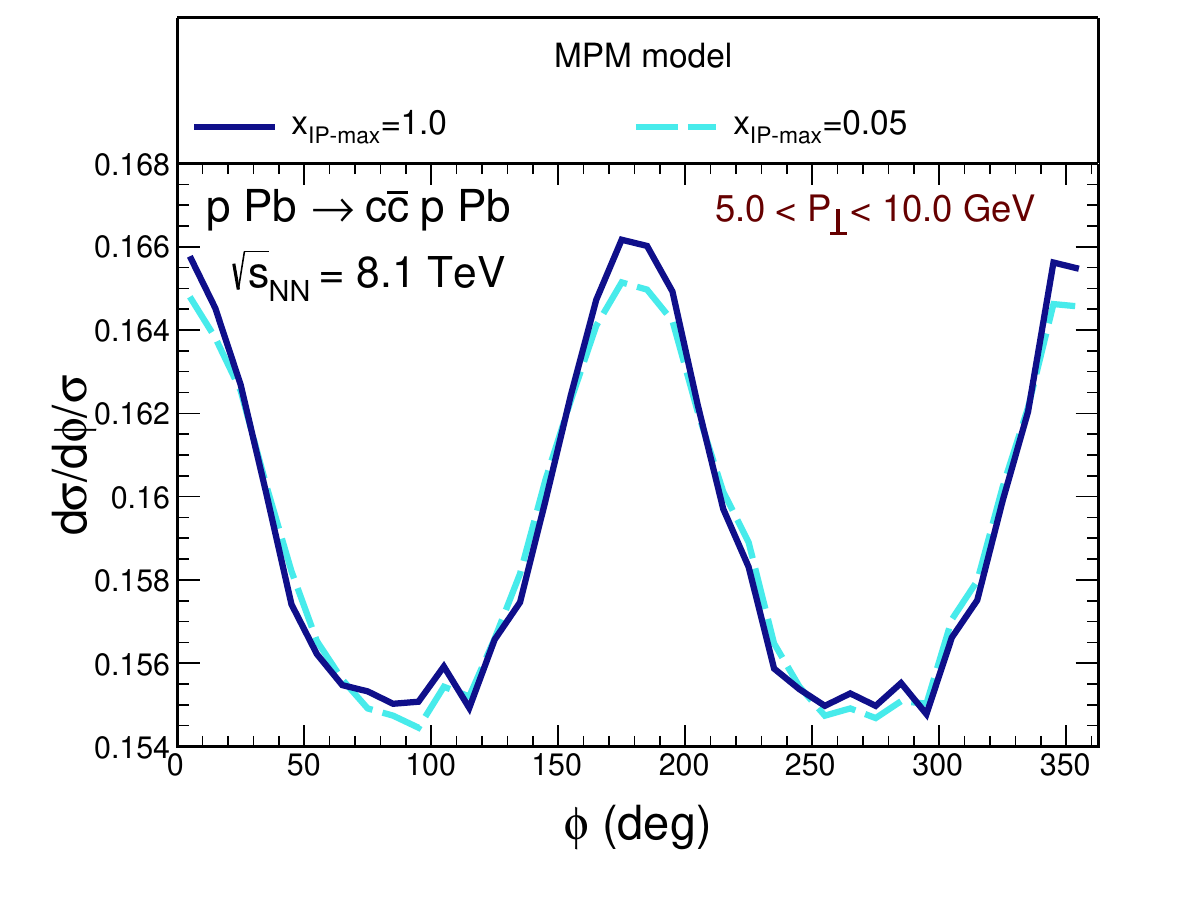}
  \caption{Distributions of MPM model in the azimuthal angle $\phi$ between $\vec P_\perp$ and $\vec \Delta_\perp$ normalised to the total cross section for $x_{\Pom}^{\rm max}= 1.0$ and $x_{\Pom}^{\rm  max}= 0.05$.}
  \label{fig:phi_xpomcut}
\end{figure}

Let us return to the issue on a cutoff in $x_\Pom$. In Figs.~\ref{fig:y_xpomcut}, \ref{fig:delta_xpomcut}, \ref{fig:phi_xpomcut} we show the effect of introducing an upper limit 
$x_\Pom^{\rm max}$ into our integration by example of three distributions. Here we used the MPM model for the GTMD. In Fig.~\ref{fig:y_xpomcut} we clearly see the expected effect of the cutoff on the distribution at large negative rapidities, beyond $y_c^{\rm LAB} \sim -4$ the cross section quickly drops. The effect of the cutoff is more marked in the high $P_\perp$ domain as seen from the right panel.
The $\Delta_\perp$ distribution shown in Fig.~\ref{fig:delta_xpomcut} is rather mildly affected, with again a stronger effect at large $P_\perp$.
Finally, the distribution in the azimuthal angle $\phi$ shown in Fig.~\ref{fig:phi_xpomcut} is practically unaffected.

As these are parton-level observables, these angular distributions are not directly measurable. For the case of dijets, one would expect soft-gluon corrections to have an impact, see e.g.~Ref.~\cite{Hatta:2021jcd}. For the more relevant case of exclusive (or inclusive diffractive) pairs of open heavy flavor mesons ($D$-mesons), a thorough study of hadronization corrections would be required. Such a study however goes beyond the scope of the present work.
Recently, the LHCb Collaboration was able to measure the inclusive $c \bar c$ dijets \cite{LHCb:2020frr} (see also \cite{Maciula:2022lzk}), whether such a measurement would also be possible in the kinematics discussed here is an open issue.
%%%%
%%%%
%%%%
%%%%
The different models give slightly different azimuthal angle distributions. Specially interesting are distributions for $5.0 < P_{\perp} < 10.0 \, \rm{GeV}$ where the localization of maxima for some models are reversed.

%%%%%%%%%%%%%%%%%%%%%%%%%%%%%%
\section{Conclusions}
\label{sec:conclusions}
%%%%%%%%%%%%%%%%%%%%%%%%%%%%%%
In this paper, we have presented several differential distributions for the diffractive photoproduction of $c \bar c$ pairs in the $pA  \to  p(c \bar c)A$ reaction at LHC energies. Our results have been obtained using different models for gluon GTMDs from the literature. Some GTMDs were obtained via the Fourier transform of the dipole $S$-matrix, $N(Y, \vec{r}_{\perp}, \vec{b}_{\perp})$. In this case, the GTMDs are regularized by an extra factor as done already in the literature for exclusive dijet photoproduction. This regularization leads to rather large uncertainties as far as normalization of the cross section is concerned. Therefore, for the distributions derived from dipole amplitudes, one has to focus rather on their shapes than on magnitudes.

In the present work, we go beyond the earlier analysis of Ref.~\cite{ReinkePelicer:2018gyh} considering realistic conditions of proton-lead collisions at the LHC and integrating over the phase space variables (such as quark rapidities and transverse momenta) in the measurable domains. We have paid special attention  to  azimuthal correlations which were proposed in the literature to test the models of small-$x$ dynamics encoded in the so-called elliptic gluon distributions. We find rather small azimuthal angle modulation in $\phi(\vec P_\perp, \vec \Delta_\perp)$. The modulations as well as the structure of maxima/minima depends on the GTMD models used in our analysis, so in principle the models can be tested in actual measurements at the LHC. For completeness, we have also presented the predictions for differential distributions in $P_\perp$, $\Delta_\perp$ (transverse momentum of the $c \bar c$ pair) and rapidity of the $c\bar c$ pair.

Regarding the possible measurement we need to repeat an important caveat. 
First, for the case of open heavy flavour meson production, a thorough analysis of hadronization corrections would be required to investigate if the 
sensitivity of our results to the GTMD remains visible at the hadron level.
For the case of dijet production, one expects  soft gluon emissions (Sudakov resummation) to have an impact, see e.g.~Ref.~\cite{Hatta:2021jcd}. We intend to come back to some of these issues in future work.
%%%%%%%%%%%%%%%%%%%%%%%%%%%%%%%%%%%%%%%%%%%%%%%%%%%%%%%%%%%%%%%%%%%%%%%%%%%%%
\acknowledgments
The authors would like to thank Yoshitaka Hatta for providing grids for numerical solution of the BK equation.
This work was partially supported by the Polish National Science Center grant UMO-2018/31/B/ST2/03537 and by the Center for Innovation and Transfer of Natural Sciences and Engineering Knowledge in Rzesz{\'o}w.
R.P.~is supported in part by the Swedish Research Council grant, contract number 2016-05996, as well as by the European Research Council (ERC) under the European Union's Horizon 2020 research and innovation programme (grant agreement No 668679).
%%%%%%%%%%%%%%%%%%%%%%%%%%%%%%%%%%%%%%%%%%%%%%%%%%%%%%%%%%%%%%%%%%%%%%%%%%%%%

\appendix
\section{Convolution of amplitude with elliptic GTMD}
\label{sec:appendix_me}

Here, we collect some steps necessary to analytically perform the azimuthal integration in the convolution of the hard amplitude with the azimuthally asymmetric part of the gluon GTMD.
The gluon GTMD is expanded as
%%%
\begin{eqnarray}
T(Y,\vec k_\perp ,\vec \Delta_\perp) &=& T_0(Y,k_\perp,\Delta_\perp) + 2 \, \cos 2(\phi_k - \phi_\Delta) \, T_\epsilon(Y,k_\perp,\Delta_\perp) \nonumber \\
&=& T_0(Y,k_\perp,\Delta_\perp) + 2 \, {2 (\vec k_\perp \cdot \vec \Delta_\perp)^2 - k_\perp^2 \Delta_\perp^2 \over k_\perp^2 \Delta_\perp^2} \, T_\epsilon(Y, k_\perp,\Delta_\perp) \, .
\end{eqnarray}
%%%%
The parts of the amplitudes that we are interested in, are
%%%%
\begin{eqnarray}
\delta \mathcal{\vec M}_{0}(\vec P_\perp ,\vec \Delta_{\perp}) &=& 2 \, \int \frac{d^2\vec k_\perp}{2\pi}\frac{\vec P_\perp -\vec k_\perp}{(\vec P_\perp -\vec k_\perp)^{2}+m_Q^2} \, {2 (\vec k_\perp \cdot \vec \Delta_\perp)^2 - k_\perp^2 \Delta_\perp^2 \over k_\perp^2 \Delta_\perp^2} \, T_\epsilon(Y,k_\perp,\Delta_\perp) \nonumber \\
&=& 2  \, \Big( 2 {\Delta_\perp^i \Delta_\perp^j \over \Delta^2_\perp} - \delta_{ij} \Big) 
\int \frac{d^2\vec k_\perp}{2\pi}\frac{\vec P_\perp -\vec k_\perp}{(\vec P_\perp -\vec k_\perp)^{2}+m_Q^2} \, {k_\perp^i k_\perp^j \over k_\perp^2} \, T_\epsilon(Y,k_\perp,\Delta_\perp) \, , \nonumber \\
\delta \mathcal{\vec M}_{1}(\vec P_\perp ,\vec \Delta_{\perp}) &=& 2  \, \Big( 2 {\Delta_\perp^i \Delta_\perp^j \over \Delta^2_\perp} - \delta_{ij} \Big) 
\int \frac{d^2\vec k_\perp}{2\pi}\frac{1}{(\vec P_\perp -\vec k_\perp)^{2}+m_Q^2} \, {k_\perp^i k_\perp^j \over k_\perp^2} \, T_\epsilon(Y,k_\perp,\Delta_\perp) \, . 
\label{eq:amplitude_elliptic}
\end{eqnarray}
%%%%
We now concentrate on the integral over azimuthal angles
%%%&
\begin{eqnarray}
\vec I_{ij} = \int_0^{2 \pi} {d \phi_k \over 2 \pi} \frac{\vec P_\perp -\vec k_\perp}{(\vec P_\perp -\vec k_\perp)^{2}+m_Q^2} {k_\perp^i k_\perp^j \over k^2_\perp} \, , I^1_{ij} = \int_0^{2 \pi} {d \phi_k \over 2 \pi} \frac{1}{(\vec P_\perp -\vec k_\perp)^{2}+m_Q^2} {k_\perp^i k_\perp^j \over k^2_\perp} \, .
\end{eqnarray}
%%%%
This vector integral will be proportional to $\vec P_\perp$. We therefore write
%%%%
\begin{eqnarray}
\vec I_{ij} = {\vec P_\perp \over P_\perp} \, I^0_{ij}(\vec P_\perp, k_\perp) \, ,
\end{eqnarray}
%%%%
with
%%%%
\begin{eqnarray}
I^0_{ij}(\vec P_\perp, k_\perp) = {1 \over P_\perp} \int_0^{2 \pi} {d \phi_k \over 2 \pi} \frac{ P^2_\perp -\vec P_\perp \cdot \vec k_\perp}{(\vec P_\perp -\vec k_\perp)^{2}+m_Q^2} \, {k_\perp^i k_\perp^j \over k^2_\perp} \, .
\end{eqnarray}
%%%%
Now we decompose $I^{0,1}_{ij}$ in terms of invariant functions of $|\vec P_\perp|$ and $|\vec k_\perp|$ and two orthogonal tensor structures, for which we choose
%%%%%
\begin{eqnarray}
I^{0,1}_{ij}(\vec P_\perp) = \Big( 2 {P^i_\perp P^j_\perp \over P^2_\perp} - \delta_{ij} \Big) \half I^{0,1}_\epsilon(P_\perp,k_\perp) + \delta_{ij} \, \half I^{0,1}_0(P_\perp,k_\perp) \, .
\end{eqnarray}
%%%%
When inserting this into Eq.~(\ref{eq:amplitude_elliptic}), the contraction with $\delta_{ij}$ vanishes, an we obtain
%%%%
\begin{eqnarray}
\delta \mathcal{\vec M}_{0}(\vec P_\perp ,\vec \Delta_{\perp}) &=& 2 \Big( 2 {(\vec P_\perp \cdot \vec \Delta_\perp)^2 \over P^2_\perp \Delta^2_\perp} -1 \Big) \, {\vec P_\perp \over P_\perp} \int_0^{\infty} k_\perp d k_\perp \, I^0_\epsilon(P_\perp ,k_\perp) \, T_\epsilon(Y,k_\perp , \Delta_\perp) \, \nonumber \\
&=& {\vec P_\perp \over P_\perp} \, 2 \cos 2 (\phi_\Delta - \phi_P) \int_0^\infty k_\perp  d k_\perp \, I^0_\epsilon(P_\perp, k_\perp) \, T_\epsilon(Y,k_\perp , \Delta_\perp) \, .
\end{eqnarray}
%%%%
We still need to find the expression for the azimuthal integral $I_\epsilon(P_\perp, k_\perp)$. Introducing
%%%%
\begin{eqnarray}
a = P^2_\perp + k^2_\perp + m^2 \, , \qquad b = 2 P_\perp k_\perp \, ,
\end{eqnarray}
%%%%
we obtain
%%%%
\begin{eqnarray}
   I^1_\epsilon(P_\perp, k_\perp) = \int_0^{2 \pi} \frac{d \phi}{2 \pi}  {\cos 2 \phi \over a - b \cos \phi} \equiv g(a,b) \, ,
\end{eqnarray}
%%%
and
%%%
\begin{eqnarray}
P_\perp I^0_\epsilon(P_\perp, k_\perp) &=& \int_0^{2 \pi} {d \phi \over 2 \pi} {P^2_\perp - \half b \cos \phi \over a - b \cos \phi} \, \cos 2 \phi \, \nonumber \\
&=&  \int_0^{2 \pi} {d \phi \over 2 \pi} {P^2_\perp - \half a + \half (a - b \cos \phi ) \over a - b \cos \phi} \, \cos 2 \phi \, \nonumber \\
&=& (P^2_\perp - \half a) \int_0^{2 \pi} {d \phi \over 2 \pi}  {\cos 2 \phi \over a - b \cos \phi} \nonumber \\
&=& \half (P^2_\perp - k^2_\perp - m^2) g(a,b) \, ,
\end{eqnarray}
%%%%
where 
%%%%
\begin{eqnarray}
g(a,b) &=& {1 \over b^2} {2 a^2 - b^2 - 2a \sqrt{a^2 - b^2} \over \sqrt{a^2 - b^2}} \nonumber \\
&=& \frac{1}{2 P_\perp^2 k_\perp^2} \Big(
\frac{(P_\perp^2 + k_\perp^2 + m_Q^2)^2 - 2 P_\perp^2 k_\perp^2}{\sqrt{(P_\perp^2 - k_\perp^2 - m_Q^2)^2 + 4 P_\perp^2 m_Q^2 }} - (P_\perp^2 + k_\perp^2 + m_Q^2) \Big) \, .
\end{eqnarray}
%%%%
Here, we used the identity 
%%%%
\begin{eqnarray}
a^2 - b^2 = (P^2_\perp + k^2_\perp + m^2)^2 - 4 P^2_\perp k^2_\perp = (P^2_\perp - k^2_\perp - m^2)^2 + 4 P^2_\perp m^2 \, .
\end{eqnarray}
%%%

\bibliography{main}

%merlin.mbs apsrev4-1.bst 2010-07-25 4.21a (PWD, AO, DPC) hacked
%Control: key (0)
%Control: author (8) initials jnrlst
%Control: editor formatted (1) identically to author
%Control: production of article title (-1) disabled
%Control: page (0) single
%Control: year (1) truncated
%Control: production of eprint (0) enabled
\begin{thebibliography}{46}%
\makeatletter
\providecommand \@ifxundefined [1]{%
 \@ifx{#1\undefined}
}%
\providecommand \@ifnum [1]{%
 \ifnum #1\expandafter \@firstoftwo
 \else \expandafter \@secondoftwo
 \fi
}%
\providecommand \@ifx [1]{%
 \ifx #1\expandafter \@firstoftwo
 \else \expandafter \@secondoftwo
 \fi
}%
\providecommand \natexlab [1]{#1}%
\providecommand \enquote  [1]{``#1''}%
\providecommand \bibnamefont  [1]{#1}%
\providecommand \bibfnamefont [1]{#1}%
\providecommand \citenamefont [1]{#1}%
\providecommand \href@noop [0]{\@secondoftwo}%
\providecommand \href [0]{\begingroup \@sanitize@url \@href}%
\providecommand \@href[1]{\@@startlink{#1}\@@href}%
\providecommand \@@href[1]{\endgroup#1\@@endlink}%
\providecommand \@sanitize@url [0]{\catcode `\\12\catcode `\$12\catcode `\&12\catcode `\#12\catcode `\^12\catcode `\_12\catcode `\%12\relax}%
\providecommand \@@startlink[1]{}%
\providecommand \@@endlink[0]{}%
\providecommand \url  [0]{\begingroup\@sanitize@url \@url }%
\providecommand \@url [1]{\endgroup\@href {#1}{\urlprefix }}%
\providecommand \urlprefix  [0]{URL }%
\providecommand \Eprint [0]{\href }%
\providecommand \doibase [0]{http://dx.doi.org/}%
\providecommand \selectlanguage [0]{\@gobble}%
\providecommand \bibinfo  [0]{\@secondoftwo}%
\providecommand \bibfield  [0]{\@secondoftwo}%
\providecommand \translation [1]{[#1]}%
\providecommand \BibitemOpen [0]{}%
\providecommand \bibitemStop [0]{}%
\providecommand \bibitemNoStop [0]{.\EOS\space}%
\providecommand \EOS [0]{\spacefactor3000\relax}%
\providecommand \BibitemShut  [1]{\csname bibitem#1\endcsname}%
\let\auto@bib@innerbib\@empty
%</preamble>
\bibitem [{\citenamefont {Ashery}(2006)}]{Ashery:2006zw}%
  \BibitemOpen
  \bibfield  {author} {\bibinfo {author} {\bibfnamefont {D.}~\bibnamefont {Ashery}},\ }\href {\doibase 10.1016/j.ppnp.2005.08.003} {\bibfield  {journal} {\bibinfo  {journal} {Prog. Part. Nucl. Phys.}\ }\textbf {\bibinfo {volume} {56}},\ \bibinfo {pages} {279} (\bibinfo {year} {2006})}\BibitemShut {NoStop}%
\bibitem [{\citenamefont {Nikolaev}\ and\ \citenamefont {Zakharov}(1994{\natexlab{a}})}]{Nikolaev:1994cd}%
  \BibitemOpen
  \bibfield  {author} {\bibinfo {author} {\bibfnamefont {N.~N.}\ \bibnamefont {Nikolaev}}\ and\ \bibinfo {author} {\bibfnamefont {B.~G.}\ \bibnamefont {Zakharov}},\ }\href {\doibase 10.1016/0370-2693(94)90876-1} {\bibfield  {journal} {\bibinfo  {journal} {Phys. Lett. B}\ }\textbf {\bibinfo {volume} {332}},\ \bibinfo {pages} {177} (\bibinfo {year} {1994}{\natexlab{a}})},\ \Eprint {http://arxiv.org/abs/hep-ph/9403281} {arXiv:hep-ph/9403281} \BibitemShut {NoStop}%
\bibitem [{\citenamefont {Nikolaev}\ \emph {et~al.}(2001)\citenamefont {Nikolaev}, \citenamefont {Sch{\"a}fer},\ and\ \citenamefont {Schwiete}}]{Nikolaev:2000sh}%
  \BibitemOpen
  \bibfield  {author} {\bibinfo {author} {\bibfnamefont {N.~N.}\ \bibnamefont {Nikolaev}}, \bibinfo {author} {\bibfnamefont {W.}~\bibnamefont {Sch{\"a}fer}}, \ and\ \bibinfo {author} {\bibfnamefont {G.}~\bibnamefont {Schwiete}},\ }\href {\doibase 10.1103/PhysRevD.63.014020} {\bibfield  {journal} {\bibinfo  {journal} {Phys. Rev. D}\ }\textbf {\bibinfo {volume} {63}},\ \bibinfo {pages} {014020} (\bibinfo {year} {2001})},\ \Eprint {http://arxiv.org/abs/hep-ph/0009038} {arXiv:hep-ph/0009038} \BibitemShut {NoStop}%
\bibitem [{\citenamefont {Ji}(2003)}]{Ji:2003ak}%
  \BibitemOpen
  \bibfield  {author} {\bibinfo {author} {\bibfnamefont {X.-d.}\ \bibnamefont {Ji}},\ }\href {\doibase 10.1103/PhysRevLett.91.062001} {\bibfield  {journal} {\bibinfo  {journal} {Phys. Rev. Lett.}\ }\textbf {\bibinfo {volume} {91}},\ \bibinfo {pages} {062001} (\bibinfo {year} {2003})},\ \Eprint {http://arxiv.org/abs/hep-ph/0304037} {arXiv:hep-ph/0304037} \BibitemShut {NoStop}%
\bibitem [{\citenamefont {Belitsky}\ \emph {et~al.}(2004)\citenamefont {Belitsky}, \citenamefont {Ji},\ and\ \citenamefont {Yuan}}]{Belitsky:2003nz}%
  \BibitemOpen
  \bibfield  {author} {\bibinfo {author} {\bibfnamefont {A.~V.}\ \bibnamefont {Belitsky}}, \bibinfo {author} {\bibfnamefont {X.-d.}\ \bibnamefont {Ji}}, \ and\ \bibinfo {author} {\bibfnamefont {F.}~\bibnamefont {Yuan}},\ }\href {\doibase 10.1103/PhysRevD.69.074014} {\bibfield  {journal} {\bibinfo  {journal} {Phys. Rev. D}\ }\textbf {\bibinfo {volume} {69}},\ \bibinfo {pages} {074014} (\bibinfo {year} {2004})},\ \Eprint {http://arxiv.org/abs/hep-ph/0307383} {arXiv:hep-ph/0307383} \BibitemShut {NoStop}%
\bibitem [{\citenamefont {Meissner}\ \emph {et~al.}(2009)\citenamefont {Meissner}, \citenamefont {Metz},\ and\ \citenamefont {Schlegel}}]{Meissner:2009ww}%
  \BibitemOpen
  \bibfield  {author} {\bibinfo {author} {\bibfnamefont {S.}~\bibnamefont {Meissner}}, \bibinfo {author} {\bibfnamefont {A.}~\bibnamefont {Metz}}, \ and\ \bibinfo {author} {\bibfnamefont {M.}~\bibnamefont {Schlegel}},\ }\href {\doibase 10.1088/1126-6708/2009/08/056} {\bibfield  {journal} {\bibinfo  {journal} {JHEP}\ }\textbf {\bibinfo {volume} {08}},\ \bibinfo {pages} {056} (\bibinfo {year} {2009})},\ \Eprint {http://arxiv.org/abs/0906.5323} {arXiv:0906.5323 [hep-ph]} \BibitemShut {NoStop}%
\bibitem [{\citenamefont {Lorc\'e}\ and\ \citenamefont {Pasquini}(2013)}]{Lorce:2013pza}%
  \BibitemOpen
  \bibfield  {author} {\bibinfo {author} {\bibfnamefont {C.}~\bibnamefont {Lorc\'e}}\ and\ \bibinfo {author} {\bibfnamefont {B.}~\bibnamefont {Pasquini}},\ }\href {\doibase 10.1007/JHEP09(2013)138} {\bibfield  {journal} {\bibinfo  {journal} {JHEP}\ }\textbf {\bibinfo {volume} {09}},\ \bibinfo {pages} {138} (\bibinfo {year} {2013})},\ \Eprint {http://arxiv.org/abs/1307.4497} {arXiv:1307.4497 [hep-ph]} \BibitemShut {NoStop}%
\bibitem [{\citenamefont {Boussarie}\ \emph {et~al.}(2023)\citenamefont {Boussarie} \emph {et~al.}}]{Boussarie:2023izj}%
  \BibitemOpen
  \bibfield  {author} {\bibinfo {author} {\bibfnamefont {R.}~\bibnamefont {Boussarie}} \emph {et~al.},\ }\href@noop {} {\  (\bibinfo {year} {2023})},\ \Eprint {http://arxiv.org/abs/2304.03302} {arXiv:2304.03302 [hep-ph]} \BibitemShut {NoStop}%
\bibitem [{\citenamefont {Hagiwara}\ \emph {et~al.}(2016)\citenamefont {Hagiwara}, \citenamefont {Hatta},\ and\ \citenamefont {Ueda}}]{Hagiwara:2016kam}%
  \BibitemOpen
  \bibfield  {author} {\bibinfo {author} {\bibfnamefont {Y.}~\bibnamefont {Hagiwara}}, \bibinfo {author} {\bibfnamefont {Y.}~\bibnamefont {Hatta}}, \ and\ \bibinfo {author} {\bibfnamefont {T.}~\bibnamefont {Ueda}},\ }\href {\doibase 10.1103/PhysRevD.94.094036} {\bibfield  {journal} {\bibinfo  {journal} {Phys. Rev. D}\ }\textbf {\bibinfo {volume} {94}},\ \bibinfo {pages} {094036} (\bibinfo {year} {2016})},\ \Eprint {http://arxiv.org/abs/1609.05773} {arXiv:1609.05773 [hep-ph]} \BibitemShut {NoStop}%
\bibitem [{\citenamefont {Kopeliovich}\ \emph {et~al.}(1981)\citenamefont {Kopeliovich}, \citenamefont {Lapidus},\ and\ \citenamefont {Zamolodchikov}}]{Kopeliovich:1981pz}%
  \BibitemOpen
  \bibfield  {author} {\bibinfo {author} {\bibfnamefont {B.~Z.}\ \bibnamefont {Kopeliovich}}, \bibinfo {author} {\bibfnamefont {L.~I.}\ \bibnamefont {Lapidus}}, \ and\ \bibinfo {author} {\bibfnamefont {A.~B.}\ \bibnamefont {Zamolodchikov}},\ }\href@noop {} {\bibfield  {journal} {\bibinfo  {journal} {JETP Lett.}\ }\textbf {\bibinfo {volume} {33}},\ \bibinfo {pages} {595} (\bibinfo {year} {1981})}\BibitemShut {NoStop}%
\bibitem [{\citenamefont {Nikolaev}\ and\ \citenamefont {Zakharov}(1992)}]{Nikolaev:1991et}%
  \BibitemOpen
  \bibfield  {author} {\bibinfo {author} {\bibfnamefont {N.}~\bibnamefont {Nikolaev}}\ and\ \bibinfo {author} {\bibfnamefont {B.~G.}\ \bibnamefont {Zakharov}},\ }\href {\doibase 10.1007/BF01597573} {\bibfield  {journal} {\bibinfo  {journal} {Z. Phys. C}\ }\textbf {\bibinfo {volume} {53}},\ \bibinfo {pages} {331} (\bibinfo {year} {1992})}\BibitemShut {NoStop}%
\bibitem [{\citenamefont {Kopeliovich}\ \emph {et~al.}(2008)\citenamefont {Kopeliovich}, \citenamefont {Pirner}, \citenamefont {Rezaeian},\ and\ \citenamefont {Schmidt}}]{Kopeliovich:2007fv}%
  \BibitemOpen
  \bibfield  {author} {\bibinfo {author} {\bibfnamefont {B.~Z.}\ \bibnamefont {Kopeliovich}}, \bibinfo {author} {\bibfnamefont {H.~J.}\ \bibnamefont {Pirner}}, \bibinfo {author} {\bibfnamefont {A.~H.}\ \bibnamefont {Rezaeian}}, \ and\ \bibinfo {author} {\bibfnamefont {I.}~\bibnamefont {Schmidt}},\ }\href {\doibase 10.1103/PhysRevD.77.034011} {\bibfield  {journal} {\bibinfo  {journal} {Phys. Rev. D}\ }\textbf {\bibinfo {volume} {77}},\ \bibinfo {pages} {034011} (\bibinfo {year} {2008})},\ \Eprint {http://arxiv.org/abs/0711.3010} {arXiv:0711.3010 [hep-ph]} \BibitemShut {NoStop}%
\bibitem [{\citenamefont {Iancu}\ and\ \citenamefont {Rezaeian}(2017)}]{Iancu:2017fzn}%
  \BibitemOpen
  \bibfield  {author} {\bibinfo {author} {\bibfnamefont {E.}~\bibnamefont {Iancu}}\ and\ \bibinfo {author} {\bibfnamefont {A.~H.}\ \bibnamefont {Rezaeian}},\ }\href {\doibase 10.1103/PhysRevD.95.094003} {\bibfield  {journal} {\bibinfo  {journal} {Phys. Rev. D}\ }\textbf {\bibinfo {volume} {95}},\ \bibinfo {pages} {094003} (\bibinfo {year} {2017})},\ \Eprint {http://arxiv.org/abs/1702.03943} {arXiv:1702.03943 [hep-ph]} \BibitemShut {NoStop}%
\bibitem [{\citenamefont {Hatta}\ \emph {et~al.}(2016)\citenamefont {Hatta}, \citenamefont {Xiao},\ and\ \citenamefont {Yuan}}]{Hatta:2016dxp}%
  \BibitemOpen
  \bibfield  {author} {\bibinfo {author} {\bibfnamefont {Y.}~\bibnamefont {Hatta}}, \bibinfo {author} {\bibfnamefont {B.-W.}\ \bibnamefont {Xiao}}, \ and\ \bibinfo {author} {\bibfnamefont {F.}~\bibnamefont {Yuan}},\ }\href {\doibase 10.1103/PhysRevLett.116.202301} {\bibfield  {journal} {\bibinfo  {journal} {Phys. Rev. Lett.}\ }\textbf {\bibinfo {volume} {116}},\ \bibinfo {pages} {202301} (\bibinfo {year} {2016})},\ \Eprint {http://arxiv.org/abs/1601.01585} {arXiv:1601.01585 [hep-ph]} \BibitemShut {NoStop}%
\bibitem [{\citenamefont {Boer}\ and\ \citenamefont {Setyadi}(2021)}]{Boer:2021upt}%
  \BibitemOpen
  \bibfield  {author} {\bibinfo {author} {\bibfnamefont {D.}~\bibnamefont {Boer}}\ and\ \bibinfo {author} {\bibfnamefont {C.}~\bibnamefont {Setyadi}},\ }\href {\doibase 10.1103/PhysRevD.104.074006} {\bibfield  {journal} {\bibinfo  {journal} {Phys. Rev. D}\ }\textbf {\bibinfo {volume} {104}},\ \bibinfo {pages} {074006} (\bibinfo {year} {2021})},\ \Eprint {http://arxiv.org/abs/2106.15148} {arXiv:2106.15148 [hep-ph]} \BibitemShut {NoStop}%
\bibitem [{\citenamefont {Altinoluk}\ \emph {et~al.}(2016)\citenamefont {Altinoluk}, \citenamefont {Armesto}, \citenamefont {Beuf},\ and\ \citenamefont {Rezaeian}}]{Altinoluk:2015dpi}%
  \BibitemOpen
  \bibfield  {author} {\bibinfo {author} {\bibfnamefont {T.}~\bibnamefont {Altinoluk}}, \bibinfo {author} {\bibfnamefont {N.}~\bibnamefont {Armesto}}, \bibinfo {author} {\bibfnamefont {G.}~\bibnamefont {Beuf}}, \ and\ \bibinfo {author} {\bibfnamefont {A.~H.}\ \bibnamefont {Rezaeian}},\ }\href {\doibase 10.1016/j.physletb.2016.05.032} {\bibfield  {journal} {\bibinfo  {journal} {Phys. Lett. B}\ }\textbf {\bibinfo {volume} {758}},\ \bibinfo {pages} {373} (\bibinfo {year} {2016})},\ \Eprint {http://arxiv.org/abs/1511.07452} {arXiv:1511.07452 [hep-ph]} \BibitemShut {NoStop}%
\bibitem [{\citenamefont {M\"antysaari}\ \emph {et~al.}(2019)\citenamefont {M\"antysaari}, \citenamefont {Mueller},\ and\ \citenamefont {Schenke}}]{Mantysaari:2019csc}%
  \BibitemOpen
  \bibfield  {author} {\bibinfo {author} {\bibfnamefont {H.}~\bibnamefont {M\"antysaari}}, \bibinfo {author} {\bibfnamefont {N.}~\bibnamefont {Mueller}}, \ and\ \bibinfo {author} {\bibfnamefont {B.}~\bibnamefont {Schenke}},\ }\href {\doibase 10.1103/PhysRevD.99.074004} {\bibfield  {journal} {\bibinfo  {journal} {Phys. Rev. D}\ }\textbf {\bibinfo {volume} {99}},\ \bibinfo {pages} {074004} (\bibinfo {year} {2019})},\ \Eprint {http://arxiv.org/abs/1902.05087} {arXiv:1902.05087 [hep-ph]} \BibitemShut {NoStop}%
\bibitem [{\citenamefont {Salazar}\ and\ \citenamefont {Schenke}(2019)}]{Salazar:2019ncp}%
  \BibitemOpen
  \bibfield  {author} {\bibinfo {author} {\bibfnamefont {F.}~\bibnamefont {Salazar}}\ and\ \bibinfo {author} {\bibfnamefont {B.}~\bibnamefont {Schenke}},\ }\href {\doibase 10.1103/PhysRevD.100.034007} {\bibfield  {journal} {\bibinfo  {journal} {Phys. Rev. D}\ }\textbf {\bibinfo {volume} {100}},\ \bibinfo {pages} {034007} (\bibinfo {year} {2019})},\ \Eprint {http://arxiv.org/abs/1905.03763} {arXiv:1905.03763 [hep-ph]} \BibitemShut {NoStop}%
\bibitem [{\citenamefont {Hagiwara}\ \emph {et~al.}(2017)\citenamefont {Hagiwara}, \citenamefont {Hatta}, \citenamefont {Pasechnik}, \citenamefont {Tasevsky},\ and\ \citenamefont {Teryaev}}]{Hagiwara:2017fye}%
  \BibitemOpen
  \bibfield  {author} {\bibinfo {author} {\bibfnamefont {Y.}~\bibnamefont {Hagiwara}}, \bibinfo {author} {\bibfnamefont {Y.}~\bibnamefont {Hatta}}, \bibinfo {author} {\bibfnamefont {R.}~\bibnamefont {Pasechnik}}, \bibinfo {author} {\bibfnamefont {M.}~\bibnamefont {Tasevsky}}, \ and\ \bibinfo {author} {\bibfnamefont {O.}~\bibnamefont {Teryaev}},\ }\href {\doibase 10.1103/PhysRevD.96.034009} {\bibfield  {journal} {\bibinfo  {journal} {Phys. Rev. D}\ }\textbf {\bibinfo {volume} {96}},\ \bibinfo {pages} {034009} (\bibinfo {year} {2017})},\ \Eprint {http://arxiv.org/abs/1706.01765} {arXiv:1706.01765 [hep-ph]} \BibitemShut {NoStop}%
\bibitem [{\citenamefont {Reinke~Pelicer}\ \emph {et~al.}(2019)\citenamefont {Reinke~Pelicer}, \citenamefont {Gr\"ave De~Oliveira},\ and\ \citenamefont {Pasechnik}}]{ReinkePelicer:2018gyh}%
  \BibitemOpen
  \bibfield  {author} {\bibinfo {author} {\bibfnamefont {M.}~\bibnamefont {Reinke~Pelicer}}, \bibinfo {author} {\bibfnamefont {E.}~\bibnamefont {Gr\"ave De~Oliveira}}, \ and\ \bibinfo {author} {\bibfnamefont {R.}~\bibnamefont {Pasechnik}},\ }\href {\doibase 10.1103/PhysRevD.99.034016} {\bibfield  {journal} {\bibinfo  {journal} {Phys. Rev. D}\ }\textbf {\bibinfo {volume} {99}},\ \bibinfo {pages} {034016} (\bibinfo {year} {2019})},\ \Eprint {http://arxiv.org/abs/1811.12888} {arXiv:1811.12888 [hep-ph]} \BibitemShut {NoStop}%
\bibitem [{\citenamefont {Staszewski}(2023)}]{Staszewski:2023chz}%
  \BibitemOpen
  \bibfield  {author} {\bibinfo {author} {\bibfnamefont {R.}~\bibnamefont {Staszewski}},\ }in\ \href@noop {} {\emph {\bibinfo {booktitle} {{30th International Workshop on Deep-Inelastic Scattering and Related Subjects}}}}\ (\bibinfo {year} {2023})\ \Eprint {http://arxiv.org/abs/2309.02097} {arXiv:2309.02097 [hep-ph]} \BibitemShut {NoStop}%
\bibitem [{\citenamefont {Trzebinski}(2019)}]{Trzebinski:2019tmk}%
  \BibitemOpen
  \bibfield  {author} {\bibinfo {author} {\bibfnamefont {M.}~\bibnamefont {Trzebinski}} (\bibinfo {collaboration} {ATLAS}),\ }in\ \href@noop {} {\emph {\bibinfo {booktitle} {{International Conference on the Structure and the Interactions of the Photon}}}}\ (\bibinfo {year} {2019})\ pp.\ \bibinfo {pages} {144--149},\ \Eprint {http://arxiv.org/abs/1909.10827} {arXiv:1909.10827 [physics.ins-det]} \BibitemShut {NoStop}%
\bibitem [{\citenamefont {Bossini}(2023)}]{Bossini:2023uwa}%
  \BibitemOpen
  \bibfield  {author} {\bibinfo {author} {\bibfnamefont {E.}~\bibnamefont {Bossini}} (\bibinfo {collaboration} {CMS, TOTEM}),\ }\href {\doibase 10.1016/j.nima.2022.167823} {\bibfield  {journal} {\bibinfo  {journal} {Nucl. Instrum. Meth. A}\ }\textbf {\bibinfo {volume} {1047}},\ \bibinfo {pages} {167823} (\bibinfo {year} {2023})}\BibitemShut {NoStop}%
\bibitem [{\citenamefont {Nemchik}\ \emph {et~al.}(1998)\citenamefont {Nemchik}, \citenamefont {Nikolaev}, \citenamefont {Predazzi}, \citenamefont {Zakharov},\ and\ \citenamefont {Zoller}}]{Nemchik:1997xb}%
  \BibitemOpen
  \bibfield  {author} {\bibinfo {author} {\bibfnamefont {J.}~\bibnamefont {Nemchik}}, \bibinfo {author} {\bibfnamefont {N.~N.}\ \bibnamefont {Nikolaev}}, \bibinfo {author} {\bibfnamefont {E.}~\bibnamefont {Predazzi}}, \bibinfo {author} {\bibfnamefont {B.~G.}\ \bibnamefont {Zakharov}}, \ and\ \bibinfo {author} {\bibfnamefont {V.~R.}\ \bibnamefont {Zoller}},\ }\href {\doibase 10.1134/1.558573} {\bibfield  {journal} {\bibinfo  {journal} {J. Exp. Theor. Phys.}\ }\textbf {\bibinfo {volume} {86}},\ \bibinfo {pages} {1054} (\bibinfo {year} {1998})},\ \Eprint {http://arxiv.org/abs/hep-ph/9712469} {arXiv:hep-ph/9712469} \BibitemShut {NoStop}%
\bibitem [{\citenamefont {Baur}\ \emph {et~al.}(2002)\citenamefont {Baur}, \citenamefont {Hencken}, \citenamefont {Trautmann}, \citenamefont {Sadovsky},\ and\ \citenamefont {Kharlov}}]{Baur:2001jj}%
  \BibitemOpen
  \bibfield  {author} {\bibinfo {author} {\bibfnamefont {G.}~\bibnamefont {Baur}}, \bibinfo {author} {\bibfnamefont {K.}~\bibnamefont {Hencken}}, \bibinfo {author} {\bibfnamefont {D.}~\bibnamefont {Trautmann}}, \bibinfo {author} {\bibfnamefont {S.}~\bibnamefont {Sadovsky}}, \ and\ \bibinfo {author} {\bibfnamefont {Y.}~\bibnamefont {Kharlov}},\ }\href {\doibase 10.1016/S0370-1573(01)00101-6} {\bibfield  {journal} {\bibinfo  {journal} {Phys. Rept.}\ }\textbf {\bibinfo {volume} {364}},\ \bibinfo {pages} {359} (\bibinfo {year} {2002})},\ \Eprint {http://arxiv.org/abs/hep-ph/0112211} {arXiv:hep-ph/0112211} \BibitemShut {NoStop}%
\bibitem [{\citenamefont {Kovchegov}\ and\ \citenamefont {Levin}(2012)}]{Kovchegov.2012}%
  \BibitemOpen
  \bibfield  {author} {\bibinfo {author} {\bibfnamefont {Y.~V.}\ \bibnamefont {Kovchegov}}\ and\ \bibinfo {author} {\bibfnamefont {E.}~\bibnamefont {Levin}},\ }\href {https://link.aps.org/doi/10.1103/PhysRevD.50.3134} {\bibfield  {journal} {\bibinfo  {journal} {Cambridge University Press}\ }\textbf {\bibinfo {volume} {33}},\ \bibinfo {pages} {ISBN:9780521112574} (\bibinfo {year} {2012})}\BibitemShut {NoStop}%
\bibitem [{\citenamefont {Barone}\ and\ \citenamefont {Predazzi}(2002)}]{Barone:2002cv}%
  \BibitemOpen
  \bibfield  {author} {\bibinfo {author} {\bibfnamefont {V.}~\bibnamefont {Barone}}\ and\ \bibinfo {author} {\bibfnamefont {E.}~\bibnamefont {Predazzi}},\ }\href@noop {} {\emph {\bibinfo {title} {{High-Energy Particle Diffraction}}}},\ \bibinfo {series} {Texts and Monographs in Physics}, Vol.\ \bibinfo {volume} {v.565}\ (\bibinfo  {publisher} {Springer-Verlag},\ \bibinfo {address} {Berlin Heidelberg},\ \bibinfo {year} {2002})\BibitemShut {NoStop}%
\bibitem [{\citenamefont {Nikolaev}\ and\ \citenamefont {Zakharov}(1994{\natexlab{b}})}]{Nikolaev:1994ce}%
  \BibitemOpen
  \bibfield  {author} {\bibinfo {author} {\bibfnamefont {N.~N.}\ \bibnamefont {Nikolaev}}\ and\ \bibinfo {author} {\bibfnamefont {B.~G.}\ \bibnamefont {Zakharov}},\ }\href {\doibase 10.1016/0370-2693(94)90877-X} {\bibfield  {journal} {\bibinfo  {journal} {Phys. Lett. B}\ }\textbf {\bibinfo {volume} {332}},\ \bibinfo {pages} {184} (\bibinfo {year} {1994}{\natexlab{b}})},\ \Eprint {http://arxiv.org/abs/hep-ph/9403243} {arXiv:hep-ph/9403243} \BibitemShut {NoStop}%
\bibitem [{\citenamefont {Nikolaev}\ \emph {et~al.}(1999)\citenamefont {Nikolaev}, \citenamefont {Pronyaev},\ and\ \citenamefont {Zakharov}}]{Nikolaev:1998wf}%
  \BibitemOpen
  \bibfield  {author} {\bibinfo {author} {\bibfnamefont {N.~N.}\ \bibnamefont {Nikolaev}}, \bibinfo {author} {\bibfnamefont {A.~V.}\ \bibnamefont {Pronyaev}}, \ and\ \bibinfo {author} {\bibfnamefont {B.~G.}\ \bibnamefont {Zakharov}},\ }\href {\doibase 10.1103/PhysRevD.59.091501} {\bibfield  {journal} {\bibinfo  {journal} {Phys. Rev. D}\ }\textbf {\bibinfo {volume} {59}},\ \bibinfo {pages} {091501} (\bibinfo {year} {1999})},\ \Eprint {http://arxiv.org/abs/hep-ph/9812212} {arXiv:hep-ph/9812212} \BibitemShut {NoStop}%
\bibitem [{\citenamefont {Ivanov}\ \emph {et~al.}(2006)\citenamefont {Ivanov}, \citenamefont {Nikolaev},\ and\ \citenamefont {Savin}}]{Ivanov:2004ax}%
  \BibitemOpen
  \bibfield  {author} {\bibinfo {author} {\bibfnamefont {I.~P.}\ \bibnamefont {Ivanov}}, \bibinfo {author} {\bibfnamefont {N.~N.}\ \bibnamefont {Nikolaev}}, \ and\ \bibinfo {author} {\bibfnamefont {A.~A.}\ \bibnamefont {Savin}},\ }\href {\doibase 10.1134/S1063779606010011} {\bibfield  {journal} {\bibinfo  {journal} {Phys. Part. Nucl.}\ }\textbf {\bibinfo {volume} {37}},\ \bibinfo {pages} {1} (\bibinfo {year} {2006})},\ \Eprint {http://arxiv.org/abs/hep-ph/0501034} {arXiv:hep-ph/0501034} \BibitemShut {NoStop}%
\bibitem [{\citenamefont {Cisek}\ \emph {et~al.}(2023)\citenamefont {Cisek}, \citenamefont {Sch\"afer},\ and\ \citenamefont {Szczurek}}]{Cisek:2022yjj}%
  \BibitemOpen
  \bibfield  {author} {\bibinfo {author} {\bibfnamefont {A.}~\bibnamefont {Cisek}}, \bibinfo {author} {\bibfnamefont {W.}~\bibnamefont {Sch\"afer}}, \ and\ \bibinfo {author} {\bibfnamefont {A.}~\bibnamefont {Szczurek}},\ }\href {\doibase 10.1016/j.physletb.2022.137595} {\bibfield  {journal} {\bibinfo  {journal} {Phys. Lett. B}\ }\textbf {\bibinfo {volume} {836}},\ \bibinfo {pages} {137595} (\bibinfo {year} {2023})},\ \Eprint {http://arxiv.org/abs/2209.06578} {arXiv:2209.06578 [hep-ph]} \BibitemShut {NoStop}%
\bibitem [{\citenamefont {Golec-Biernat}\ and\ \citenamefont {W{\"u}sthoff}(1998)}]{Golec-Biernat:1998zce}%
  \BibitemOpen
  \bibfield  {author} {\bibinfo {author} {\bibfnamefont {K.~J.}\ \bibnamefont {Golec-Biernat}}\ and\ \bibinfo {author} {\bibfnamefont {M.}~\bibnamefont {W{\"u}sthoff}},\ }\href {\doibase 10.1103/PhysRevD.59.014017} {\bibfield  {journal} {\bibinfo  {journal} {Phys. Rev. D}\ }\textbf {\bibinfo {volume} {59}},\ \bibinfo {pages} {014017} (\bibinfo {year} {1998})},\ \Eprint {http://arxiv.org/abs/hep-ph/9807513} {arXiv:hep-ph/9807513} \BibitemShut {NoStop}%
\bibitem [{\citenamefont {Moriggi}\ \emph {et~al.}(2020)\citenamefont {Moriggi}, \citenamefont {Peccini},\ and\ \citenamefont {Machado}}]{Moriggi:2020zbv}%
  \BibitemOpen
  \bibfield  {author} {\bibinfo {author} {\bibfnamefont {L.~S.}\ \bibnamefont {Moriggi}}, \bibinfo {author} {\bibfnamefont {G.~M.}\ \bibnamefont {Peccini}}, \ and\ \bibinfo {author} {\bibfnamefont {M.~V.~T.}\ \bibnamefont {Machado}},\ }\href {\doibase 10.1103/PhysRevD.102.034016} {\bibfield  {journal} {\bibinfo  {journal} {Phys. Rev. D}\ }\textbf {\bibinfo {volume} {102}},\ \bibinfo {pages} {034016} (\bibinfo {year} {2020})},\ \Eprint {http://arxiv.org/abs/2005.07760} {arXiv:2005.07760 [hep-ph]} \BibitemShut {NoStop}%
\bibitem [{\citenamefont {Balitsky}(1996)}]{Balitsky:1995ub}%
  \BibitemOpen
  \bibfield  {author} {\bibinfo {author} {\bibfnamefont {I.}~\bibnamefont {Balitsky}},\ }\href {\doibase 10.1016/0550-3213(95)00638-9} {\bibfield  {journal} {\bibinfo  {journal} {Nucl. Phys. B}\ }\textbf {\bibinfo {volume} {463}},\ \bibinfo {pages} {99} (\bibinfo {year} {1996})},\ \Eprint {http://arxiv.org/abs/hep-ph/9509348} {arXiv:hep-ph/9509348} \BibitemShut {NoStop}%
\bibitem [{\citenamefont {Kovchegov}(1999)}]{Kovchegov:1999yj}%
  \BibitemOpen
  \bibfield  {author} {\bibinfo {author} {\bibfnamefont {Y.~V.}\ \bibnamefont {Kovchegov}},\ }\href {\doibase 10.1103/PhysRevD.60.034008} {\bibfield  {journal} {\bibinfo  {journal} {Phys. Rev. D}\ }\textbf {\bibinfo {volume} {60}},\ \bibinfo {pages} {034008} (\bibinfo {year} {1999})},\ \Eprint {http://arxiv.org/abs/hep-ph/9901281} {arXiv:hep-ph/9901281} \BibitemShut {NoStop}%
\bibitem [{\citenamefont {Golec-Biernat}\ and\ \citenamefont {Stasto}(2003)}]{Golec-Biernat:2003naj}%
  \BibitemOpen
  \bibfield  {author} {\bibinfo {author} {\bibfnamefont {K.~J.}\ \bibnamefont {Golec-Biernat}}\ and\ \bibinfo {author} {\bibfnamefont {A.~M.}\ \bibnamefont {Stasto}},\ }\href {\doibase 10.1016/j.nuclphysb.2003.07.011} {\bibfield  {journal} {\bibinfo  {journal} {Nucl. Phys. B}\ }\textbf {\bibinfo {volume} {668}},\ \bibinfo {pages} {345} (\bibinfo {year} {2003})},\ \Eprint {http://arxiv.org/abs/hep-ph/0306279} {arXiv:hep-ph/0306279} \BibitemShut {NoStop}%
\bibitem [{\citenamefont {Berger}\ and\ \citenamefont {Stasto}(2011{\natexlab{a}})}]{Berger:2010sh}%
  \BibitemOpen
  \bibfield  {author} {\bibinfo {author} {\bibfnamefont {J.}~\bibnamefont {Berger}}\ and\ \bibinfo {author} {\bibfnamefont {A.}~\bibnamefont {Stasto}},\ }\href {\doibase 10.1103/PhysRevD.83.034015} {\bibfield  {journal} {\bibinfo  {journal} {Phys. Rev. D}\ }\textbf {\bibinfo {volume} {83}},\ \bibinfo {pages} {034015} (\bibinfo {year} {2011}{\natexlab{a}})},\ \Eprint {http://arxiv.org/abs/1010.0671} {arXiv:1010.0671 [hep-ph]} \BibitemShut {NoStop}%
\bibitem [{\citenamefont {Berger}\ and\ \citenamefont {Stasto}(2011{\natexlab{b}})}]{Berger:2011ew}%
  \BibitemOpen
  \bibfield  {author} {\bibinfo {author} {\bibfnamefont {J.}~\bibnamefont {Berger}}\ and\ \bibinfo {author} {\bibfnamefont {A.~M.}\ \bibnamefont {Stasto}},\ }\href {\doibase 10.1103/PhysRevD.84.094022} {\bibfield  {journal} {\bibinfo  {journal} {Phys. Rev. D}\ }\textbf {\bibinfo {volume} {84}},\ \bibinfo {pages} {094022} (\bibinfo {year} {2011}{\natexlab{b}})},\ \Eprint {http://arxiv.org/abs/1106.5740} {arXiv:1106.5740 [hep-ph]} \BibitemShut {NoStop}%
\bibitem [{\citenamefont {Gubser}(2011)}]{Gubser:2011qva}%
  \BibitemOpen
  \bibfield  {author} {\bibinfo {author} {\bibfnamefont {S.~S.}\ \bibnamefont {Gubser}},\ }\href {\doibase 10.1103/PhysRevD.84.085024} {\bibfield  {journal} {\bibinfo  {journal} {Phys. Rev. D}\ }\textbf {\bibinfo {volume} {84}},\ \bibinfo {pages} {085024} (\bibinfo {year} {2011})},\ \Eprint {http://arxiv.org/abs/1102.4040} {arXiv:1102.4040 [hep-th]} \BibitemShut {NoStop}%
\bibitem [{\citenamefont {McLerran}\ and\ \citenamefont {Venugopalan}(1994)}]{McLerran:1993ka}%
  \BibitemOpen
  \bibfield  {author} {\bibinfo {author} {\bibfnamefont {L.~D.}\ \bibnamefont {McLerran}}\ and\ \bibinfo {author} {\bibfnamefont {R.}~\bibnamefont {Venugopalan}},\ }\href {\doibase 10.1103/PhysRevD.49.3352} {\bibfield  {journal} {\bibinfo  {journal} {Phys. Rev. D}\ }\textbf {\bibinfo {volume} {49}},\ \bibinfo {pages} {3352} (\bibinfo {year} {1994})},\ \Eprint {http://arxiv.org/abs/hep-ph/9311205} {arXiv:hep-ph/9311205} \BibitemShut {NoStop}%
\bibitem [{\citenamefont {Kowalski}\ and\ \citenamefont {Teaney}(2003)}]{Kowalski:2003hm}%
  \BibitemOpen
  \bibfield  {author} {\bibinfo {author} {\bibfnamefont {H.}~\bibnamefont {Kowalski}}\ and\ \bibinfo {author} {\bibfnamefont {D.}~\bibnamefont {Teaney}},\ }\href {\doibase 10.1103/PhysRevD.68.114005} {\bibfield  {journal} {\bibinfo  {journal} {Phys. Rev. D}\ }\textbf {\bibinfo {volume} {68}},\ \bibinfo {pages} {114005} (\bibinfo {year} {2003})},\ \Eprint {http://arxiv.org/abs/hep-ph/0304189} {arXiv:hep-ph/0304189} \BibitemShut {NoStop}%
\bibitem [{\citenamefont {M\"antysaari}\ \emph {et~al.}(2021)\citenamefont {M\"antysaari}, \citenamefont {Roy}, \citenamefont {Salazar},\ and\ \citenamefont {Schenke}}]{Mantysaari:2020lhf}%
  \BibitemOpen
  \bibfield  {author} {\bibinfo {author} {\bibfnamefont {H.}~\bibnamefont {M\"antysaari}}, \bibinfo {author} {\bibfnamefont {K.}~\bibnamefont {Roy}}, \bibinfo {author} {\bibfnamefont {F.}~\bibnamefont {Salazar}}, \ and\ \bibinfo {author} {\bibfnamefont {B.}~\bibnamefont {Schenke}},\ }\href {\doibase 10.1103/PhysRevD.103.094026} {\bibfield  {journal} {\bibinfo  {journal} {Phys. Rev. D}\ }\textbf {\bibinfo {volume} {103}},\ \bibinfo {pages} {094026} (\bibinfo {year} {2021})},\ \Eprint {http://arxiv.org/abs/2011.02464} {arXiv:2011.02464 [hep-ph]} \BibitemShut {NoStop}%
\bibitem [{\citenamefont {Gon\c{c}alves}\ \emph {et~al.}(2020)\citenamefont {Gon\c{c}alves}, \citenamefont {Sampaio~dos Santos},\ and\ \citenamefont {Sena}}]{Goncalves:2019jdl}%
  \BibitemOpen
  \bibfield  {author} {\bibinfo {author} {\bibfnamefont {V.~P.}\ \bibnamefont {Gon\c{c}alves}}, \bibinfo {author} {\bibfnamefont {G.}~\bibnamefont {Sampaio~dos Santos}}, \ and\ \bibinfo {author} {\bibfnamefont {C.~R.}\ \bibnamefont {Sena}},\ }\href {\doibase 10.1016/j.nuclphysa.2020.121862} {\bibfield  {journal} {\bibinfo  {journal} {Nucl. Phys. A}\ }\textbf {\bibinfo {volume} {1000}},\ \bibinfo {pages} {121862} (\bibinfo {year} {2020})},\ \Eprint {http://arxiv.org/abs/1911.03453} {arXiv:1911.03453 [hep-ph]} \BibitemShut {NoStop}%
\bibitem [{\citenamefont {Hatta}\ \emph {et~al.}(2021)\citenamefont {Hatta}, \citenamefont {Xiao}, \citenamefont {Yuan},\ and\ \citenamefont {Zhou}}]{Hatta:2021jcd}%
  \BibitemOpen
  \bibfield  {author} {\bibinfo {author} {\bibfnamefont {Y.}~\bibnamefont {Hatta}}, \bibinfo {author} {\bibfnamefont {B.-W.}\ \bibnamefont {Xiao}}, \bibinfo {author} {\bibfnamefont {F.}~\bibnamefont {Yuan}}, \ and\ \bibinfo {author} {\bibfnamefont {J.}~\bibnamefont {Zhou}},\ }\href {\doibase 10.1103/PhysRevD.104.054037} {\bibfield  {journal} {\bibinfo  {journal} {Phys. Rev. D}\ }\textbf {\bibinfo {volume} {104}},\ \bibinfo {pages} {054037} (\bibinfo {year} {2021})},\ \Eprint {http://arxiv.org/abs/2106.05307} {arXiv:2106.05307 [hep-ph]} \BibitemShut {NoStop}%
\bibitem [{\citenamefont {Aaij}\ \emph {et~al.}(2021)\citenamefont {Aaij} \emph {et~al.}}]{LHCb:2020frr}%
  \BibitemOpen
  \bibfield  {author} {\bibinfo {author} {\bibfnamefont {R.}~\bibnamefont {Aaij}} \emph {et~al.} (\bibinfo {collaboration} {LHCb}),\ }\href {\doibase 10.1007/JHEP02(2021)023} {\bibfield  {journal} {\bibinfo  {journal} {JHEP}\ }\textbf {\bibinfo {volume} {02}},\ \bibinfo {pages} {023} (\bibinfo {year} {2021})},\ \Eprint {http://arxiv.org/abs/2010.09437} {arXiv:2010.09437 [hep-ex]} \BibitemShut {NoStop}%
\bibitem [{\citenamefont {Maciula}\ and\ \citenamefont {Szczurek}(2023)}]{Maciula:2022lzk}%
  \BibitemOpen
  \bibfield  {author} {\bibinfo {author} {\bibfnamefont {R.}~\bibnamefont {Maciula}}\ and\ \bibinfo {author} {\bibfnamefont {A.}~\bibnamefont {Szczurek}},\ }\href {\doibase 10.1103/PhysRevD.107.034002} {\bibfield  {journal} {\bibinfo  {journal} {Phys. Rev. D}\ }\textbf {\bibinfo {volume} {107}},\ \bibinfo {pages} {034002} (\bibinfo {year} {2023})},\ \Eprint {http://arxiv.org/abs/2210.08890} {arXiv:2210.08890 [hep-ph]} \BibitemShut {NoStop}%
\end{thebibliography}%
\end{document}